\documentclass{aa}
\usepackage{hyperref}
\hypersetup{colorlinks,linkcolor={blue},citecolor={Green},urlcolor={blue}}
\usepackage[dvipsnames]{xcolor}
\usepackage{txfonts}
\usepackage{times}
\usepackage{caption}
\usepackage{natbib}

\begin{document}

    \title{How does the solar chromospheric activity look like under different inclination angles?}

    \author{G. Vanden Broeck\inst{1,2} \and S. Bechet\inst{1} \and G. Rauw\inst{2} \and F. Clette\inst{3}}

    \institute{Department of Solar Physics and Space Weather, Royal Observatory of Belgium (ROB), Av. Circulaire 3, 1180 Uccle, Belgium \\ email: \href{mailto:gregory.vandenbroeck@observatory.be}{gregory.vandenbroeck@observatory.be}
    \and Space sciences, Technologies and Astrophysics Research (STAR) Institute, Université de Liège, Allée du 6 Août, 19c, Bât B5c, 4000 Liège, Belgium
    \and Institut d’Astronomie et d’Astrophysique (IAA), Université Libre de Bruxelles, CP 226, Boulevard du Triomphe, 1050 Bruxelles, Belgium}


    \abstract
    {Chromospheric observations in the Ca {\sc ii} lines are essential to study the magnetic activity of stars. The chromospheric plages, main contributors to the Ca {\sc ii} K emission, are distributed between mid-latitude and the Equator and never close to the Poles. Therefore, we suspect that the inclination angle of the solar rotation axis has an impact on the observable chromospheric emission. Until now, the effect of such inclination on chromospheric emission has not been extensively studied through direct solar observations.}
    {We reproduce the solar images from any inclination in order to study the effect of the inclination axis on the solar variability by using direct observations of the Sun in the Ca {\sc ii} K line. In the context of the solar/stellar connection, while the Sun is observed from Earth from its near-Equator point of view and the others stars are observed most of the time under unknown inclinations, our results can improve our understanding of the magnetic activity of other solar-type stars.}
    {More than 2700 days of observations since the beginning of the Ca {\sc ii} K observations with USET, in July 2012, were used in our analysis. For each observation day, we produce synoptic maps to map the entire solar surface during a full solar rotation. Then by choosing a given inclination, we generate solar-disk views, representing the segmented brightest structures of the chromosphere (plages and enhanced network), as seen under this inclination. The area fraction are extracted from the masks for each inclination and we compare the evolution of those time series to quantify the impact of the inclination angle.}
    {We find a variation of the area fraction between an Equator-on view and a Pole-on view. Our results show an important impact of the viewing angle on the detection of modulation due to the solar rotation. With the dense temporal sampling of USET data, the solar rotation is detectable up to an inclination of about $|i| = 70^{\circ}$ and the solar-cycle modulation is clearly detected for all inclinations, though with a reduced amplitude in polar views. When applying a sparse temporal sampling typical for time series of solar-like stars, the rotational modulation is no longer detected, whatever the inclination, due entirely to the under-sampling. On the other hand, we find that the activity-cycle modulation remains detectable, even for Pole-on inclinations, as long as the sampling contains at least 20 observations per year and the cycle amplitude reaches at least 30\% of the solar-cycle amplitude.}
    {The inclination of the rotation axis of stars relative to our line of sight is most of the time unknown. Based on solar observations, we have shown that the impact of this inclination is important on the detection of the rotation period but negligible on the detection of the activity cycle period. For other stars, the time series have usually more complicated and scarcer samplings due to restricted target visibility and this leads to a decrease of the signal of the chromospheric activity cycle. However, our results suggest that the inclination is unlikely to be the primary factor contributing to the relative scarcity of well-established cycles.}

    \maketitle
%

\section{Introduction}\label{sec:introduction}

    Chromospheric activity of a large number of stars is monitored in the Ca {\sc ii} K and H lines \citep{Radick-2018, BoroSaikia-2018, Mittag-2023}. Over the past sixty years, significant progress has been made in the study of cool stellar chromospheres. It started in 1966 with the Mount Wilson HK project \citep{Wilson-1978}, recording chromospheric measurements of approximately 2.000 Sun-like stars. Understanding how the Sun compares to Sun-like stars in terms of variability and magnetic activity cycles can provide profound insights into the mechanisms driving stellar magnetic activity. The Mount Wilson HK project led to the creation of the Mount Wilson S-index, defined as the ratio of the total flux in the Ca {\sc ii} K and H line cores to the total flux in two pseudo-continuum regions located near the K and H lines. This index is widely used to study the magnetic activity of stars. In particular the stellar cycle and rotation can be calculated based on the detection of temporal modulations in the time series of the S-index \citep{Hempelman-2016}. For instance, the Mount Wilson monitoring program demonstrated that the Sun is not unique in exhibiting periodic activity cycles \citep{Baliunas-1998}: such behavior is common among solar-like stars (60\% of the Wilson's sample stars). However, some studies have demonstrated that most Sun-like stars (with age, mass, temperature and chromospheric activity similar to the Sun) exhibit different photometric variability on the activity cycle timescale than the Sun \citep{Lockwood-1990, Lockwood-2007, Reinhold-2020}. The impact of the viewing angle on the solar irradiance variability was first proposed by \cite{Schatten-1993}. Indeed, the magnetic structures are distributed between the Equator and mid-latitudes and the position of the observer relative to the rotation axis affects their brightness contrasts. While stars are usually observed without information on their inclinations, it is crucial to quantitatively evaluate this dependence to understand how solar variability is comparable to other Sun-like stars.
    
    The rotation periods of stars were successfully estimated from S-index time series, but no correction for the effect of inclination on the observed level of variability was applied \citep{Vaughan-1981, Noyes-1984, Baliunas-1985, Wright-2004}. More recently, \cite{Radick-2018} studied the variation of the chromospheric emission of the Sun and other Sun-like stars but they also did not apply a correction for the effect of inclination. The impact of the viewing angle on the observable magnetic activity of the Sun has been mainly studied based on synthetic images of the Sun obtained with rather simplified models and numerical simulations \citep{Schatten-1993, Knaack-2001, Shapiro-2014, Borgniet-2015, Meunier-2019, Meunier-2024, Nemec-2020, Sowmya-2021_I}. For example, \cite{Shapiro-2014}, \cite{Knaack-2001} and \cite{Sowmya-2021_I} analysed the influence of the inclination on solar irradiance and chromospheric activity by using physics-based model to calculate relative flux variations. \cite{Borgniet-2015} and \cite{Meunier-2019} presented similar studies but using times series of radial velocity, chromospheric emission and photometry to analyse the effects of activity (via spots, faculae and inhibition of convective blueshift) on exoplanet detectability. Those studies showed through simulations of different inclinations that the inclination of the stellar rotation axis has a strong impact on those time series. Here we propose to study this impact on the determination of the temporal modulations through real solar observations in the Ca {\sc ii} K line.

    Ground-based solar observations allow to see the surface of one side of the Sun, the far-side of the Sun being not observable. However, based on images recorded during a full solar rotation, we can map the entire solar surface into a synoptic map. Then by an appropriate projection, we can generate images of the Sun for various viewing angles. Finally we can build time series for different inclinations and study the impact of inclination axis on the detection of modulations. More specifically we consider the effect on the time series of the plages and enhanced network area fraction in the Ca {\sc ii} K line. 
    
\begin{figure}[t]
        \centering
        \includegraphics[width=0.9\hsize]{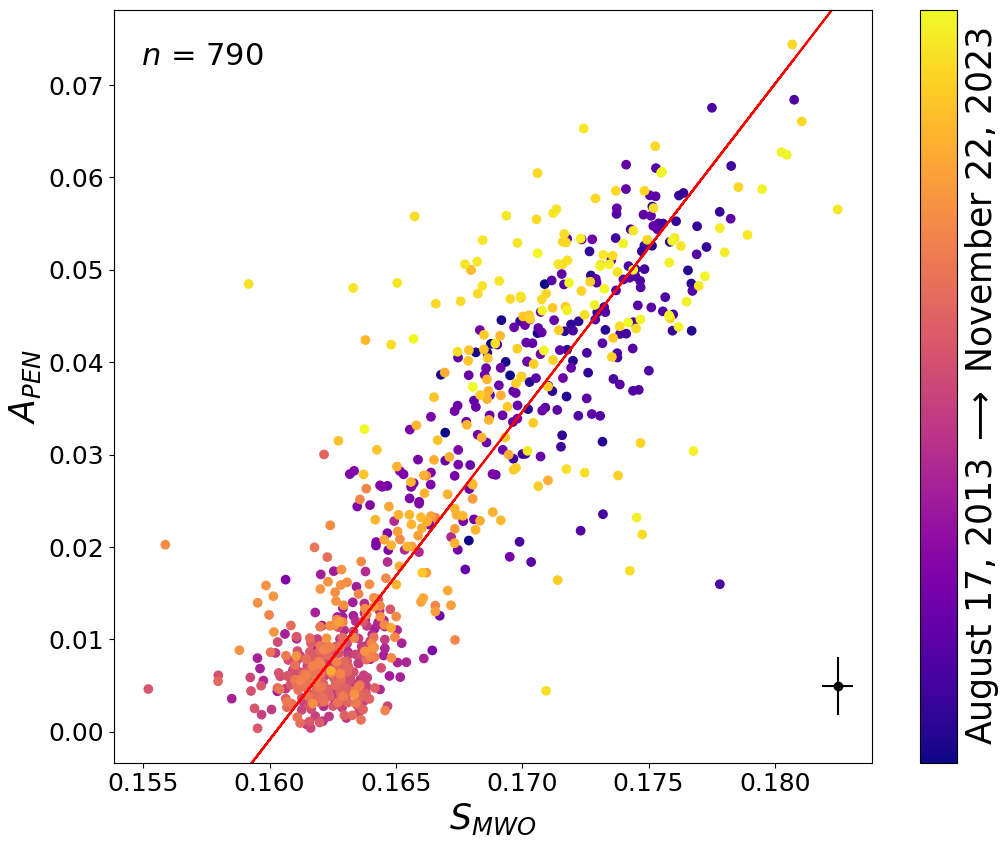}
        \caption{Correlation between daily values of the solar S-index from TIGRE (in the Mount Wilson scale) and the USET area fraction of plages and enhanced network, $A_{PEN}$. The parameter $n$ represents the number of data, and the chronological order is color-coded. The first and last date of the data are given next to the color bar. Additionally, a linear fit was performed to the data (red solid line). A mean error bar is displayed on the bottom right corner to have an idea of the uncertainties on the data. Image taken from \cite{VandenBroeck-2024}.}
        \label{fig:USET_TIGRE}
    \end{figure}

    Indeed it has been shown that those structures are the main cause of the modulation in the Ca {\sc ii} K emission and their area fraction is a good proxy for the S-index. Using Ca {\sc ii} K images with USET (Uccle Solar Equatorial Table) and S-index collected from TIGRE (Telescopio Internacional de Guanajuato Robotico Espectroscopico), \cite{VandenBroeck-2024} compared the area fraction of plages and enhanced network with the S-index based on data obtained over the same time interval and for observations made on the same dates. As seen in Fig. \ref{fig:USET_TIGRE}, they found a linear correlation between the indices, described by the following equation:
    
    \begin{equation}
        \centering
        A_{PEN} = (3.55 \pm 0.06) \ S_{\textrm{MWO}} - (0.57 \pm 0.01)
        \label{eq:USET-TIGRE-equation}
    \end{equation} 
    where $A_{PEN}$ is the area fraction of the plages and enhanced network from USET images, and $S_{\textrm{MWO}}$ the solar S-index from TIGRE in the Mt. Wilson scale. Since a well-defined linear transformation was found to convert the S-index from TIGRE to the Mount Wilson scale \citep{Mittag-2016}, the area fraction of plages and enhanced network offers a reliable diagnostic of chromospheric activity. This area fraction can then be used to generalize the results obtained for the Sun to the population of Sun-like stars. 
       
    Section \ref{sec:dataset} briefly describes the instrument and data used for our analysis. In section \ref{sec:data_processing}, we explain the method of data processing: the segmentation of the chromospheric structures, the creation of the synoptic maps and the production of the solar-disk views under different inclination angles. Our results about the detection of the temporal modulation for various inclinations are presented in section \ref{sec:temporal_modulations}. Finally, we summarize and discuss our results in section \ref{sec:Conclusions}.

\section{Dataset}\label{sec:dataset}

    In this study we use the synoptic images acquired by the USET station (Uccle Solar Equatorial Table) from the Royal Observatory of Belgium (ROB), located in Uccle, South of Brussels \citep{2023-USET}. Synoptic images are appropriate for this study as the full-disk images cover the whole surface of the Sun which is needed to map the entire solar surface. In particular we consider the full-disk daily images in the Ca {\sc ii} K line. The dataset covers a long time period, from July 11, 2012 to November 28, 2023, thus spanning 11.38 years, which is requested to search for long temporal modulations. After performing an automatic quality selection to keep the best image recorded per day \citep{VandenBroeck-2024}, the resulting series consists of 2725 images, covering more than 70\% of the analysed period.
    
    In addition, the station carries three other solar telescopes (White-Light, 656.3 nm H-alpha and sunspot drawings) to monitor simultaneously the photosphere and the chromosphere. The averaged total number of observation days is 260 per year. The gaps are essentially due to bad weather conditions.
        
    The optical set-up consists in a refractor of 925 mm focal length and 132 mm aperture. The filter is thermo-regulated and its central wavelength is $\lambda = 3933.67 \AA$, with a bandwidth of 2.7$\AA$. The images are acquired with a $2048 \times 2048$ CCD with a dynamic range of 12 bits. The instrumental set-up is the same since 2012, except for the introduction of an additional neutral filter on July 10, 2013. The acquisition cadence can go up to 4 frames per second in case of transient events to record, and the daily synoptic cadence is 15 minutes.

\section{Data processing}\label{sec:data_processing}

    After correcting the raw USET Ca {\sc ii} K images for instrumental and atmospheric effects, we segment the brightest chromospheric features to generate binary masks. Then, using successive daily disk masks, we assemble synoptic maps covering each a full Carrington rotation, thus mapping the whole solar surface. We can then apply a spherical projection to those synthetic maps to create solar-disk views reproducing the projected area of bright chromospheric structures as seen under different inclinations of the rotation axis relative to the line of sight. Finally, by summing those apparent areas over the full disk for each date, we create a time series of the total area fraction spanning the duration of the USET dataset.

    In the next sections, we define the inclination as the angle between the solar Equator and the observer's line of sight. The Equator-on view refers to an inclination $i = 0^{\circ}$ while the Poles-on view correspond to an inclinations $i = 90^{\circ}$ for the North Pole-on view and $i = -90^{\circ}$ for the South Pole-on view.
    
    \subsection{Chromospheric structures segmentation}\label{subsec:segmentation}

\begin{figure}[t]
            \centering
            \includegraphics[width=0.49\hsize]{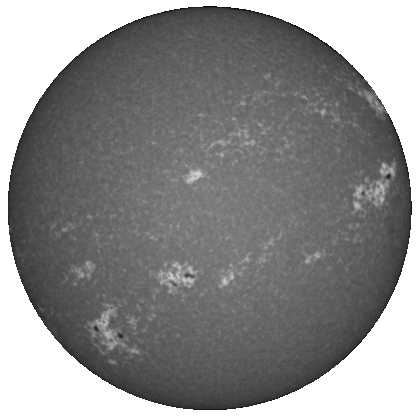}
            \includegraphics[width=0.49\hsize]{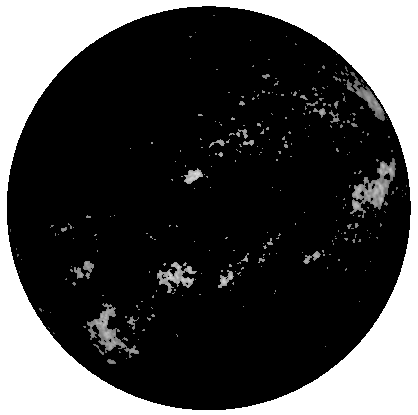}
            \caption{Example of the segmentation process. Left: recentered raw image from the 29th of October 2013. Right image: result of the segmentation.}
            \label{fig:Plages_segmentation}
        \end{figure}

        The solar chromosphere exhibits various structures that differ in size and brightness. In this study, we have segmented the brightest structures, namely the plages, which are the chromospheric counterparts of the faculae, and the enhanced network, considered as small regions of decaying plages \citep{2023-Singh}. More details on this segmentation method can be found in \cite{VandenBroeck-2024} and an example of this process is shown in Fig. \ref{fig:Plages_segmentation}.
    
    \subsection{Synoptic segmented map construction}\label{subsec:synmap_construction}
    
\begin{figure}[t]
            \centering
            \includegraphics[width=\hsize]{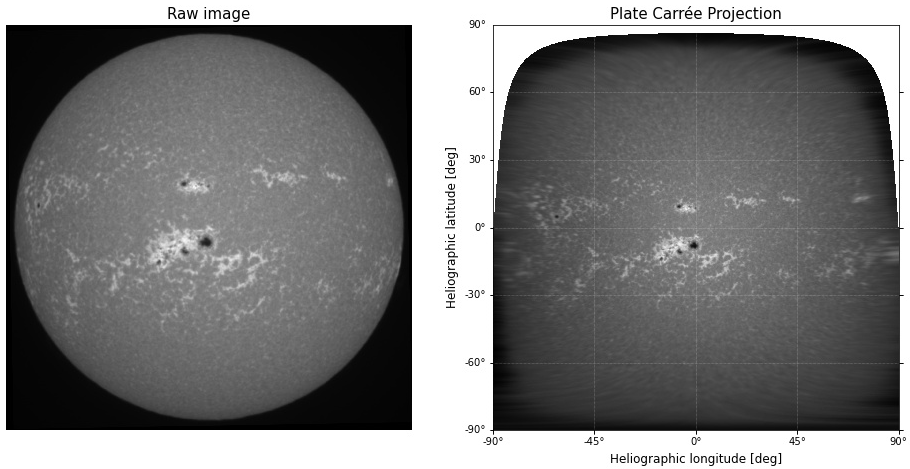}
            \caption{Example of a CAR projection. Left: Raw solar image. Right: The corresponding image after the projection.}
            \label{fig:Solar_projection}
        \end{figure}

        To map the entire solar surface, a projection is needed to convert spherical coordinates (latitude and longitude) into a flat representation. For this, we use the Plate Carrée (CAR) projection where the meridians and the parallels are equally spaced, forming a grid of squares from East to West and from North to South. With this projection, it is particularly easy to associate the heliographic coordinates of a structure on the Sun to its position in pixels on the map \citep{Calabretta-2002}. Fig. \ref{fig:Solar_projection} shows an example of the CAR projection of a solar image. The white part of the image is due to the fact that the rotation axis of the Sun is tilted with respect to the ecliptic. This tilt goes from $-7.25^{\circ}$ to $7.25^{\circ}$. Therefore, except when this inclination is equal to $0^{\circ}$, there is always a small hidden area around either the North or the South Pole. Those parts of the images are filled with zero values.
    
\begin{figure}[t]
            \includegraphics[width=\hsize]{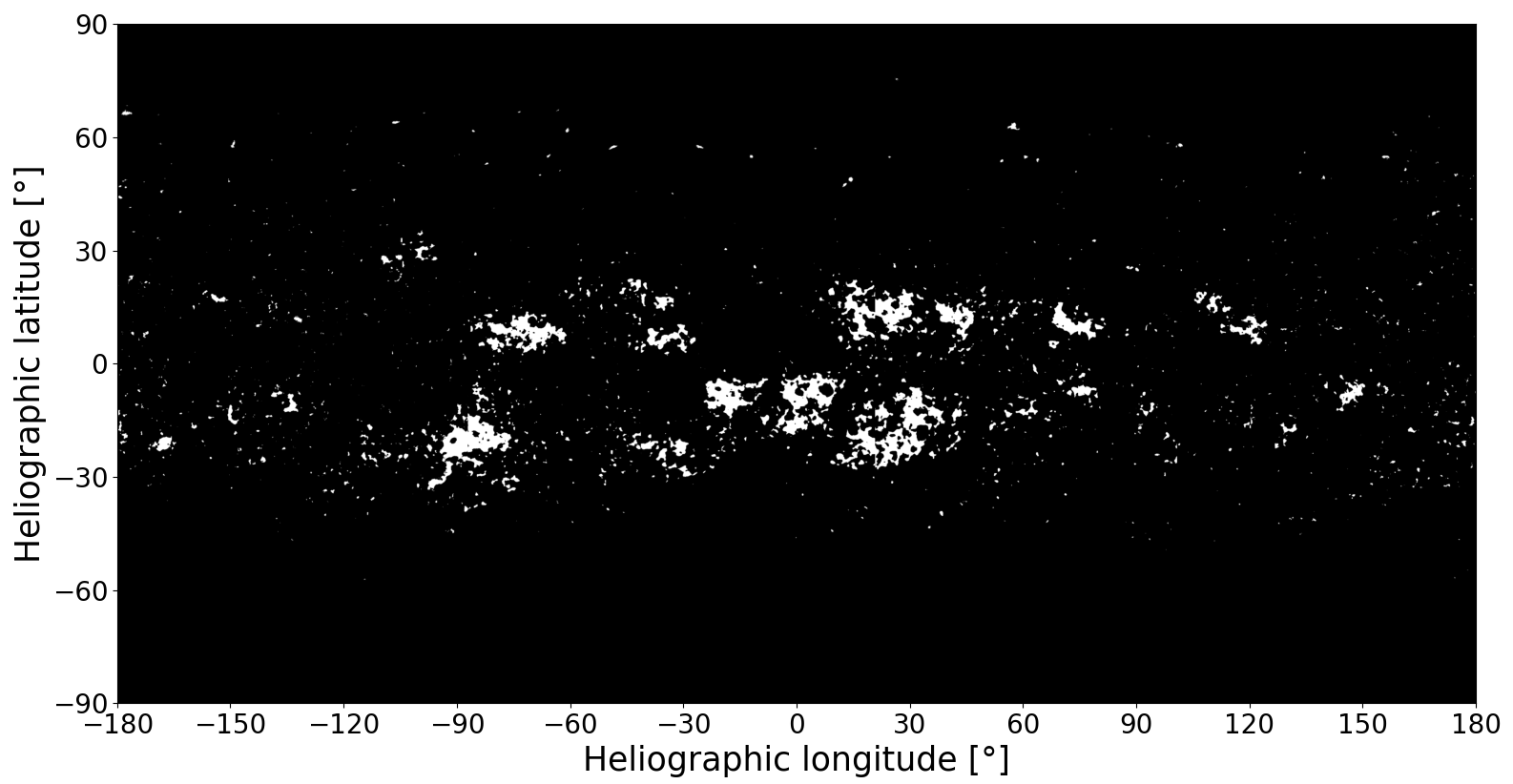}
            \caption{Example of a segmented synoptic map build around the 7th of July 2014 illustrating the distribution of the plages and enhanced network during a full solar rotation. The x-axis represents the number of degrees of longitudes from the Carrington longitude on 7th of July 2014 (center of the image).}
            \label{fig:Synoptic_map}
        \end{figure}

        A synoptic map is made of consecutive strips from successive solar images, spanning a whole solar rotation. In each image, the strip is centered on the central meridian and its pixel size corresponds to a time interval depending on the observing time of the previous and the following available images. This process is applied for each day of the dataset. Finally, by doing a calculation for the boundary conditions to build a map of $180^{\circ}$ before and $180^{\circ}$ after a given day, we get a synoptic map representing the entire surface of the Sun around this given date (see Fig. \ref{fig:Synoptic_map}).        

    \subsection{Reconstructed solar-disk views for different inclinations}\label{subsec:images_reconstruction}

\begin{figure}[t]
            \includegraphics[width=0.6\hsize]{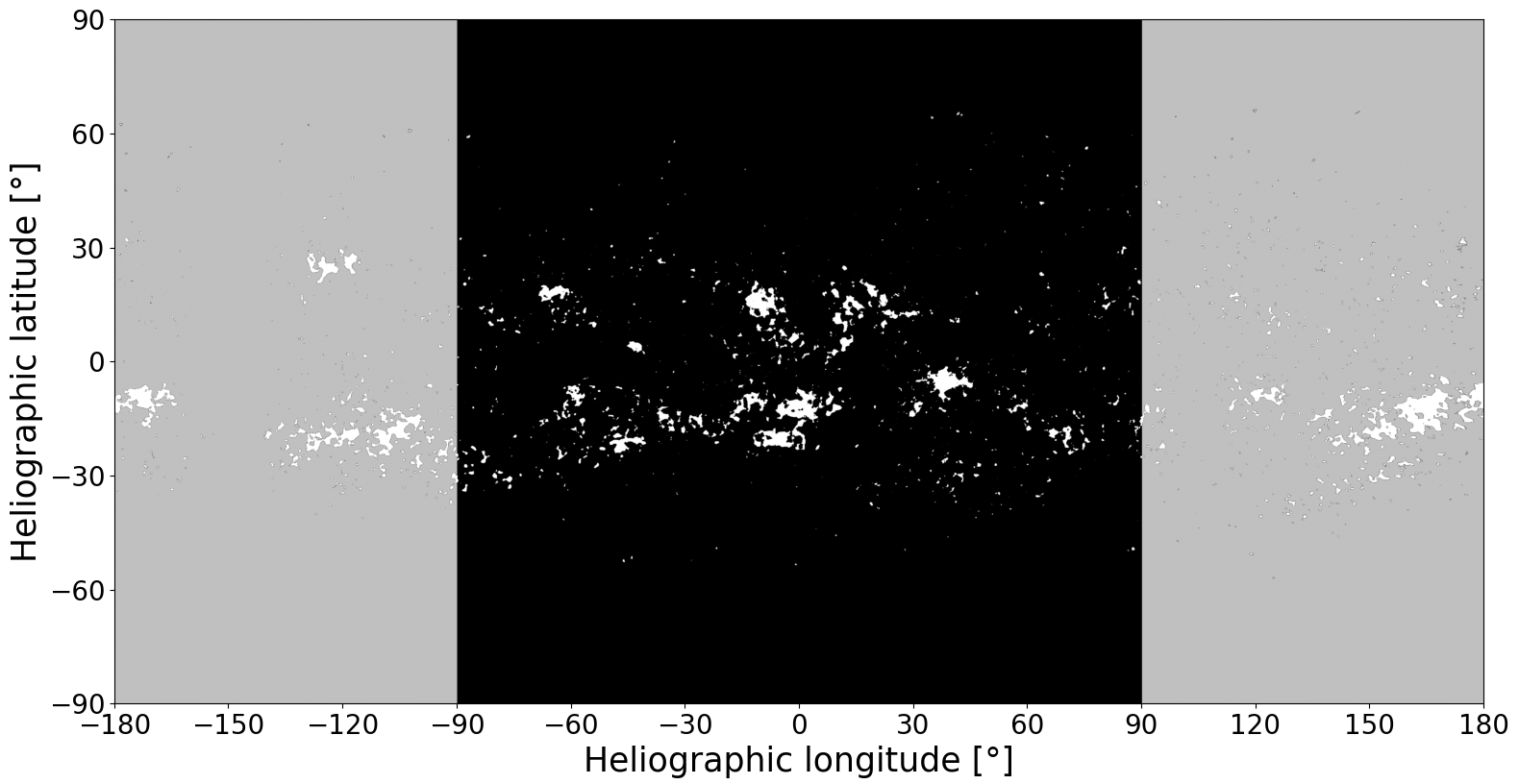}
            \includegraphics[width=0.32\hsize]{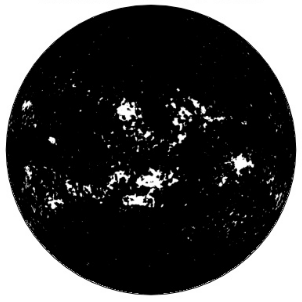}
            \put(-5,68){\small{$i = 0^{\circ}$}}
        
            \includegraphics[width=0.6\hsize]{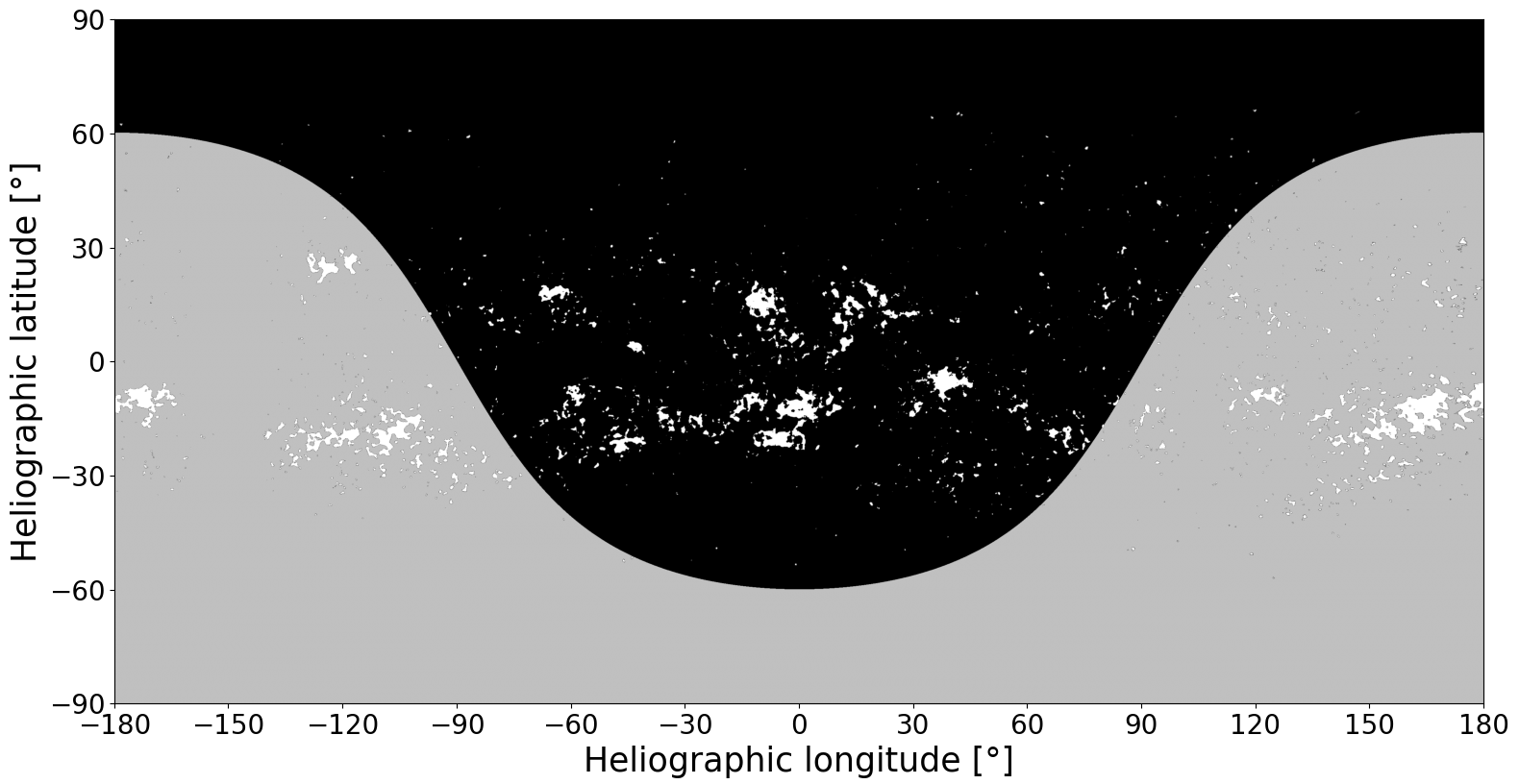}
            \includegraphics[width=0.32\hsize]{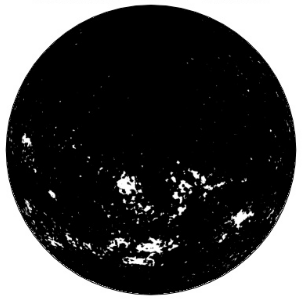}
            \put(-5,68){\small{$i = 30^{\circ}$}}
                            
            \includegraphics[width=0.6\hsize]{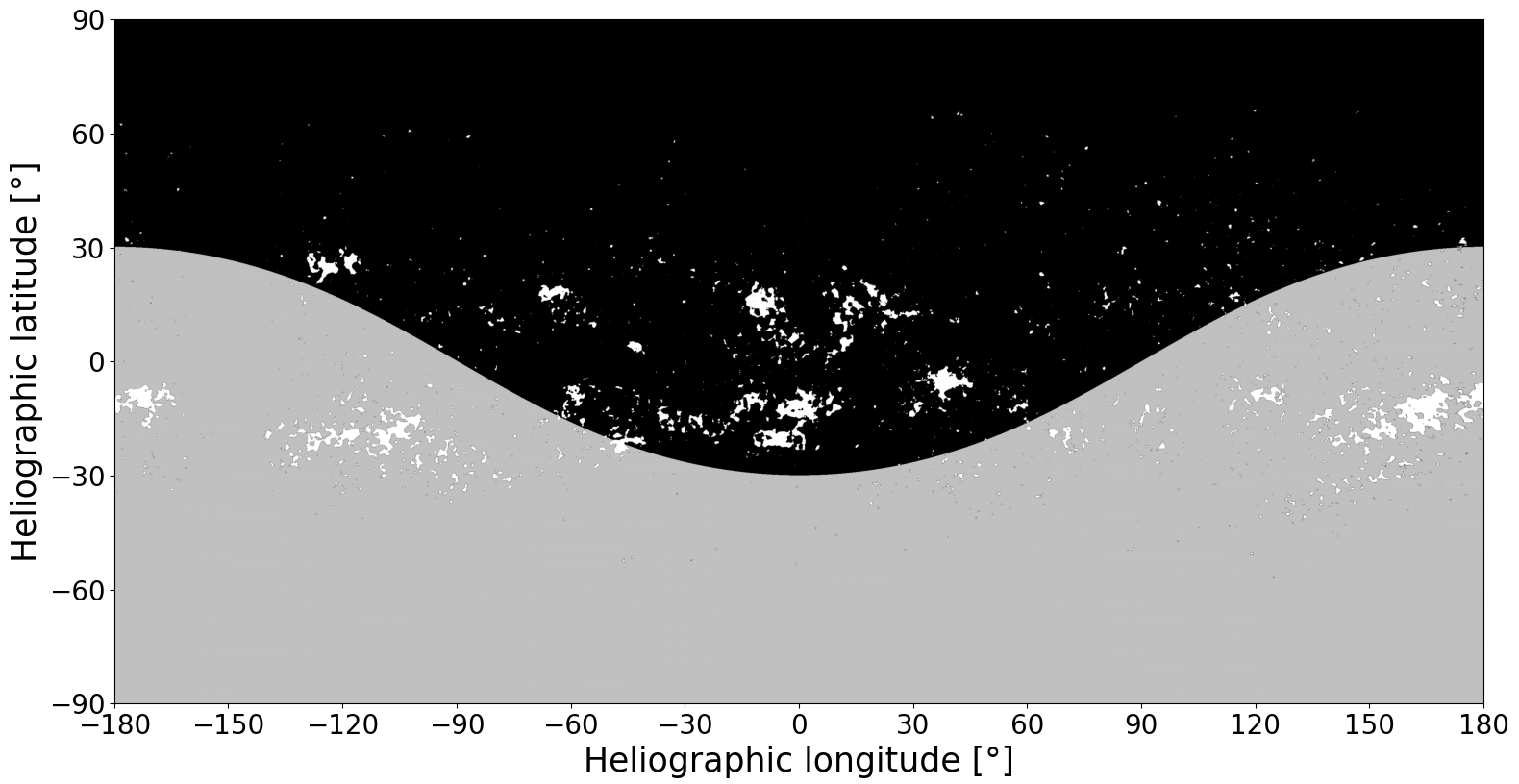}
            \includegraphics[width=0.32\hsize]{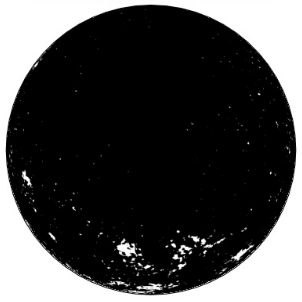}
            \put(-5,68){\small{$i = 60^{\circ}$}}

            \includegraphics[width=0.6\hsize]{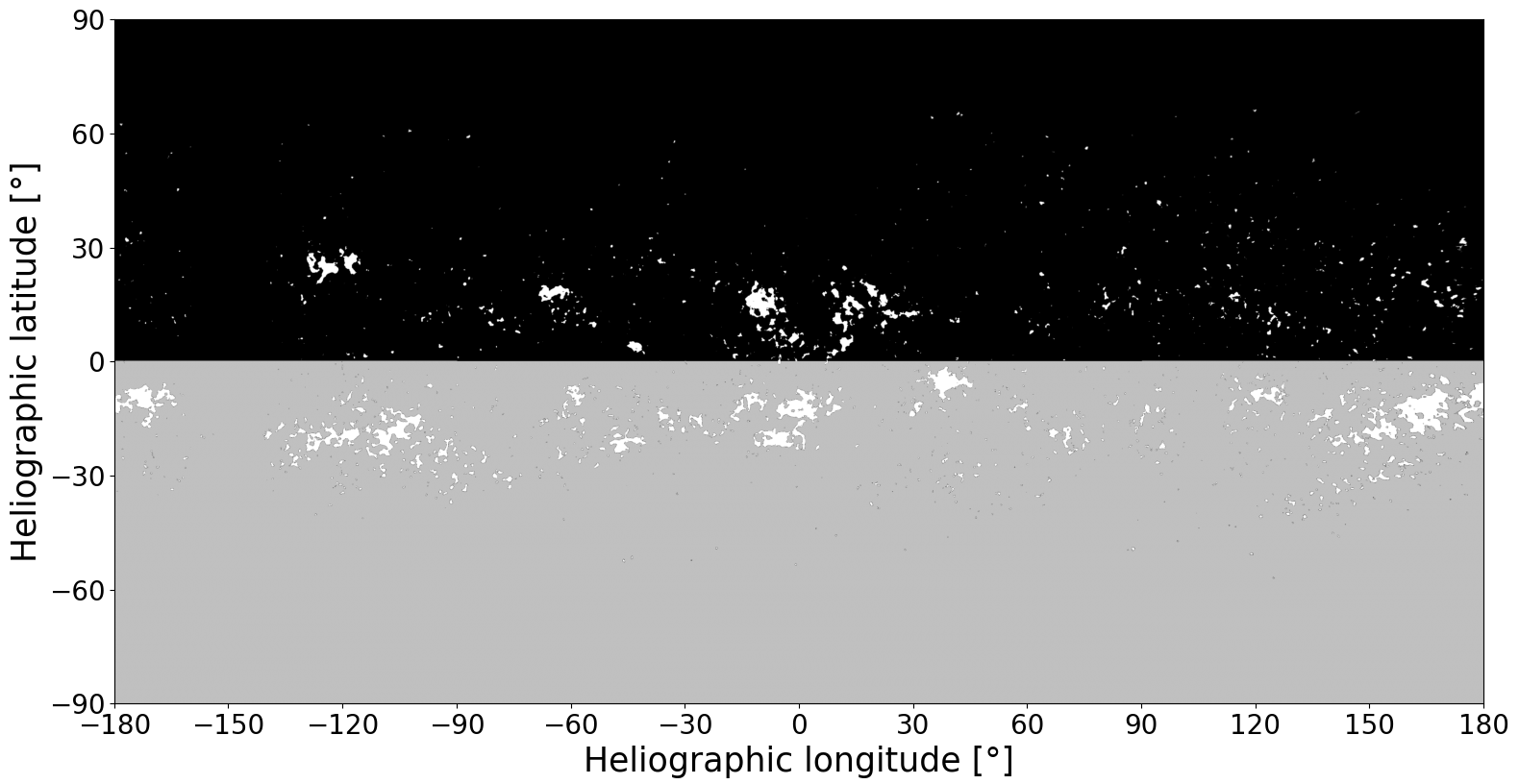}
            \includegraphics[width=0.32\hsize]{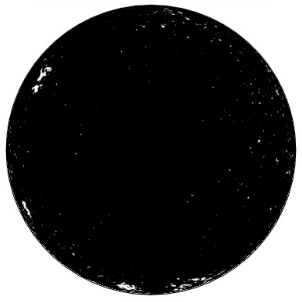}
            \put(-5,68){\small{$i = 90^{\circ}$}}
                
            \caption{Generation of solar-disk views for different inclination angles, indicated in the upper right corner of the panels and representing the number of degrees relative to the Equator-on view (inclination of $0^{\circ}$). Left: Synoptic map illustrating the distribution of the entire solar surface around the 8th of June 2014. The shaded areas (grey part) mark the far-side of the Sun. Right: The corresponding solar-disk. Inclination of $90^{\circ}$ corresponds to the Sun's North Pole-on view.}
            \label{fig:Solar_images_construction}
        \end{figure}

         For each observed date in the original USET series, using the corresponding whole-Sun segmented synoptic map, we create solar-disk views at different inclinations by using an orthographic projection with various centers. In particular we specify the central latitude and longitude in the orthographic function from the Cartopy library, designed for cartographic projection and geospatial data visualization \citep{Cartopy}. The generation of solar-disk views is illustrated in Fig. \ref{fig:Solar_images_construction} for inclinations $i$ of $0^{\circ}$, $30^{\circ}$, $60^{\circ}$ and $90^{\circ}$ (from the Equator-on view to the North Pole-on view). Appendix \ref{Annexe_A} provides the generated solar-disk views for all the inclinations from the North Pole-on view to the South Pole-on view. The gray areas on the synoptic maps in Fig. \ref{fig:Solar_images_construction} represent the far-side of the Sun. The chromospheric plages are distributed up to $50^{\circ}$ of latitude relative to the Equator \citep{Devi-2021}. Therefore, as we move to a Pole-on observation, those structures will be distributed closer to the limbs which makes them appear smaller.

    \subsection{Temporal evolution of the area fraction for different inclinations}\label{subsec:temporal_evolution}
    
\begin{figure}[t]
            \centering
            \includegraphics[width=\hsize]{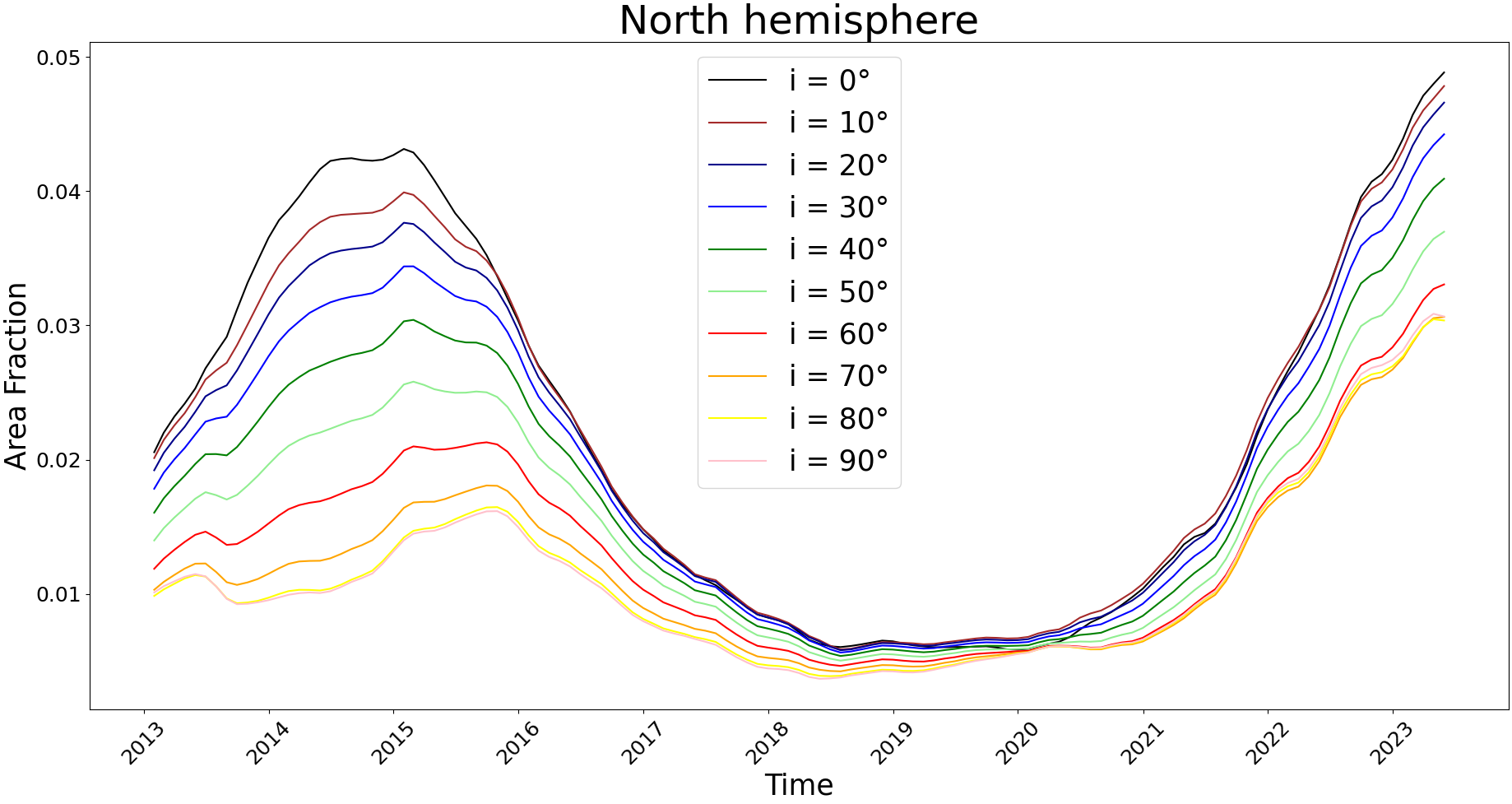}
            \includegraphics[width=\hsize]{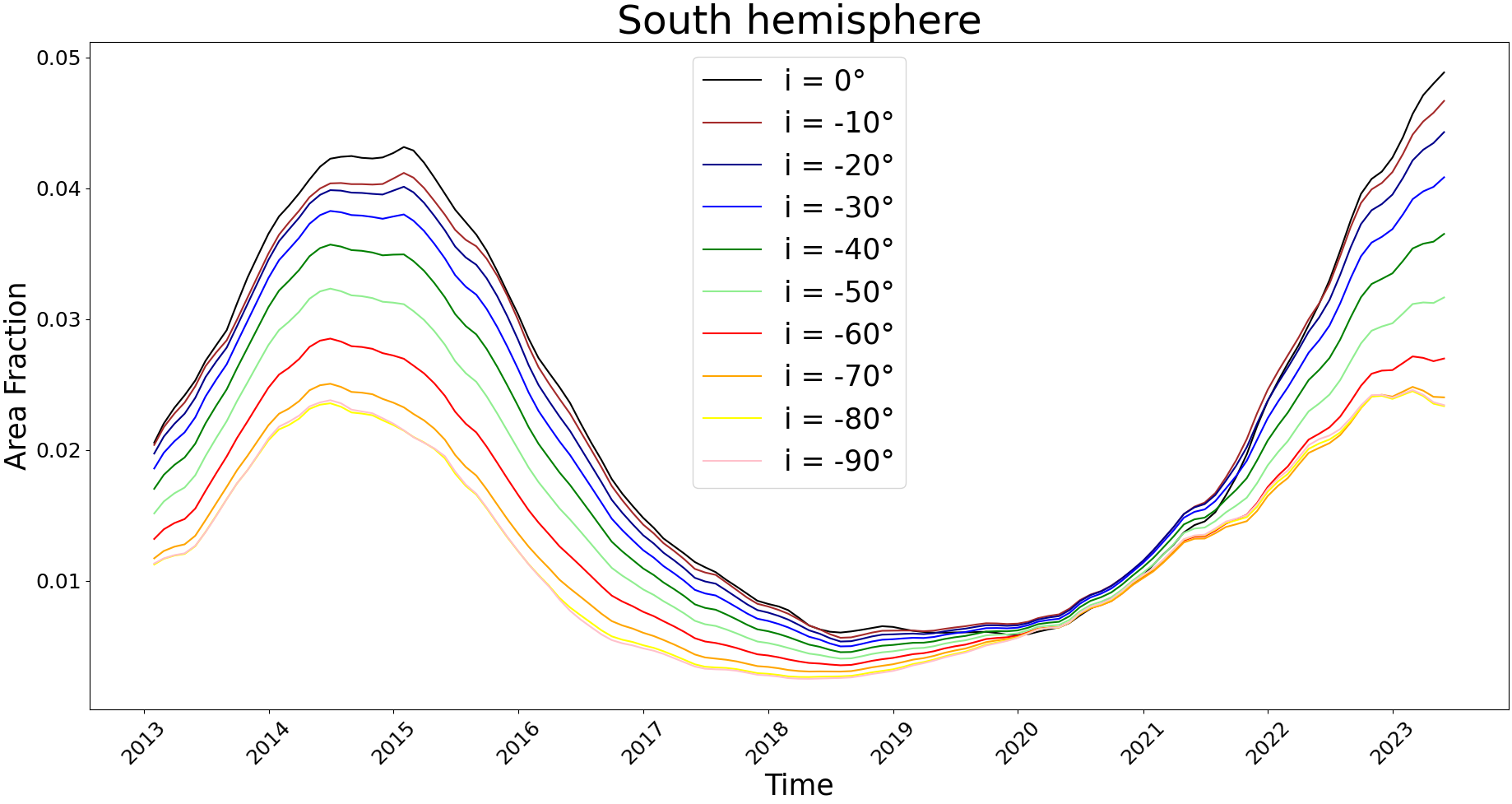}
            \caption{Evolution of plages and enhanced network area fraction for different inclinations. Top panel: Inclinations from the Equator-on view ($i = 0^{\circ}$ of latitude) to the North Pole-on view ($i = 90^{\circ}$ of latitude). Bottom panel: Inclinations from the Equator-on view to the South Pole-on view ($i = -90^{\circ}$ of latitude). The data are the monthly averaged data smoothed with a 13-months sliding window. The colors stand for different inclinations.}
            \label{fig:Plages_area_evolution_Inclinations}
        \end{figure}

        Based on those generated solar-disk views (Fig. \ref{fig:Solar_images_inclinations}), we build temporal series for each inclination by extracting the area fraction of the plages and enhanced network, called $A_{PEN}$ following the method described in \cite{VandenBroeck-2024}. For the whole USET dataset, we derive the area fraction for various inclinations from the Equator-on view ($i = 0^{\circ}$ of latitude) to the Pole-on view ($i = 90^{\circ}$ and $i = -90^{\circ}$) as represented in Fig. \ref{fig:Plages_area_evolution_Inclinations}. The plotted data are the monthly averaged data smoothed with a 13-months sliding window. There is an obvious lower peak-to-peak amplitude in the modulation of the area fraction when we move away from the Equator-on point of view, in both hemispheres. At the minimum of the solar cycle, the variation of $A_{PEN}$ between the Equator-on view and the Poles-on view is not significant. However, at the maximum, this variation is of 63\% between the Equator-on and North Pole-on views, and 45\% between the Equator-on and the South Pole-on views. This is related to the apparent projected area of magnetic structures that are reduced by the foreshortening as observed from the Poles. Despite the large change in amplitude at solar cycle maximum, the modulation associated with the solar cycle remains observable for both the North and South Pole-on views. In the next section, we will analyse quantitatively the impact of this variation on the detection of the temporal modulations.
        
\begin{figure}[t]
            \centering
            \includegraphics[width=\hsize]{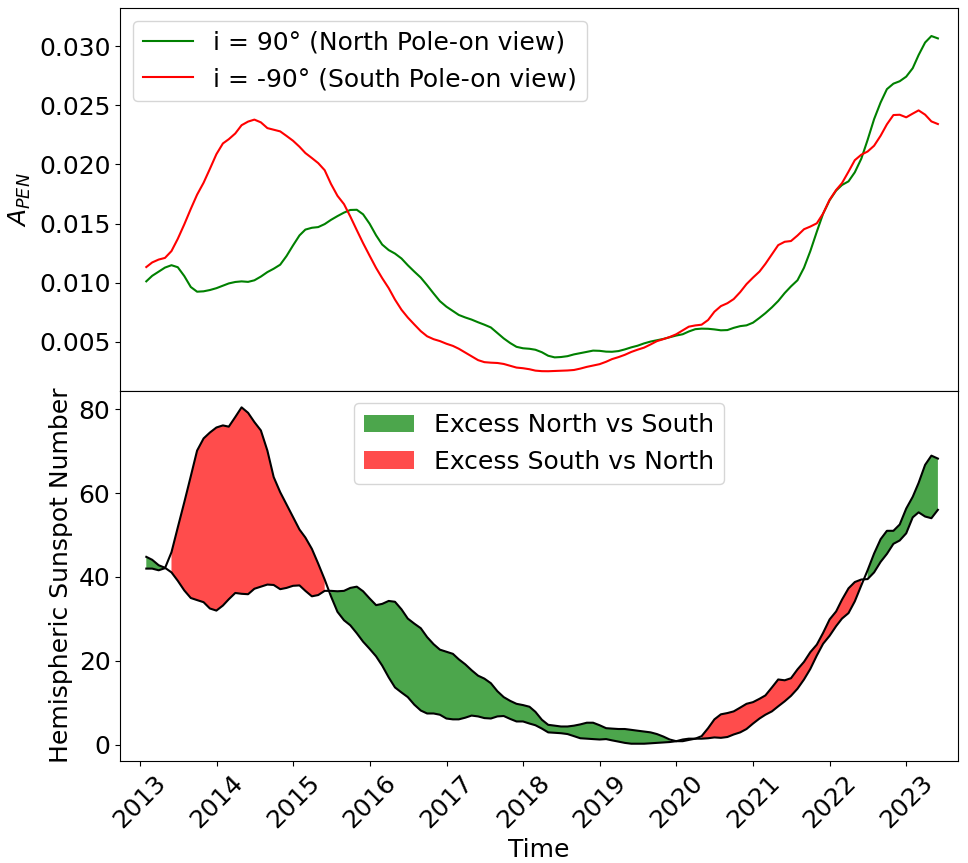}
            \caption{Top panel: Comparison between the area fraction of plages and enhanced network, $A_{PEN}$, seen from the Northern hemisphere (green) and from the Southern hemisphere (red). The curves are the monthly averaged data smoothed with a 13-month sliding window. Bottom panel: International Sunspot Number, hemispheric 13-month smoothed number. Green parts represent an excess of activity in the Northern hemisphere while red parts represent an excess in the Southern hemisphere. Credit: \href{https://sidc.be/SILSO/home}{SILSO} (Royal Observatory of Belgium).}
            \label{fig:Plages_area_evolution_North_vs_South_&_SN_hemispheric}
        \end{figure}

        In addition, we observe that the curves for the two hemispheres peak at different times. Top panel of Fig. \ref{fig:Plages_area_evolution_North_vs_South_&_SN_hemispheric} shows the $A_{PEN}$ time series for the polar view for each hemisphere. For the Southern hemisphere (red curve), the solar maximum is clearly visible around mid-2014 while for the Northern hemisphere, the solar maximum happens later, at the end of 2015. It is well known that the solar activity presents significant asymmetries. It has been studied in detail in a variety of observations and activity indices, such as sunspot groups and areas, sunspot numbers, but also on plages and flare occurrence \citep{El-Borie-2021, Veronig-2021}. The panels in Fig. \ref{fig:Plages_area_evolution_North_vs_South_&_SN_hemispheric} indeed show that, when separating the solar hemispheres, our time series of plages and enhanced network area fraction follow the same long-term evolution as the photospheric Sunspot Number, with the same variations of the North-South asymmetry. In section \ref{sec:temporal_modulations}, we will see that a corresponding asymmetry is naturally also found in the effects of inclination on our solar reconstructions.
     
\section{Detection of temporal modulations out of the ecliptic}\label{sec:temporal_modulations}    
    
    In this section we analyse the time series of $A_{PEN}$ for various inclinations more quantitatively. In particular we use Fourier power spectra to look for the presence of periodic modulations on the solar cycle and solar rotation timescale. The discrete Fourier power spectrum method of \cite{Heck-1985} and \cite{Gosset-2001} was applied to each of the time series extracted for the 19 values of inclinations from $i = -90^{\circ}$ to $i = +90^{\circ}$ in steps of $10^{\circ}$. This Fourier methodology explicitly accounts for the uneven temporal sampling of astronomical time series such as those analysed here. To assess the significance level of the peaks in the power spectrum, we used a bootstrapping method where the times of observations were kept fixed and the measured area fractions were redistributed randomly among the times of observations. For each reshuffled time series we computed a power spectrum and determined the highest value of the power. The reshuffling process was repeated a thousand times for each inclination. The distribution of the highest peaks in the power spectra was used to determine the threshold corresponding to a 99\% significance level, that is, 1\% of the power spectra of the reshuffled time series have a higher power than this threshold.
    
    \begin{figure}[t]
        \centering
        \includegraphics[width=\hsize]{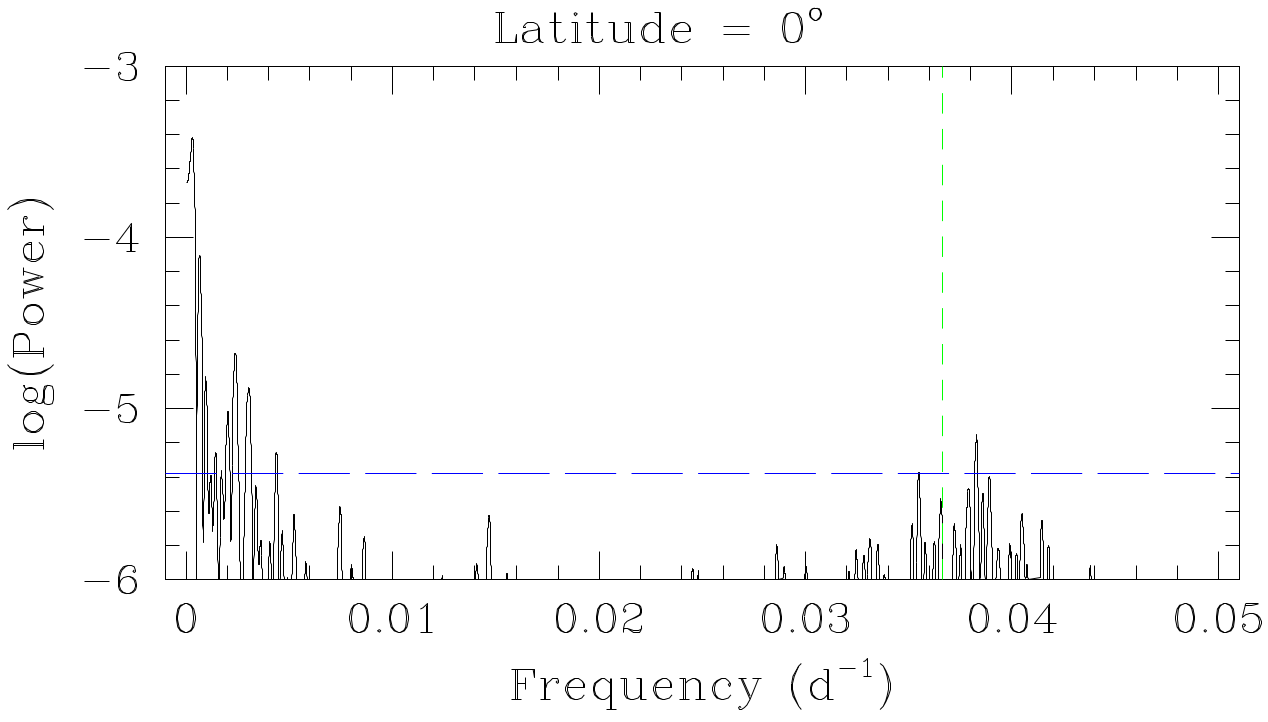}
        \caption{Fourier power spectrum of the fractional area of the plages and enhanced network in the case of an inclination $i = 0^{\circ}$ (Equator-on view). The panel illustrates the power spectrum over the frequency range between 0 and 0.05 d$^{-1}$. The green dashed line yields $\nu_{\rm Car}$, the frequency associated with the Carrington rotation period. The long-dashed blue horizontal line yields the 99\% significance level.}
        \label{fig:power0deg}
    \end{figure}
        
    Fig. \ref{fig:power0deg} illustrates the Fourier power spectrum for an inclination of $i = 0^{\circ}$. The Fourier power spectra for the other inclinations are provided in appendix \ref{Annexe_B}. As one can see, the strongest peak in the power spectrum occurs at low frequencies and is due to the long-term variations resulting from the 11-year solar cycle. Secondly, there are a group of peaks close to the Carrington rotation frequency $\nu_{\rm Car}$, indicated by the dashed green line, associated with the solar rotation. 

    \subsection{Effect on the detection of the solar rotation}\label{subsec:effect_rotation_timescale}
        
        \begin{figure}[t]
            \centering
            \includegraphics[width=\hsize]{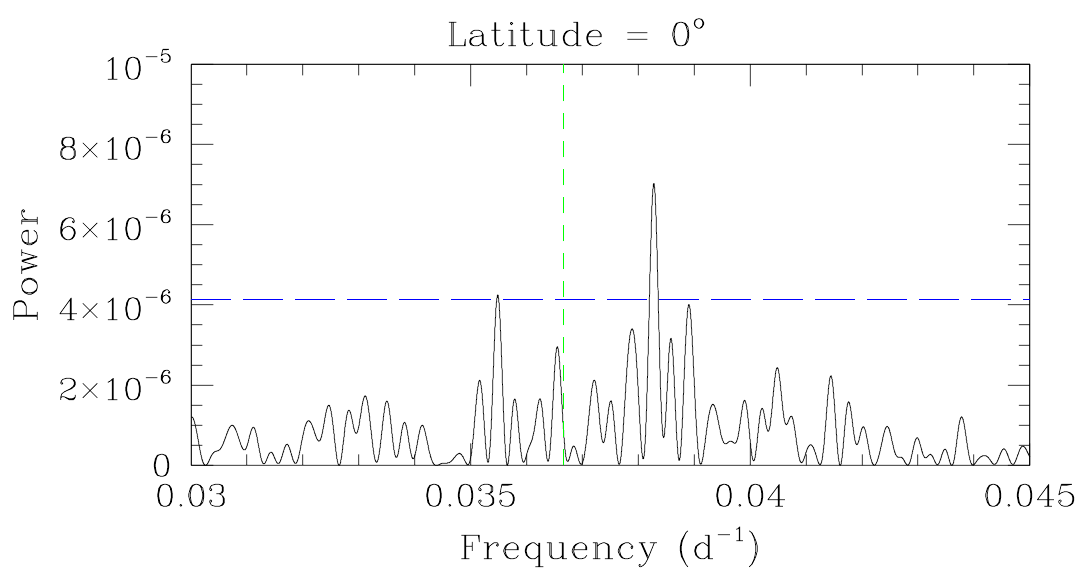}
            \caption{Fourier power spectrum zoomed on the rotation frequency in the case of a view with an inclination $i = 0^{\circ}$ (Equator-on view). The panel illustrates the power spectrum over the frequency range between 0.03 and 0.045 d$^{-1}$. The green dashed line yields $\nu_{\rm Car}$, while the long-dashed blue horizontal line yields the 99\% significance level.}
            \label{fig:fourier_rot_0}
        \end{figure}

        In order to study the solar rotation detection, we consider a zoom close the Carrington frequency $\nu_{\rm Car}$ as illustrated in Fig. \ref{fig:fourier_rot_0} for the view in the Equator plane. The other inclinations are shown in the appendix \ref{Annexe_C} by steps of $10^{\circ}$. First of all, for every inclination but essentially in the Northern hemisphere (positive inclinations), multiple peaks are detected close to $\nu_{\rm Car}$, the highest power being recorded at a frequency of 0.0383 d$^{-1}$ (period of 26.1342 d), near the frequency of the equatorial synodic rotation period.
        
        The highly complex structure of the peaks around the rotational frequency reflects the finite lifetime of the modulations which were most prominently seen during three distinct episodes of our USET time series (see \citealt{VandenBroeck-2024}). This finite lifetime leads to a beating between the actual rotational period and the characteristic times corresponding to the separation between consecutive episodes over which the rotational modulation is observed. In addition, shifts in phase between the modulations at these different epochs play a role in the relative strengths of the various subpeaks of the structure.
        
        Finally, Spörer's law which corresponds to the variation of heliographic latitudes of the active regions formation during the solar cycle also affects the shape of the Fourier power spectrum around the rotational frequency. Indeed, because of the differential rotation, structures are rotating faster close to the Equator. Hence the detection of the modulation will be spread over a range of frequencies.
        
        The rotational modulation arises from asymmetries in the longitudinal distribution of plages as explained in \cite{VandenBroeck-2024}. Therefore, the frequency of the rotational modulation is set by the latitude at which such asymmetries appear, provided that their visibility changes with the rotation phase. For an inclination $i > 0^{\circ}$ (respectively $i < 0^{\circ}$), it is the regions between about $0^{\circ}$ and $-90^{\circ}+i$ (resp. $0^{\circ}$ and $90^{\circ}+i$) for which the visibility changes most during the rotational cycle. As a result, for inclinations far away from an Equator-on view, the most relevant frequency can be associated with active regions at latitudes in the other hemisphere as they undergo a substantial modulation of their visibility.

        \begin{figure}[t]
            \centering
            \includegraphics[width=\hsize]{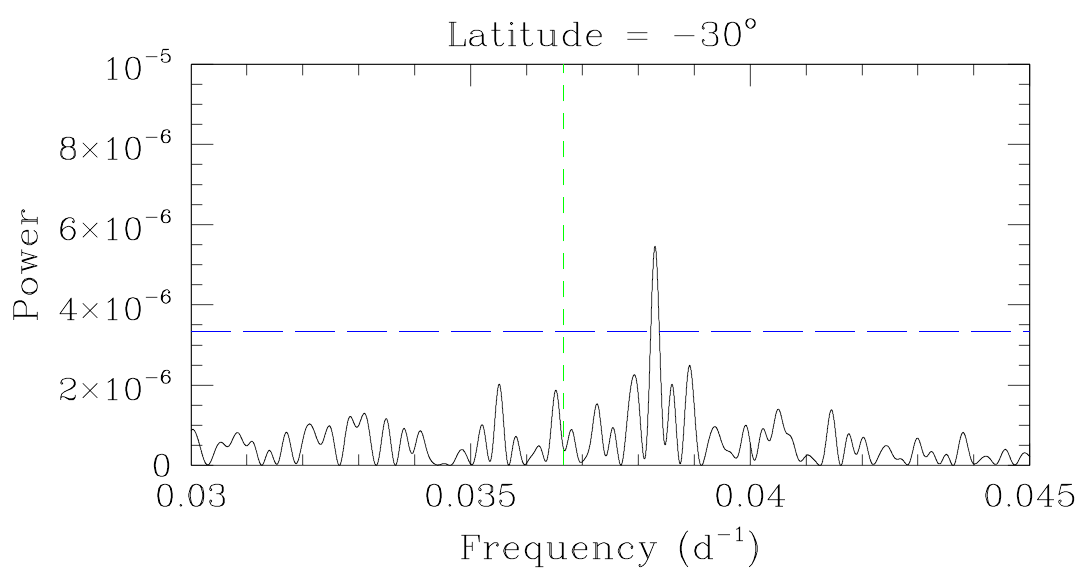}
            \includegraphics[width=\hsize]{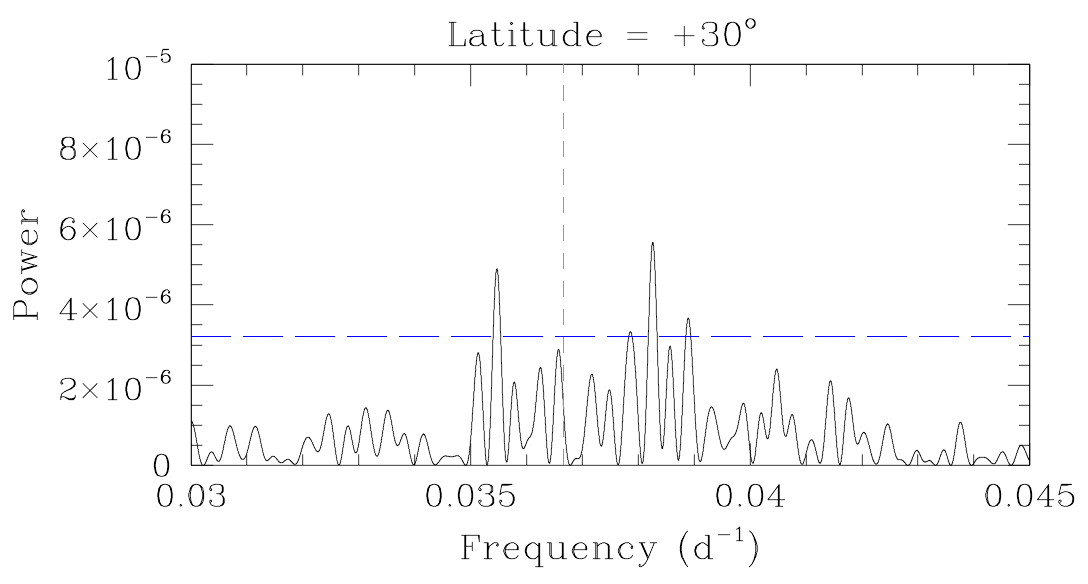}
            \caption{Fourier power spectrum zoomed on the rotation frequency in the case of a view with an inclination $i = -30^{\circ}$ (top) and $i = 30^{\circ}$ (bottom). The panel illustrates the power spectrum over the frequency range between 0.03 and 0.045 d$^{-1}$. The green dashed line yields $\nu_{\rm Car}$, while the long-dashed blue horizontal line yields the 99\% significance level.}
            \label{fig:fourier_rot_-30_30}
        \end{figure}

        For inclinations between $-20^{\circ}$ and $+20^{\circ}$ (see Fig. \ref{fig:Fourier_rot}), the power spectra remain essentially identical. From $|i| \geq 30^{\circ}$ on, the amplitudes of the rotational modulation decrease as the absolute value of the inclination increases as illustrated in Fig. \ref{fig:fourier_rot_-30_30} for a view with an inclination of $30^{\circ}$. We note a difference in behaviour between the Northern and the Southern inclinations: the group of peaks remain more important while moving towards the North Pole-on view ($i = +90^{\circ}$). In the Southern hemisphere (negative inclinations) only one peak remains important. This behaviour suggests that asymmetries in the longitudinal distribution of plages were stronger and thus more detectable in the Southern hemisphere.
        
        \begin{figure}[t]
            \centering
            \includegraphics[width=\hsize]{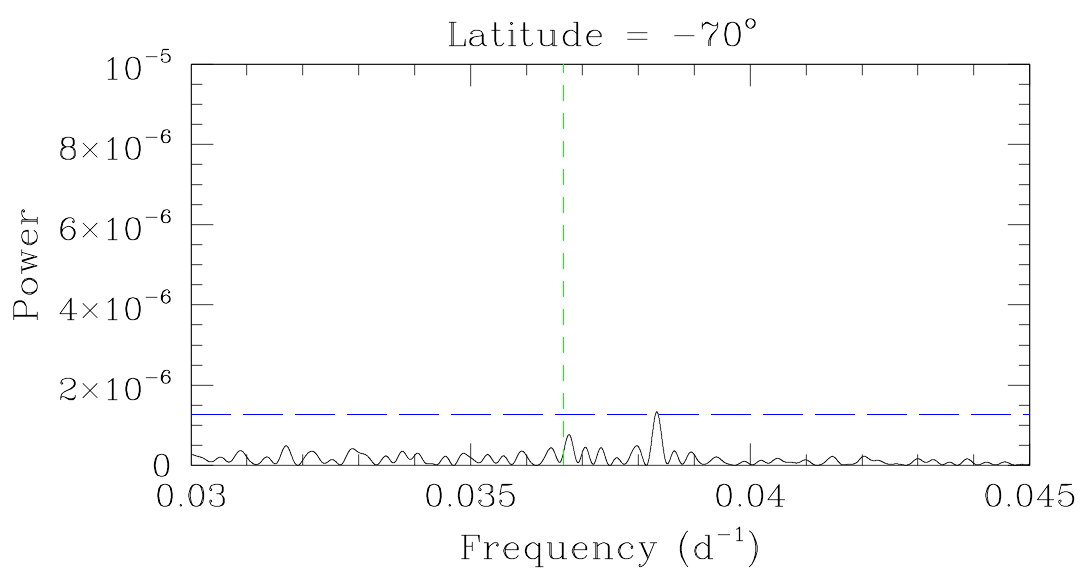}
            \includegraphics[width=\hsize]{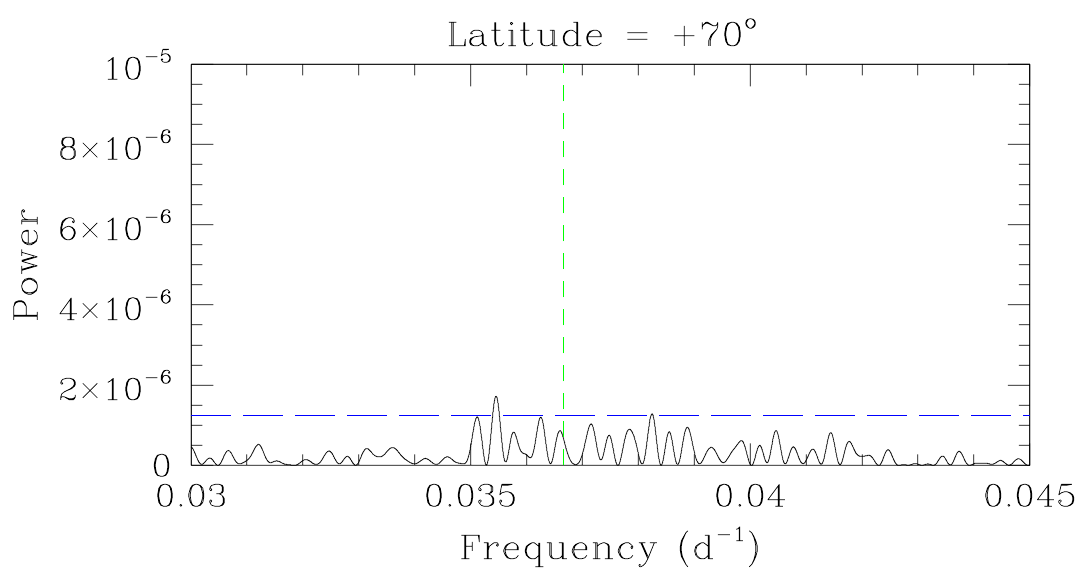}
            \caption{Fourier power spectrum zoomed on the rotation frequency in the case of a view with an inclination $i = -70^{\circ}$ (top) and $i = 70^{\circ}$ (bottom). The panel illustrates the power spectrum over the frequency range between 0.03 and 0.045 d$^{-1}$. The green dashed line yields $\nu_{\rm Car}$, while the long-dashed blue horizontal line yields the 99\% significance level.}
            \label{fig:fourier_rot_-70_70}
        \end{figure}
    
        Finally, as observed in Fig. \ref{fig:fourier_rot_-70_70}, for $|i| \geq 70^{\circ}$, the rotational modulation is no longer remarkable in the power spectrum and would certainly be missed in noisy and less densely sampled time series of other stars. Moreover, we observe a difference between both hemispheres. While the power spectrum for the Southern hemisphere (inclination $i = -70^{\circ}$) presents one clearly visible peak, the power spectrum for the Northern hemisphere (inclination $i = 70^{\circ}$) displays several peaks with the same power. An explanation of this behavior is the asymmetry of the distribution of active regions between both hemispheres, as observed in Fig. \ref{fig:Plages_area_evolution_North_vs_South_&_SN_hemispheric}. The distribution is not symmetric relative to the Equator, so that the visibility of the active regions will be affected depending on the viewing angle. Indeed, for an inclination $i = +70^{\circ}$, magnetic structures in the Northern hemisphere down to a latitude of $+20^{\circ}$ will be visible over the entire rotation cycle, although with a changing aspect angle (sometimes closer to limb, sometimes closer to the centre of the disk). Therefore, these structures will not result in a strong rotational variation. Rotational modulation instead arises from active regions located at more Southern latitudes. In our example, it is the negative latitudes, down to $-20^{\circ}$, that will have the strongest impact. In our time series, the Southern hemisphere hosts more active regions than the Northern hemisphere during the solar maximum (see Fig. \ref{fig:Plages_area_evolution_North_vs_South_&_SN_hemispheric}). Together with the finite lifetime of these active regions, the North/South asymmetry leads to the presence of multiple peaks in the power spectrum near the Carrington rotation frequency for $i = +70^{\circ}$.

    \subsection{Effect on the detection of the solar cycle}\label{subsec:effect_cycle_timescale}

\begin{figure}[t]
            \centering
            \includegraphics[width=\hsize]{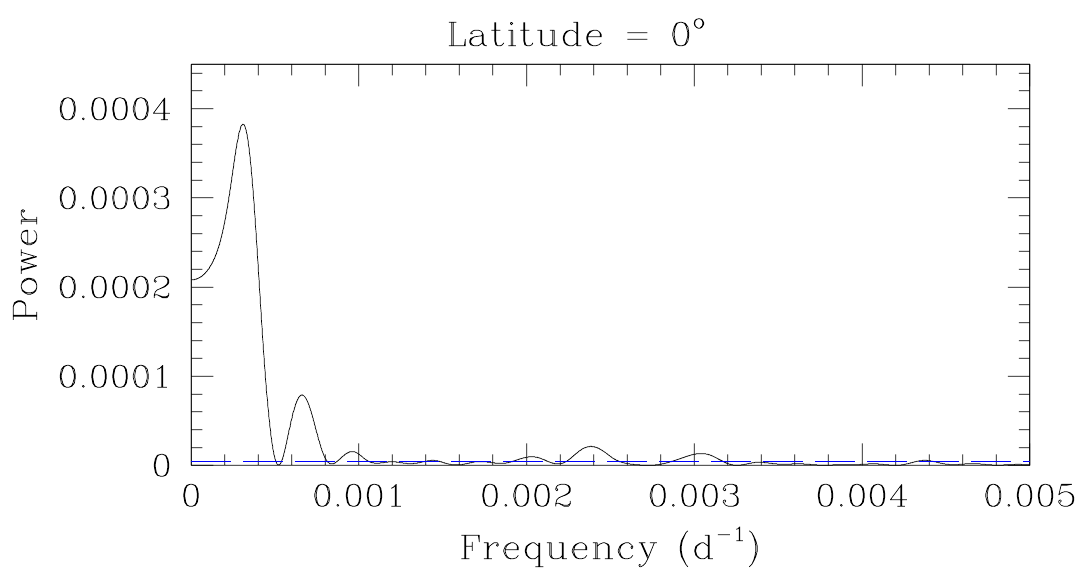}
            \caption{Fourier power spectrum zoomed at very low frequencies in the case of a view with an inclination $i = 0^{\circ}$ (Equator-on view). The panel illustrates the power spectrum over the frequency range below 0.005 d$^{-1}$. The long-dashed blue horizontal line yields the 99\% significance level.}
            \label{fig:fourier_cyc_0}
        \end{figure}

        To study the effect on the solar cycle timescale, we consider the power spectrum zoomed at very low frequencies. Fig. \ref{fig:fourier_cyc_0} illustrates this power spectrum for the view from the Equator ($i = 0^{\circ}$). We can clearly see the peak of detection of the solar cycle modulation located at the frequency of $\pm$ 0.00025 d$^{-1}$ corresponding to a period of $\sim$ 10.95 years, analogous to the typical 11-years solar cycle. 

\begin{figure}[t]
            \centering
            \includegraphics[width=\hsize]{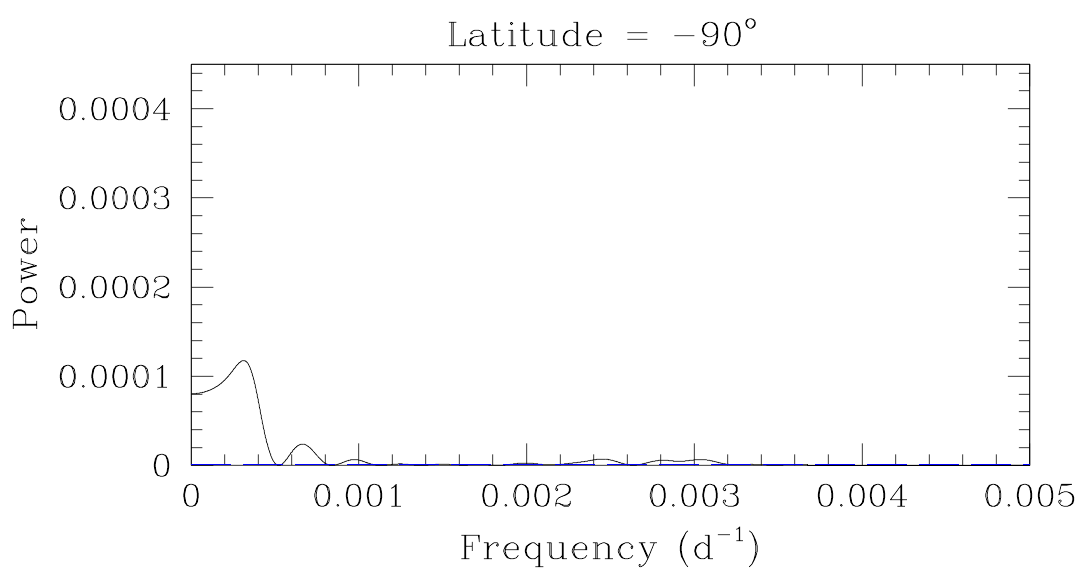}
            \includegraphics[width=\hsize]{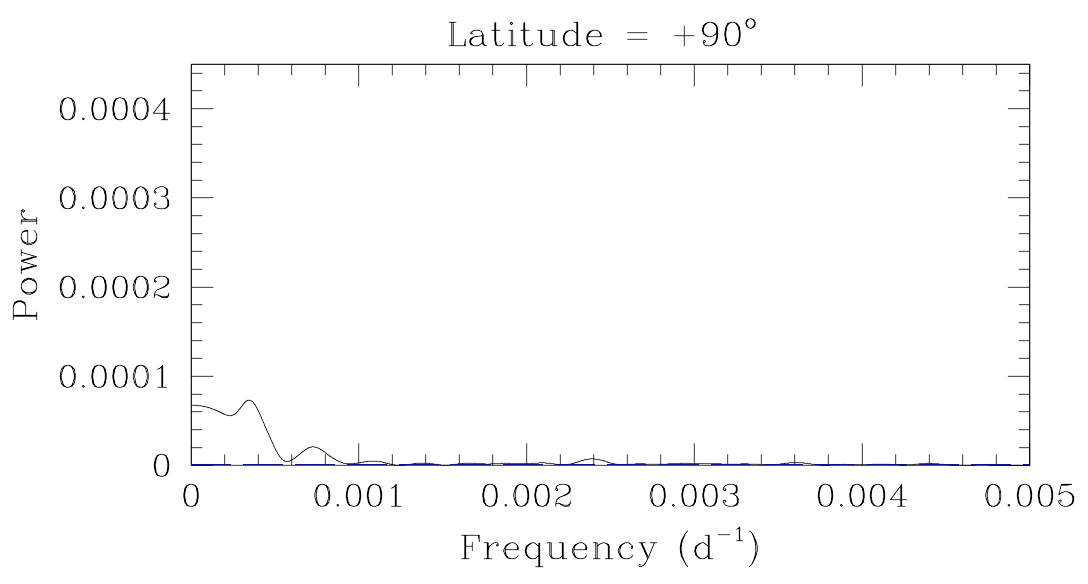}
            \caption{Fourier power spectrum zoomed at very low frequencies for the Pole-on views with an inclination $i = -90^{\circ}$ (top) and $i = 90^{\circ}$ (bottom).. The panel illustrates the power spectrum over the frequency range below 0.005 d$^{-1}$. The long-dashed blue horizontal line yields the 99\% significance level.}
            \label{fig:fourier_cyc_-90_90}
        \end{figure}

        Fig. \ref{fig:fourier_cyc_-90_90} illustrates the Fourier power spectrum over the same frequency range for the Pole-on views ($i = -90^{\circ}$ and $i = 90^{\circ}$) and Fig. \ref{fig:Fourier_cyc} in appendix \ref{Annexe_D} for all inclinations by steps of $10^{\circ}$. The reduction of the power when the absolute value of the inclination increases is somewhat stronger for positive inclinations, but in both cases (positive or negative inclinations) the long-term modulation remains visible up to $|i| = 90^{\circ}$, although with a significantly reduced power compared to an Equator-on view.

    \subsection{Effect of the sampling}\label{subsec:effect_sampling}
            
        The USET Ca {\sc ii} K observations of the Sun benefit from a denser sampling than observations of other solar-like stars. Indeed, whilst the Sun can be observed during the whole year, the visibility of most stars is restricted to periods of typically six months. Moreover, whilst USET is dedicated to observations of the Sun, telescopes used for the study of chromospherically active stars usually have to share the observing time between a number of targets, resulting in a lower cadence of observations than for the USET data. This situation could bias the discussion of the detectability of the cyclic modulation as a function of inclination.

\begin{figure}[t]
            \centering
            \includegraphics[width=\hsize]{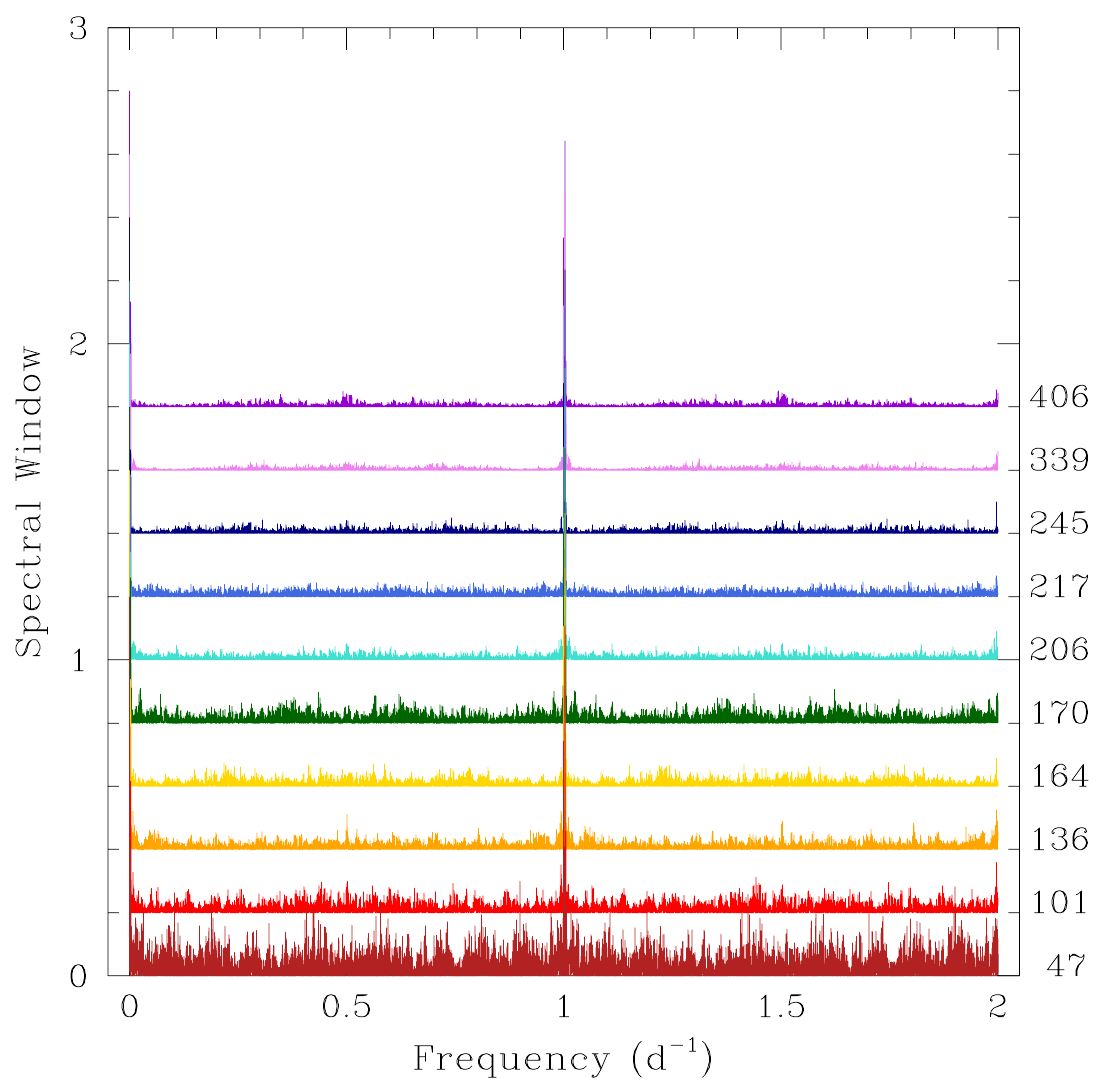}
            \includegraphics[width=\hsize]{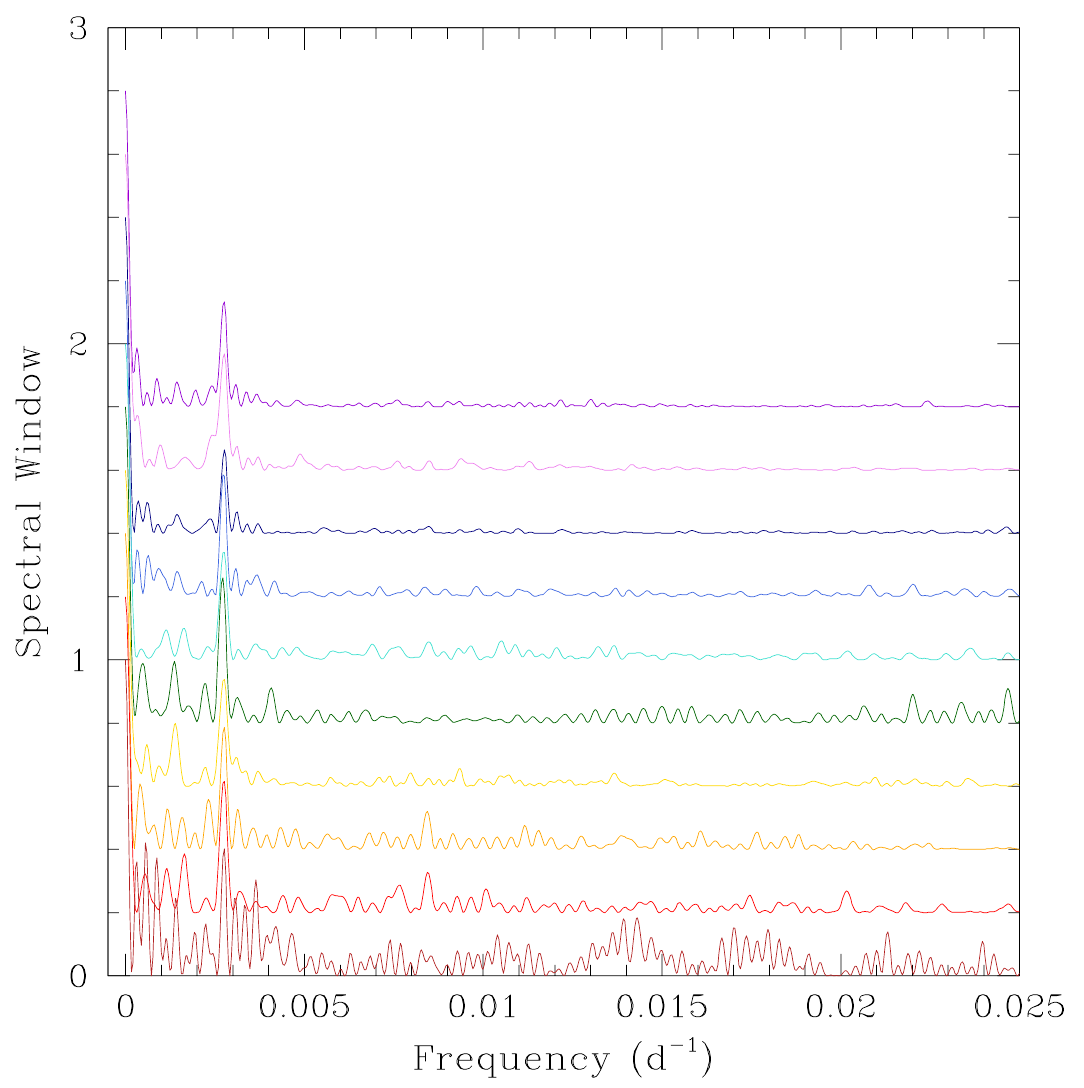}
            \caption{Spectral window of actual time series of TIGRE observations of solar-like stars. The top panel illustrates the spectral window up to 2 d$^{-1}$, whilst the bottom panel yields a zoom on the low-frequency domain, illustrating the appearance of a yearly alias. The numbers on the right of the top panel indicate the number of observations collected over the interval between 2013 and 2023. The least frequently observed star was observed less than 50 times while the most-intensively observed one has over 400 observations.}
            \label{fig:spectralwindow}
        \end{figure}

        To account for this effect, we consider the sample of solar-like stars of \cite{Hempelman-2016} which are monitored with the TIGRE telescope to search for rotational modulations and activity cycles in their S-index. We extracted the actual sampling of the TIGRE observations of these stars over the time interval from 2013 until the end of 2023 from the TIGRE data archive. The number of observations ranges from less than 50 for the least-frequently observed star to over 400 for the most-intensively observed targets. The mean and median number of observations for an individual star are both around 200 spread over this 10 years period. Fig. \ref{fig:spectralwindow} illustrates the spectral windows computed for a subset of these time series. The spectral windows are dominated by the 1 d$^{-1}$ alias. As one could expect, the spectral windows are cleaner when the number of data points increases. However, the most important difference compared to the time series of solar observations concerns the occurrence of a yearly alias at 0.00274 d$^{-1}$ which can be seen by zooming on the spectral windows (bottom panel of Fig. \ref{fig:spectralwindow}). This latter feature stems from the above-discussed visibility constraints and can clearly be expected to impact the detectability of the long-term cycles. However, it should be emphasized that, due to its short duration, our time series is less favorable for the study of the long-duration cycle than some stellar time series. Indeed our dataset only covers a single cycle. Consequently, the width of the peak in the Fourier power spectrum leads to a significant uncertainty on the actual period of the activity cycle. For other stars, even with a relatively sparse sampling, typically more than one activity cycle has been observed, allowing therefore a more accurate determination of the frequency associated with the long cycle.

\begin{figure}[t]
            \centering
            \includegraphics[width=\hsize]{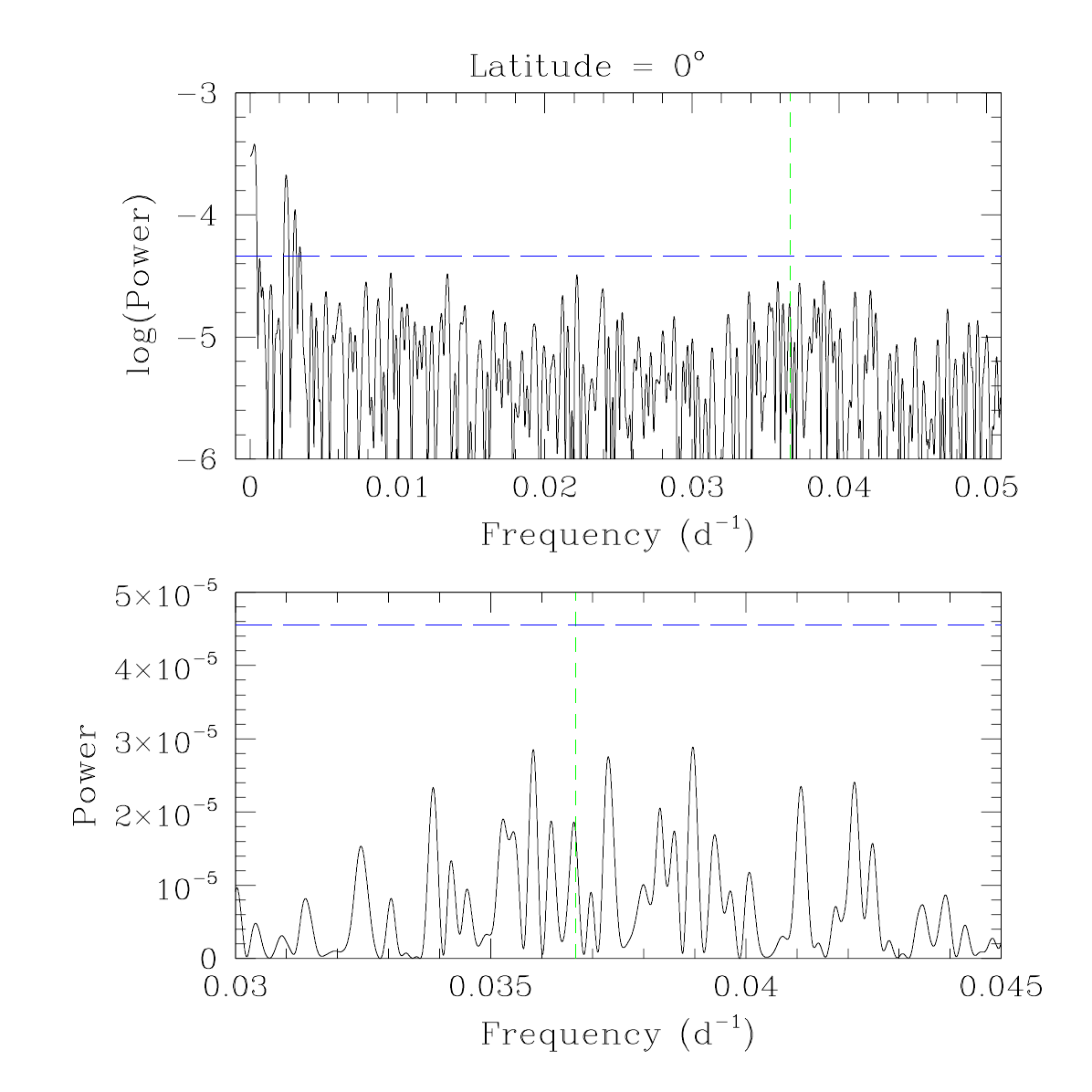}
            \caption{Fourier power spectrum of the resampled USET time series assuming 206 observations spread over ten years for an Equator-on view. The top panel illustrates the logarithm of the power spectrum for frequencies below 0.05 d$^{-1}$. The bottom panel provides a zoom on the power around $\nu_{\rm Car}$ (given by the short-dashed green vertical line). The long-dashed blue horizontal line yields the 99\% significance level.}
            \label{fig:starpower0}
        \end{figure}

        We have then sampled the time series of the USET plages and enhanced network areas for different inclinations according to the sampling of the TIGRE observations of solar-like stars. Hereafter, we focus our discussion on the sampling corresponding to a number of 206 observations spread over ten years. Figure \ref{fig:starpower0} illustrates the Fourier power spectrum for an Equator-on inclination for frequencies below 0.05 d$^{-1}$ as well as around $\nu_{\rm Car}$. Whilst the long-term cycle still provides the highest peak in the power spectrum, we note the presence of a strong yearly alias. With the sampling assumed here, the strongest peak remains the one associated with the long-term cycle. The situation is much worse as far as the detection of the rotational frequency is concerned. The sampling no longer allows an unambiguous identification of the dominant frequency that was found in the actual USET data. Indeed, there are now at least three peaks of equal strength around $\nu_{\rm Car}$, although all of them have a power well below the 99\% significance level. But what is even worse is that there are a number of peaks at very different frequencies (e.g., near 0.009, 0.013 or 0.022 d$^{-1}$) which have a power that is equal or higher than that of the peaks near the actual rotation frequency. In a real stellar time series, one would thus not be able to identify the right frequency among those peaks. The situation remains essentially the same for other inclination angles (Fig. \ref{fig:starpower1} in appendix \ref{Annexe_E}). We thus conclude that a sampling of about 200 observations spread over ten years would not allow a clear detection of the rotational modulation. This is not surprising given the fact that the visibility of the rotational modulation in the Sun's plages and enhanced network area varies significantly with time as shown in \cite{VandenBroeck-2024}. A patchy sampling can thus easily miss those episodes where the rotational modulation would be well detected. For the long-term cycle, we observe that the peak associated with the true frequency remains the dominant one.
        
\begin{figure}[t]
            \centering
            \includegraphics[width=0.49\hsize]{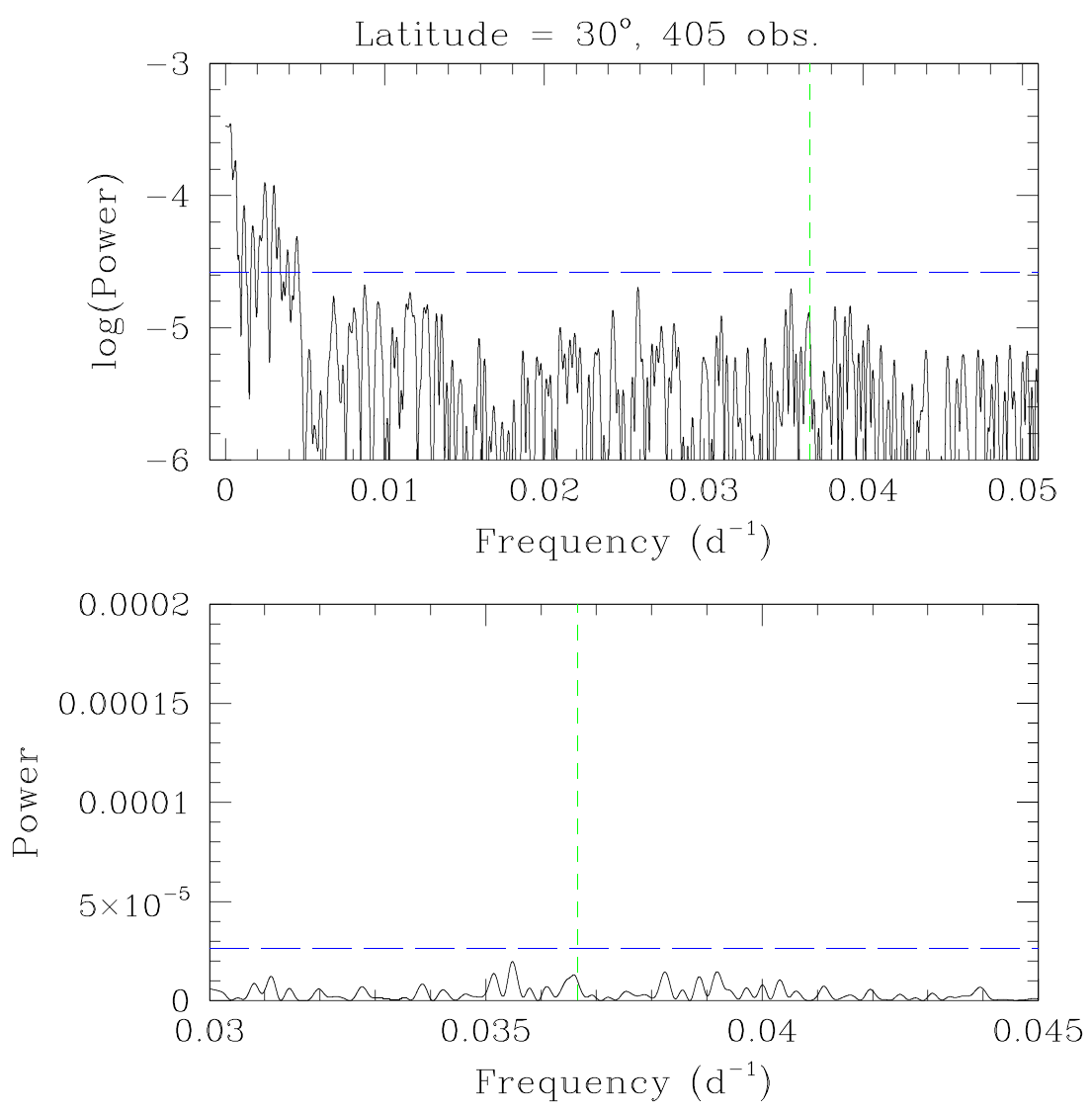}
            \includegraphics[width=0.49\hsize]{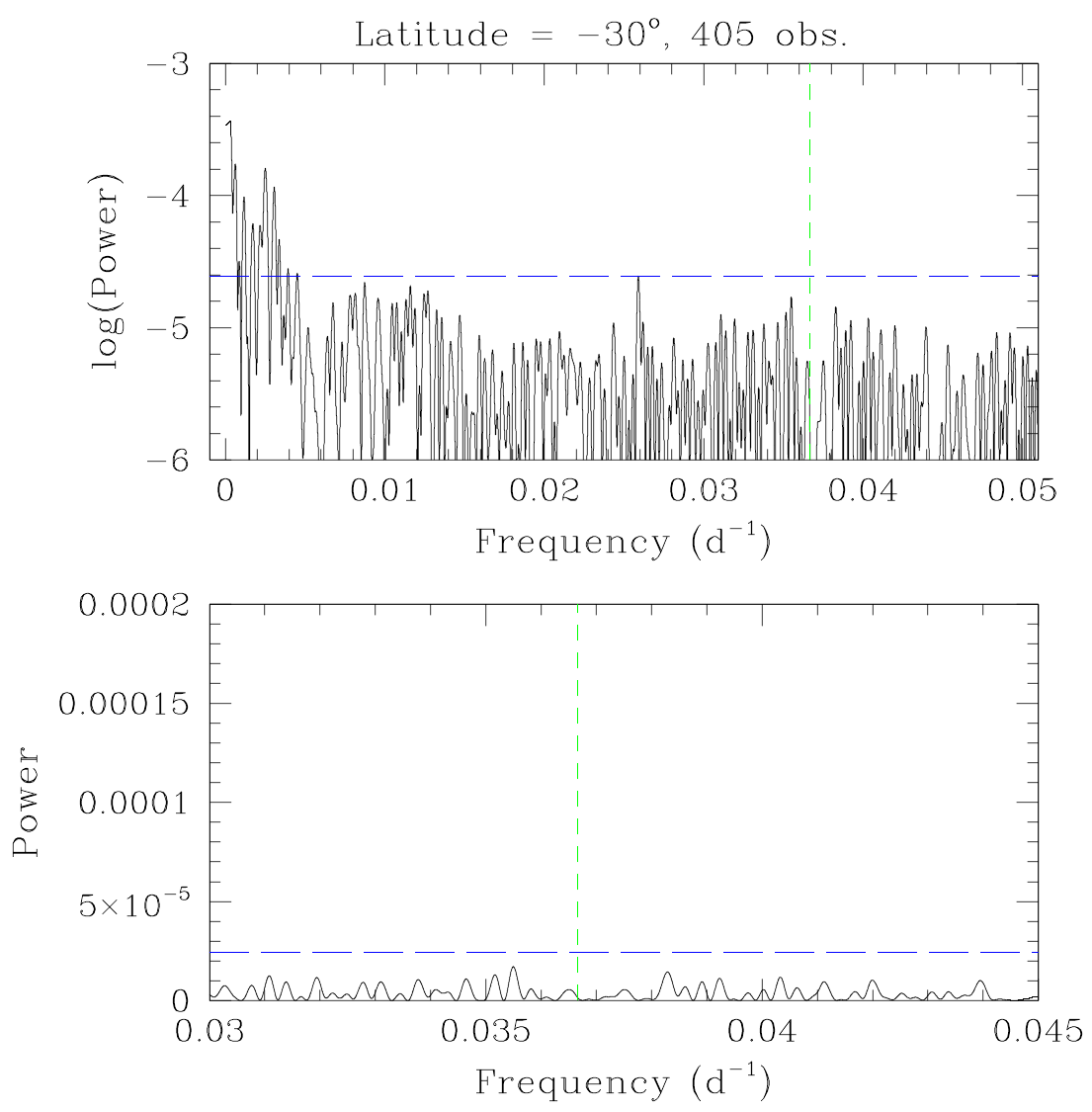}\vspace{0.7cm}
            \includegraphics[width=0.49\hsize]{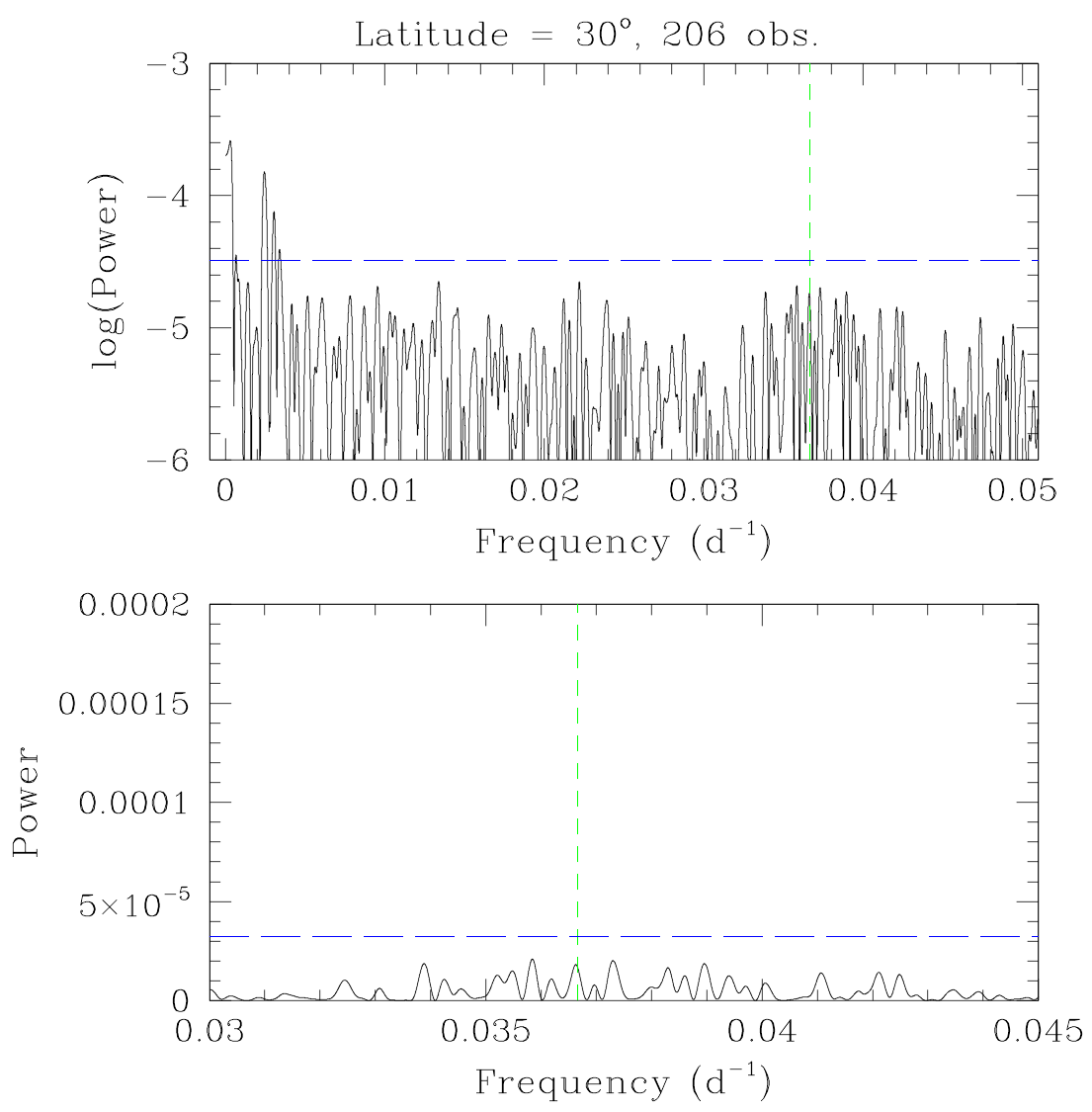}
            \includegraphics[width=0.49\hsize]{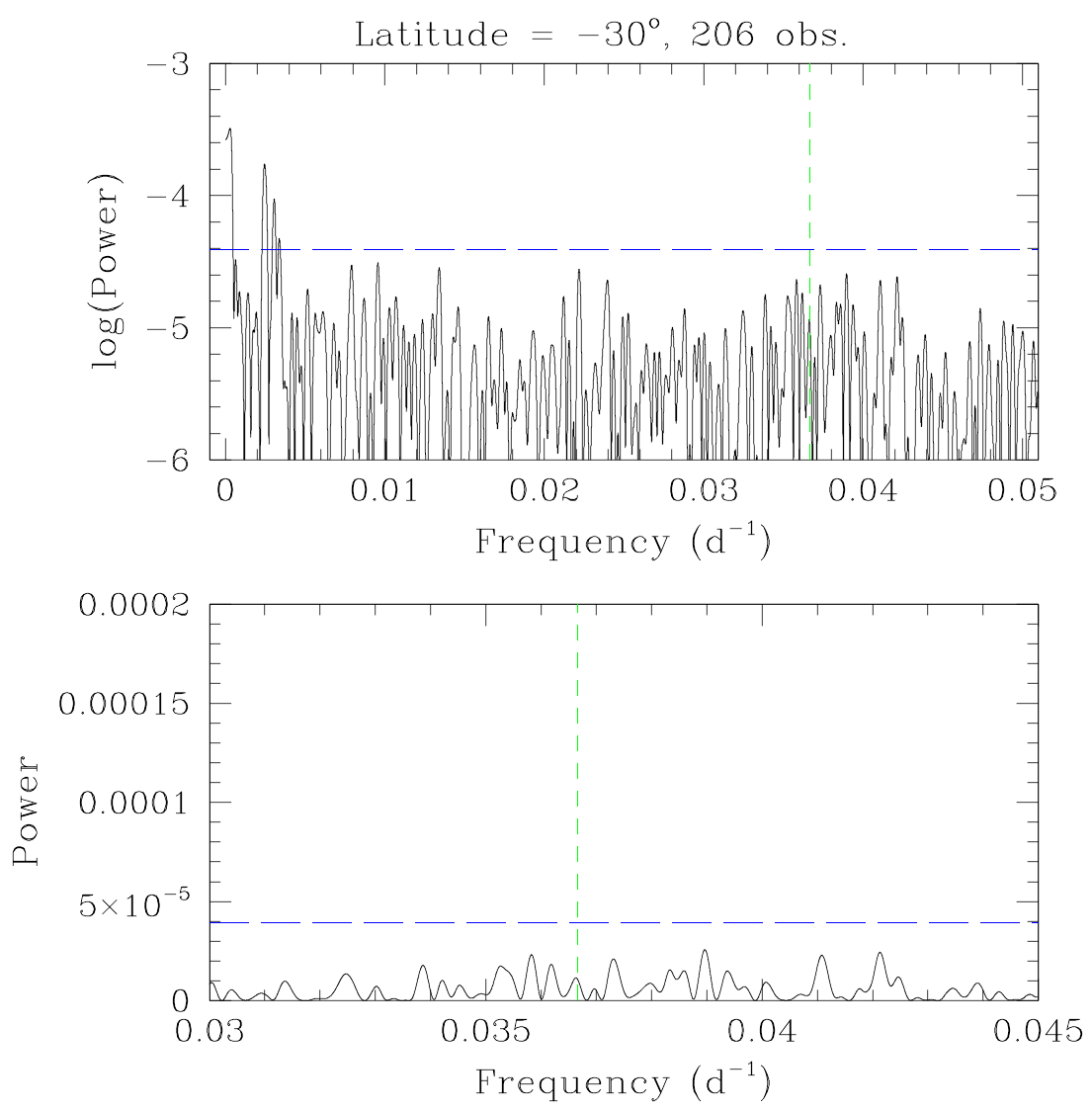}\vspace{0.7cm}
            \includegraphics[width=0.49\hsize]{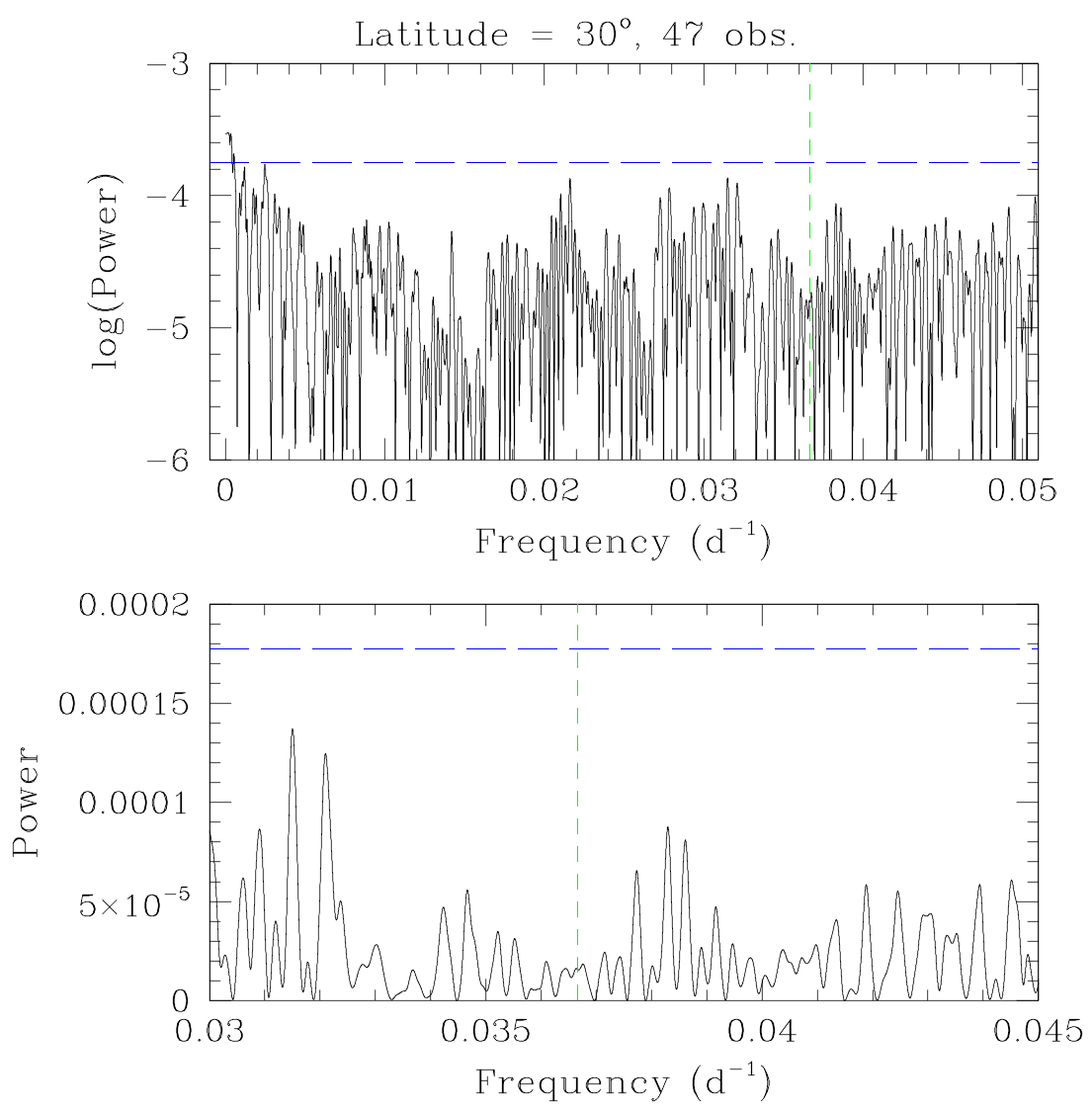}
            \includegraphics[width=0.49\hsize]{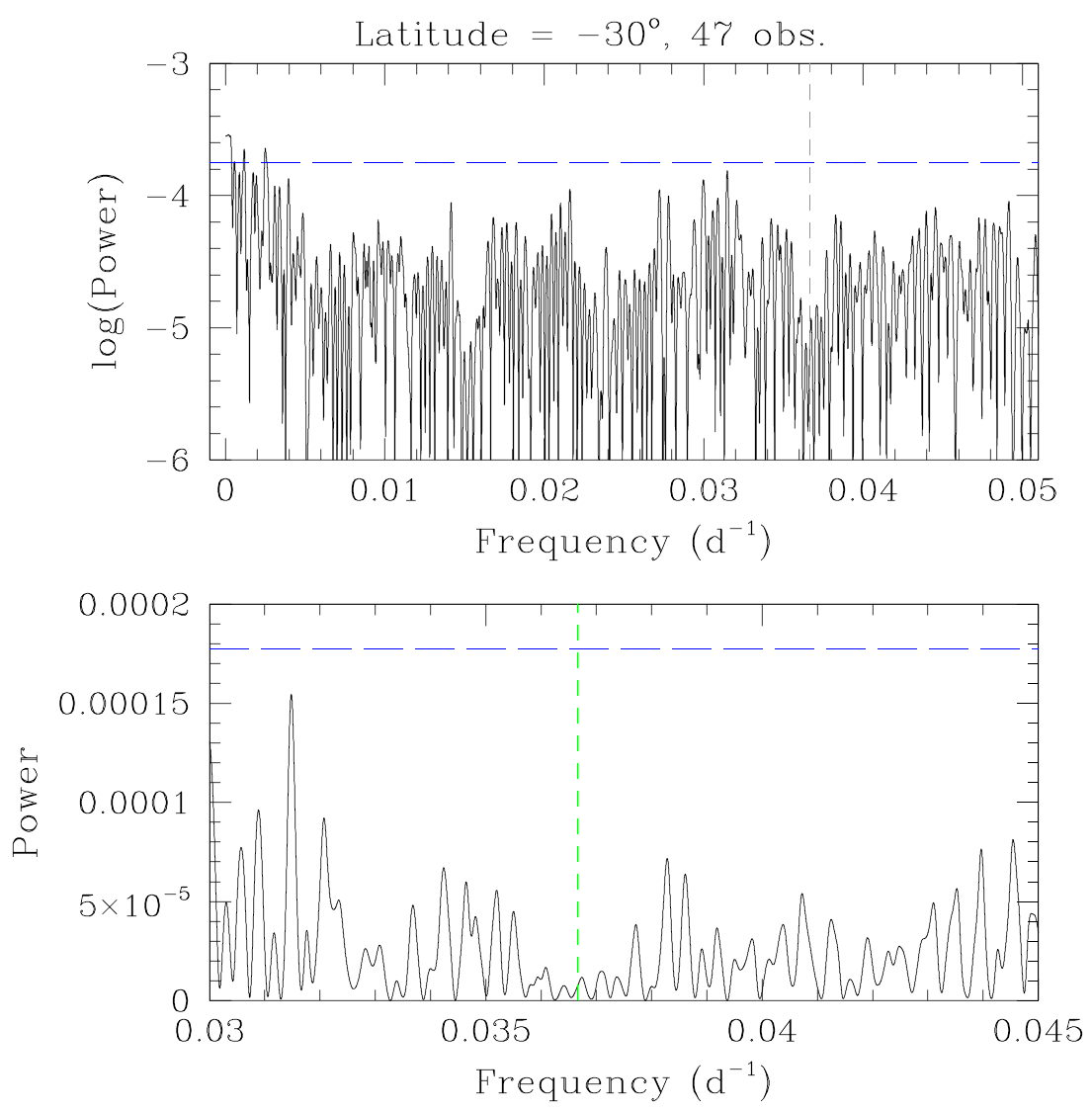}
            \caption{Fourier power spectrum of the resampled USET time series for a decreasing observing cadence (from top to bottom) and for inclination angles of $+30^{\circ}$ (left column) and $-30^{\circ}$ (right column row). For each case, top panel illustrates the logarithm of the power spectrum for frequencies below 0.05 d$^{-1}$ and the bottom panel provides a zoom on the power around $\nu_{\rm Car}$ (given by the short-dashed green vertical line). The long-dashed blue horizontal line yields the 99\% significance level.}
            \label{fig:sampling}
        \end{figure}

        As a next step, we have resampled the USET plages and enhanced network area time series for inclinations of $+30^{\circ}$ and $-30^{\circ}$ according to the observing cadence of ten representative stars of the TIGRE sample of \cite{Hempelman-2016}. We used those stars for which the spectral windows of their time series are displayed in Fig. \ref{fig:spectralwindow}. The results of this exercise are illustrated in Fig. \ref{fig:sampling} for the sampling used for three stars with 405, 206 and 47 observations over ten years. 
        
        Although the long-term cycle is detected in all cases, one can clearly see that the contrast of the peak with respect to the 99\% significance level strongly decreases when the sampling gets sparser, as expected. The ratio between the power of the highest peak and the 99\% significance level decreases from 14 for the densest sampling to 8 for the intermediate case and 1.6 for the sparsest case. Hence, we conclude that a long-term cycle with an amplitude identical to that of the Sun would remain detectable with a sampling of $\sim 20$ observations per year, provided that the data cover a sufficiently long time interval. As expected, the significance level decreases when the number of observations decreases, becoming marginal for the sparsest sampling (a handful of observations per year). Conversely, as we already concluded hereabove, in all cases, the observing strategy fails to detect the rotational modulation.
        
\begin{figure}[t]
            \centering
            \includegraphics[width=\hsize]{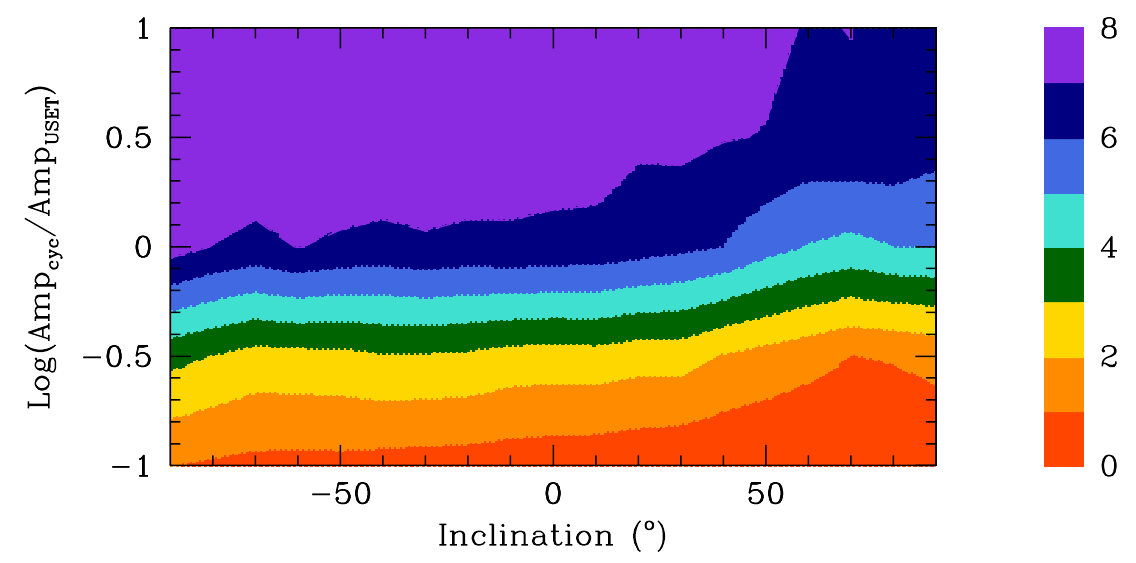}
            \caption{Detectability of the long-term cycle as a function of the inclination and the scaled amplitude of the long-term cycle. The colour-scale to the right indicates the ratio between the power of the peak associated with the long-term cycle in the Fourier power spectrum and the 99\% significance level.}
            \label{fig:power_amp}
        \end{figure}

        Finally, to assess the impact of the amplitude of the cyclic variations on their detectability with a typical sampling of solar-like stars, we performed another set of simulations. We first adjusted the long-term cycle variations for each set of simulated out-of-ecliptic USET data (i.e., for each value of the inclination) by a polynomial of degree six as we had done for the actual USET observations in \cite{VandenBroeck-2024}. Subtracting this polynomial from the simulated time series yields a proxy of the shorter term variations. We then scaled the amplitude of the adjusted long-term cycle by a factor between 0.1 and 10.0 (in logarithmic steps of $\log{{\rm Amp}_{\rm cyc}/{\rm Amp}_{\rm USET}} = -0.5$. These scaled long-term variations were then added back to the shorter term variations to simulate situations of solar-like stars with different ratios between the amplitudes of the short and long-term variations. These simulated time series were then resampled with our reference observing cadence assuming 206 observations spread over ten years. For each resampled synthetic time series, we performed a Fourier analysis and determined the 99\% significance level via our re-shuffling method. Fig. \ref{fig:power_amp} illustrates the results of this exercise. The colours indicate the ratio between the power of the strongest peak in the Fourier spectrum that is associated with the long-term cycle and the 99\% significance level. As one could expect, the detectability of the long-term cycle becomes marginal (ratio below 1, red colour in Fig. \ref{fig:power_amp}) when the amplitude of the cycle is scaled down by a factor 0.1. We note that the visibility of the peak displays some variations with the inclination. At higher (positive) inclinations, the detectability is lower whatever the value of the scaling parameter. This results from the North-South asymmetry that we have found in the USET data. Overall, we find that with the sampling adopted here, the peak due to the long-term cycle is detected at a level at least 3 times above the 99\% significance level provided that the amplitude of the cyclic variations remains at a level of at least 33\% of the amplitude seen in the USET data. For lower values, the detections become uncertain also because the highest peaks in the Fourier spectra are no longer necessarily associated with the frequency of the long-term cycle.     
        
\section{Discussions and conclusions}\label{sec:Conclusions}

    Based on full-disk images of the solar chromosphere in the Ca {\sc ii} K line from the USET station, we have mapped the solar surface with synoptic maps. We have segmented the brightest structures and produced a time series of their area fraction. Then we have used an appropriate projection to represent the solar surface as it would be seen under various viewing angles. We have computed the area fraction for different inclinations and its variation goes up to 63\% between the Equator-on view and the Poles-on view during the maximum of the solar cycle.
    
    As the plages and enhanced network area fraction is a good proxy for the S-index \citep{VandenBroeck-2024}, this study can be used to make a connection between the Sun and the other Sun-like stars. In particular it could be used to better understand the detection of temporal modulation for the other stars that are not necessarily viewed from the Equator plane. To reach this goal, we have analysed the impact of the viewing angle on the area fraction time series by using Fourier power spectrum. 

    Our results show an important impact of the viewing angle on the detection of modulation due to the solar rotation. If the observations use the same (dense) sampling as the USET data, the rotation is detectable up to an inclination of $|i| = 70^{\circ}$. For higher values the rotation is not visible anymore in the signal and would be missed. This behaviour could be explained by the fact that from a Pole-on view, an asymmetry in the plages distribution will be either permanently visible or not visible at all during a full solar rotation cycle.

    On the long term, the chromospheric activity cycles of Sun-like stars should remain detectable even for stars seen under a near Pole-on viewing angle. Positive and negative inclinations give different results due to the asymmetry of the solar activity in the Northern and Southern hemispheres, as already seen for other magnetic structures. 
    
    For other stars, the actual detectability will also depend on the sampling of the time series and on the quality of the data. Indeed, our USET time series benefits from a long continuous time series with a dense sampling. Actual time series of stellar S-indices have usually more complicated and scarcer samplings including months-long gaps due to restricted target visibility. We have analysed the effect of the sampling on the detection of the periodic modulations by extracting the actual sampling of TIGRE observations of Sun-like stars over a period of ten years. The least frequently observed star was observed less than 50 times while the most-intensively observed one has over 400 observations. We demonstrated that a more realistic sampling leads to the vanishing of the rotational modulation detection. However, the modulation due to the activity cycle remains visible at nearly all inclinations if the sampling contains at least 20 observations per year and as long as the amplitude of the cyclic variation is at least 30\% of the solar cycle amplitude. This conclusion is made on the assumption of a typical solar activity case (i.e. of $\sim$ 11 years). In the case of stars with a much shorter or much longer cycle than the solar cycle, the period of the activity cycle will be very difficult to determine because of the sparsity of the sampling. Indeed, for significantly longer cycles, we would need a much longer homogeneous dataset, which is hard to collect for stellar observations. For shorter cycles \citep[of a few months, e.g.,][]{Mittag-2019}, the sparse sampling assumed here would clearly fail to correctly identify the exact cycle duration. An extension of our present study would benefit from a dataset covering a longer time-span, as the impact of the inclination axis might vary with the solar cycle amplitude \citep{Sowmya-2021_I}.

\begin{acknowledgements}
      The authors wish to thank Sowmya Krishnamurthy from the Max Planck Institute for Solar Research System, in Germany, for the very productive discussions that enabled them to carry out the work on the orthographic projection. Grégory Vanden Broeck was supported by a PhD grant awarded by the Royal Observatory of Belgium. The USET instruments are built and operated with the financial support of the Solar-Terrestrial Center of Excellence (STCE).
    
\end{acknowledgements}

\bibliographystyle{aa}
\begin{small}
    \bibliography{biblio}

\begin{thebibliography}{33}
\expandafter\ifx\csname natexlab\endcsname\relax\def\natexlab#1{#1}\fi

\bibitem[{{Baliunas} {et~al.}(1998){Baliunas}, {Donahue}, {Soon}, \&
  {Henry}}]{Baliunas-1998}
{Baliunas}, S.~L., {Donahue}, R.~A., {Soon}, W., \& {Henry}, G.~W. 1998, in
  Astronomical Society of the Pacific Conference Series, Vol. 154, Cool Stars,
  Stellar Systems, and the Sun, ed. R.~A. {Donahue} \& J.~A. {Bookbinder}, 153

\bibitem[{{Baliunas} {et~al.}(1985){Baliunas}, {Horne}, {Porter}, {Duncan},
  {Frazer}, {Lanning}, {Misch}, {Mueller}, {Noyes}, {Soyumer}, {Vaughan}, \&
  {Woodard}}]{Baliunas-1985}
{Baliunas}, S.~L., {Horne}, J.~H., {Porter}, A., {et~al.} 1985, \apj, 294, 310

\bibitem[{{Bechet} \& {Clette}(2002)}]{2023-USET}
{Bechet}, S. \& {Clette}, F. 2002, USET images L1centered,
  \url{https://doi.org/10.24414/nc7j-b391}, published by Royal Observatory of
  Belgium (ROB)

\bibitem[{{Borgniet} {et~al.}(2015){Borgniet}, {Meunier}, \&
  {Lagrange}}]{Borgniet-2015}
{Borgniet}, S., {Meunier}, N., \& {Lagrange}, A.~M. 2015, \aap, 581, A133

\bibitem[{{Boro Saikia} {et~al.}(2018){Boro Saikia}, {Marvin}, {Jeffers},
  {Reiners}, {Cameron}, {Marsden}, {Petit}, {Warnecke}, \&
  {Yadav}}]{BoroSaikia-2018}
{Boro Saikia}, S., {Marvin}, C.~J., {Jeffers}, S.~V., {et~al.} 2018, \aap, 616,
  A108

\bibitem[{Calabretta \& Greisen(2002)}]{Calabretta-2002}
Calabretta, M.~R. \& Greisen, E.~W. 2002, \aap, 395, 1077

\bibitem[{{Devi} {et~al.}(2021){Devi}, {Singh}, {Chandra}, {Priyal}, \&
  {Joshi}}]{Devi-2021}
{Devi}, P., {Singh}, J., {Chandra}, R., {Priyal}, M., \& {Joshi}, R. 2021,
  \solphys, 296, 49

\bibitem[{El-Borie {et~al.}(2021)El-Borie, El-Taher, Thabet, Ibrahim, \&
  Bishara}]{El-Borie-2021}
El-Borie, M., El-Taher, A., Thabet, A., Ibrahim, S., \& Bishara, A. 2021,
  Chinese Journal of Physics, 72, 1

\bibitem[{{Gosset} {et~al.}(2001){Gosset}, {Royer}, {Rauw}, {Manfroid}, \&
  {Vreux}}]{Gosset-2001}
{Gosset}, E., {Royer}, P., {Rauw}, G., {Manfroid}, J., \& {Vreux}, J.~M. 2001,
  \mnras, 327, 435

\bibitem[{{Heck} {et~al.}(1985){Heck}, {Manfroid}, \& {Mersch}}]{Heck-1985}
{Heck}, A., {Manfroid}, J., \& {Mersch}, G. 1985, \aaps, 59, 63

\bibitem[{{Hempelmann} {et~al.}(2016){Hempelmann}, {Mittag}, {Gonzalez-Perez},
  {Schmitt}, {Schr{\"o}der}, \& {Rauw}}]{Hempelman-2016}
{Hempelmann}, A., {Mittag}, M., {Gonzalez-Perez}, J.~N., {et~al.} 2016, \aap,
  586, A14

\bibitem[{{Knaack} {et~al.}(2001){Knaack}, {Fligge}, {Solanki}, \&
  {Unruh}}]{Knaack-2001}
{Knaack}, R., {Fligge}, M., {Solanki}, S.~K., \& {Unruh}, Y.~C. 2001, \aap,
  376, 1080

\bibitem[{{Lockwood} \& {Skiff}(1990)}]{Lockwood-1990}
{Lockwood}, G.~W. \& {Skiff}, B.~A. 1990, in NASA Conference Publication, Vol.
  3086, NASA Conference Publication, 8--15

\bibitem[{{Lockwood} {et~al.}(2007){Lockwood}, {Skiff}, {Henry}, {Henry},
  {Radick}, {Baliunas}, {Donahue}, \& {Soon}}]{Lockwood-2007}
{Lockwood}, G.~W., {Skiff}, B.~A., {Henry}, G.~W., {et~al.} 2007, \apjs, 171,
  260

\bibitem[{{Met Office}(2010 - 2015)}]{Cartopy}
{Met Office}. 2010 - 2015, Cartopy: a cartographic python library with a
  Matplotlib interface, Exeter, Devon

\bibitem[{{Meunier} {et~al.}(2019){Meunier}, {Lagrange}, {Boulet}, \&
  {Borgniet}}]{Meunier-2019}
{Meunier}, N., {Lagrange}, A.~M., {Boulet}, T., \& {Borgniet}, S. 2019, \aap,
  627, A56

\bibitem[{{Meunier} {et~al.}(2024){Meunier}, {Lagrange}, {Dumusque}, \&
  {Sulis}}]{Meunier-2024}
{Meunier}, N., {Lagrange}, A.~M., {Dumusque}, X., \& {Sulis}, S. 2024, \aap,
  687, A303

\bibitem[{{Mittag} {et~al.}(2019){Mittag}, {Schmitt}, {Hempelmann}, \&
  {Schr{\"o}der}}]{Mittag-2019}
{Mittag}, M., {Schmitt}, J.~H.~M.~M., {Hempelmann}, A., \& {Schr{\"o}der},
  K.~P. 2019, \aap, 621, A136

\bibitem[{{Mittag} {et~al.}(2023){Mittag}, {Schmitt}, \&
  {Schr{\"o}der}}]{Mittag-2023}
{Mittag}, M., {Schmitt}, J.~H.~M.~M., \& {Schr{\"o}der}, K.~P. 2023, \aap, 674,
  A116

\bibitem[{{Mittag} {et~al.}(2016){Mittag}, {Schr{\"o}der}, {Hempelmann},
  {Gonz{\'a}lez-P{\'e}rez}, \& {Schmitt}}]{Mittag-2016}
{Mittag}, M., {Schr{\"o}der}, K.~P., {Hempelmann}, A.,
  {Gonz{\'a}lez-P{\'e}rez}, J.~N., \& {Schmitt}, J.~H.~M.~M. 2016, \aap, 591,
  A89

\bibitem[{{N{\`e}mec} {et~al.}(2020){N{\`e}mec}, {Shapiro}, {Krivova},
  {Solanki}, {Tagirov}, {Cameron}, \& {Dreizler}}]{Nemec-2020}
{N{\`e}mec}, N.~E., {Shapiro}, A.~I., {Krivova}, N.~A., {et~al.} 2020, \aap,
  636, A43

\bibitem[{{Noyes} {et~al.}(1984){Noyes}, {Hartmann}, {Baliunas}, {Duncan}, \&
  {Vaughan}}]{Noyes-1984}
{Noyes}, R.~W., {Hartmann}, L.~W., {Baliunas}, S.~L., {Duncan}, D.~K., \&
  {Vaughan}, A.~H. 1984, \apj, 279, 763

\bibitem[{{Radick} {et~al.}(2018){Radick}, {Lockwood}, {Henry}, {Hall}, \&
  {Pevtsov}}]{Radick-2018}
{Radick}, R.~R., {Lockwood}, G.~W., {Henry}, G.~W., {Hall}, J.~C., \&
  {Pevtsov}, A.~A. 2018, \apj, 855, 75

\bibitem[{{Reinhold} {et~al.}(2020){Reinhold}, {Shapiro}, {Solanki}, {Montet},
  {Krivova}, {Cameron}, \& {Amazo-G{\'o}mez}}]{Reinhold-2020}
{Reinhold}, T., {Shapiro}, A.~I., {Solanki}, S.~K., {et~al.} 2020, Science,
  368, 518

\bibitem[{{Schatten}(1993)}]{Schatten-1993}
{Schatten}, K.~H. 1993, \jgr, 98, 18907

\bibitem[{{Shapiro} {et~al.}(2014){Shapiro}, {Solanki}, {Krivova}, {Schmutz},
  {Ball}, {Knaack}, {Rozanov}, \& {Unruh}}]{Shapiro-2014}
{Shapiro}, A.~I., {Solanki}, S.~K., {Krivova}, N.~A., {et~al.} 2014, \aap, 569,
  A38

\bibitem[{{Singh} {et~al.}(2023){Singh}, {Priyal}, {Ravindra}, {Bertello}, \&
  {Pevtsov}}]{2023-Singh}
{Singh}, J., {Priyal}, M., {Ravindra}, B., {Bertello}, L., \& {Pevtsov}, A.
  2023, Research in Astronomy and Astrophysics, 23, 045016

\bibitem[{{Sowmya} {et~al.}(2021){Sowmya}, {Shapiro}, {Witzke}, {N{\`e}mec},
  {Chatzistergos}, {Yeo}, {Krivova}, \& {Solanki}}]{Sowmya-2021_I}
{Sowmya}, K., {Shapiro}, A.~I., {Witzke}, V., {et~al.} 2021, \apj, 914, 21

\bibitem[{{Vanden Broeck} {et~al.}(2024){Vanden Broeck}, {Bechet}, {Clette},
  {Rauw}, {Schröder}, \& {Mittag}}]{VandenBroeck-2024}
{Vanden Broeck}, G., {Bechet}, S., {Clette}, F., {et~al.} 2024, \aap

\bibitem[{{Vaughan} {et~al.}(1981){Vaughan}, {Baliunas}, {Middelkoop},
  {Hartmann}, {Mihalas}, {Noyes}, \& {Preston}}]{Vaughan-1981}
{Vaughan}, A.~H., {Baliunas}, S.~L., {Middelkoop}, F., {et~al.} 1981, \apj,
  250, 276

\bibitem[{{Veronig} {et~al.}(2021){Veronig}, {Jain}, {Podladchikova},
  {P{\"o}tzi}, \& {Clette}}]{Veronig-2021}
{Veronig}, A.~M., {Jain}, S., {Podladchikova}, T., {P{\"o}tzi}, W., \&
  {Clette}, F. 2021, \aap, 652, A56

\bibitem[{{Wilson}(1978)}]{Wilson-1978}
{Wilson}, O.~C. 1978, \apj, 226, 379

\bibitem[{{Wright} {et~al.}(2004){Wright}, {Marcy}, {Butler}, \&
  {Vogt}}]{Wright-2004}
{Wright}, J.~T., {Marcy}, G.~W., {Butler}, R.~P., \& {Vogt}, S.~S. 2004, \apjs,
  152, 261

\end{thebibliography}
\end{small}

\phantomsection
\addcontentsline{toc}{section}{Annexes}

\begin{appendix}
    \onecolumn

\section{Generation of solar-disk views for different inclinations.}\label{Annexe_A}

\begin{figure}[h]

    \centering
    
    \includegraphics[width=0.113\hsize]{images/Sun_20140608_0.png}
    \put(-90,28){\Large{$0^{\circ}$}} \hspace{1.5cm}
    \includegraphics[width=0.113\hsize]{images/Sun_20140608_0.png}
    \put(20,28){\Large{$0^{\circ}$}} \\
    \includegraphics[width=0.113\hsize]{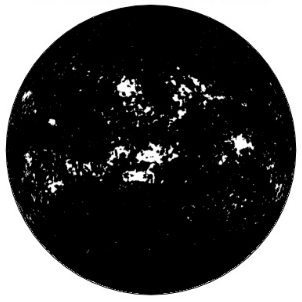}
    \put(-98,28){\Large{$-10^{\circ}$}} \hspace{1.5cm}
    \includegraphics[width=0.113\hsize]{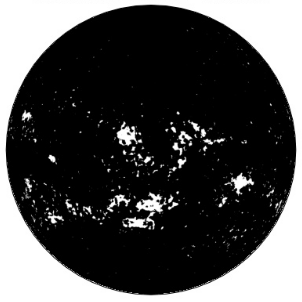}
    \put(15,28){\Large{$10^{\circ}$}} \\
    \includegraphics[width=0.113\hsize]{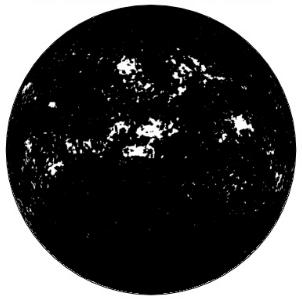}
    \put(-98,28){\Large{$-20^{\circ}$}} \hspace{1.5cm}
    \includegraphics[width=0.113\hsize]{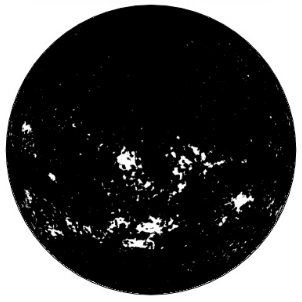}
    \put(15,28){\Large{$20^{\circ}$}} \\
    \includegraphics[width=0.113\hsize]{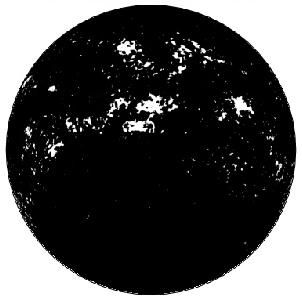} 
    \put(-98,28){\Large{$-30^{\circ}$}} \hspace{1.5cm}
    \includegraphics[width=0.113\hsize]{images/Sun_20140608_30.png}
    \put(15,28){\Large{$30^{\circ}$}} \\
    \includegraphics[width=0.113\hsize]{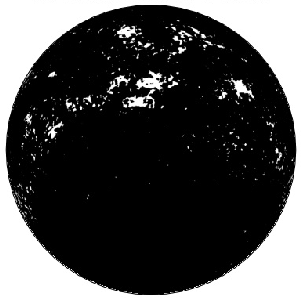}
    \put(-98,28){\Large{$-40^{\circ}$}} \hspace{1.5cm}
    \includegraphics[width=0.113\hsize]{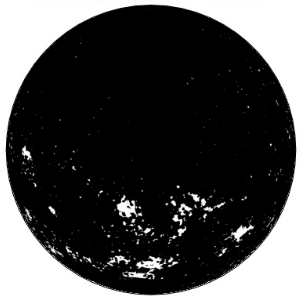}
    \put(15,28){\Large{$40^{\circ}$}} \\
    \includegraphics[width=0.113\hsize]{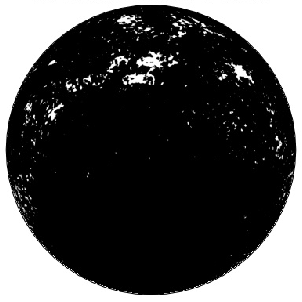}
    \put(-98,28){\Large{$-50^{\circ}$}} \hspace{1.5cm}
    \includegraphics[width=0.113\hsize]{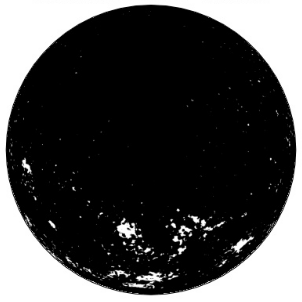}
    \put(15,28){\Large{$50^{\circ}$}} \\
    \includegraphics[width=0.113\hsize]{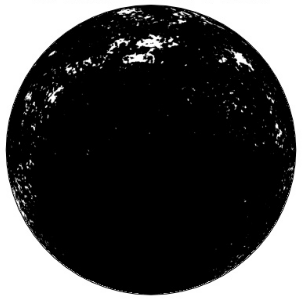}
    \put(-98,28){\Large{$-60^{\circ}$}} \hspace{1.5cm}
    \includegraphics[width=0.113\hsize]{images/Sun_20140608_60.png}
    \put(15,28){\Large{$60^{\circ}$}} \\
    \includegraphics[width=0.113\hsize]{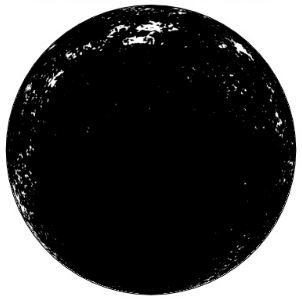}
    \put(-98,28){\Large{$-70^{\circ}$}} \hspace{1.5cm}
    \includegraphics[width=0.113\hsize]{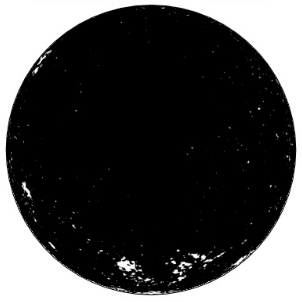}
    \put(15,28){\Large{$70^{\circ}$}} \\
    \includegraphics[width=0.113\hsize]{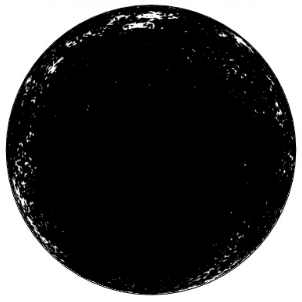}
    \put(-98,28){\Large{$-80^{\circ}$}} \hspace{1.5cm}
    \includegraphics[width=0.113\hsize]{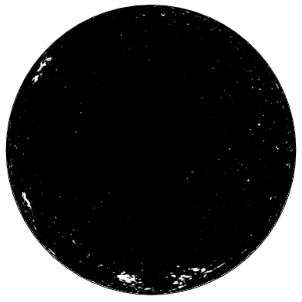}
    \put(15,28){\Large{$80^{\circ}$}} \\
    \includegraphics[width=0.113\hsize]{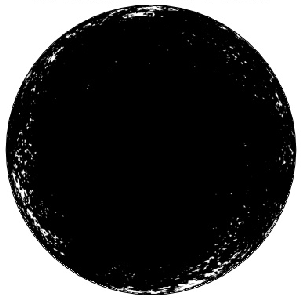}
    \put(-98,28){\Large{$-90^{\circ}$}} \hspace{1.5cm}
    \includegraphics[width=0.113\hsize]{images/Sun_20140608_90.png}
    \put(15,28){\Large{$90^{\circ}$}}
    
    \caption{Distribution of the segmented structures with the generated solar-disk views on 1st of August 2014 for inclinations from $0^{\circ}$ of latitude to $90^{\circ}$ and $-90^{\circ}$ of latitude by steps of $10^{\circ}$. Inclination angles are specified next to each image and represent the number of degree relative to the Equator-on view ($i = 0^{\circ}$). Top images illustrate the view from the Equator and it goes to the South Pole view (left) and to the North Pole view (right).}
    \label{fig:Solar_images_inclinations}
\end{figure}

    \onecolumn

\section{Fourier power spectra for different inclinations}\label{Annexe_B}

\begin{figure}[h]
    \vspace{-0.3cm}
    \centering 
    \includegraphics[width=0.4\hsize]{images/Fourier/power0deg.png} \\
    \includegraphics[width=0.4\hsize]{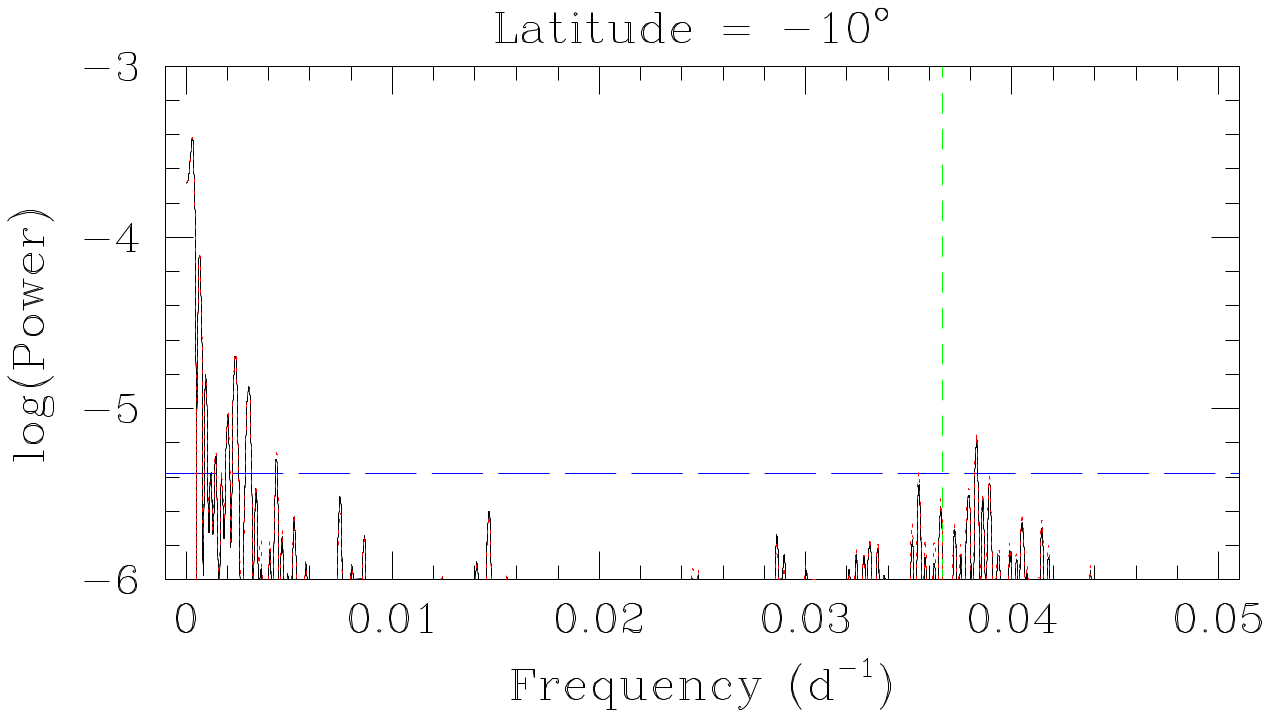}
    \hfill
    \includegraphics[width=0.4\hsize]{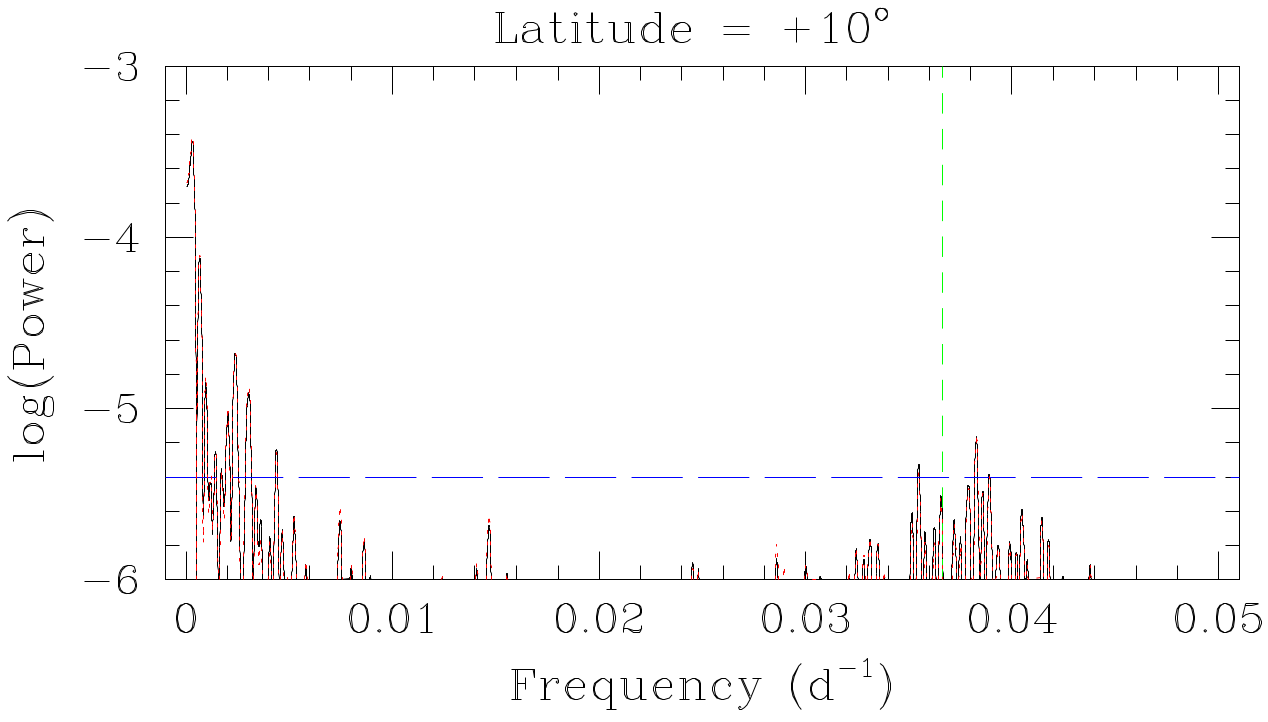}
    \hfill
    \includegraphics[width=0.4\hsize]{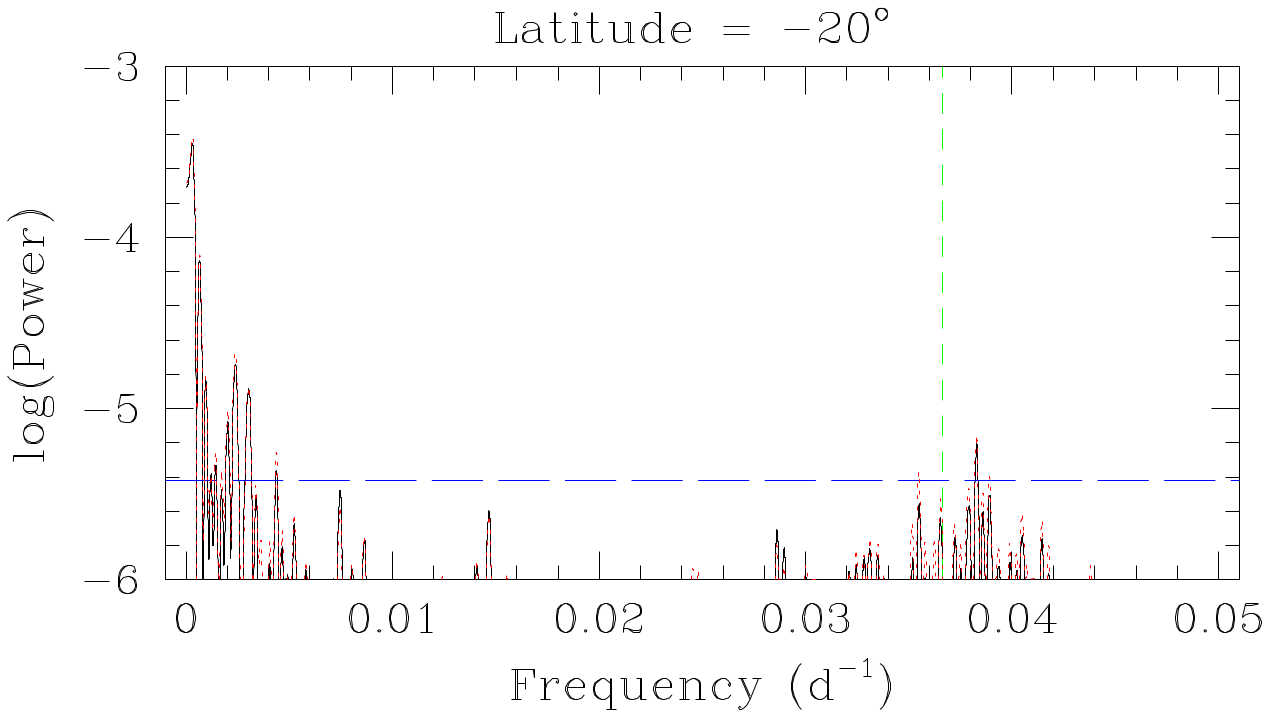}
    \hfill
    \includegraphics[width=0.4\hsize]{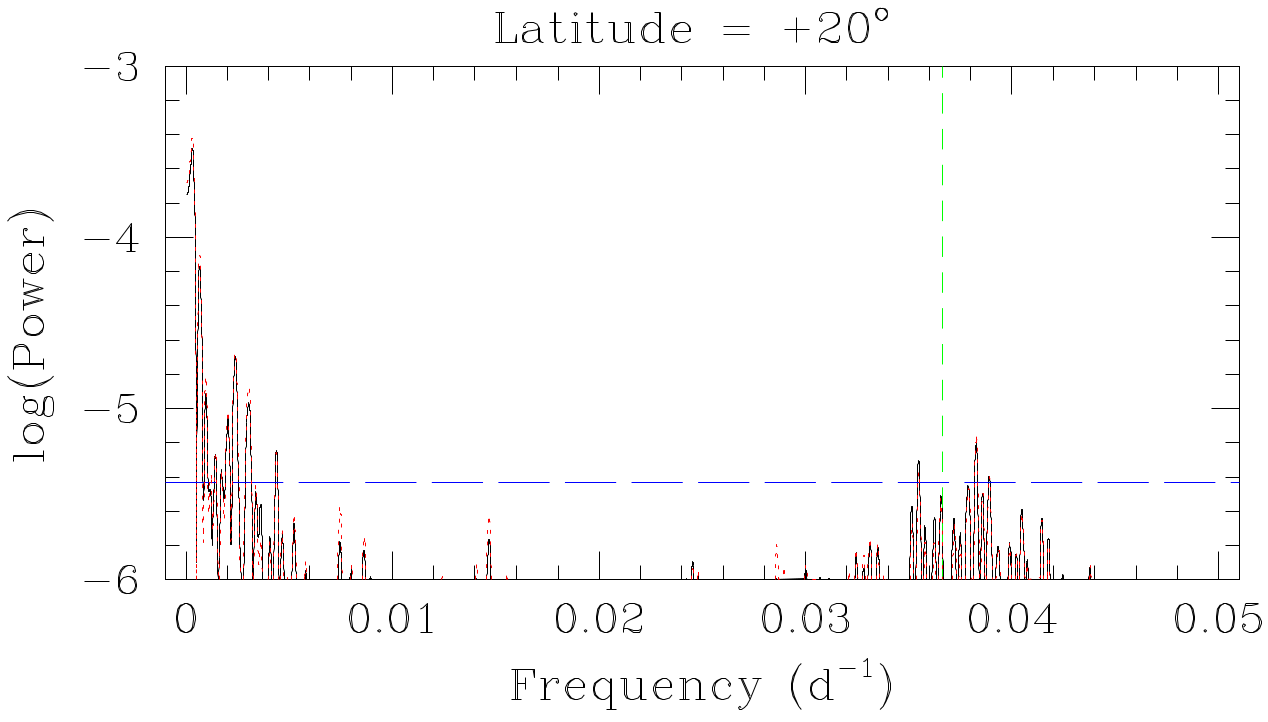}
    \hfill
    \includegraphics[width=0.4\hsize]{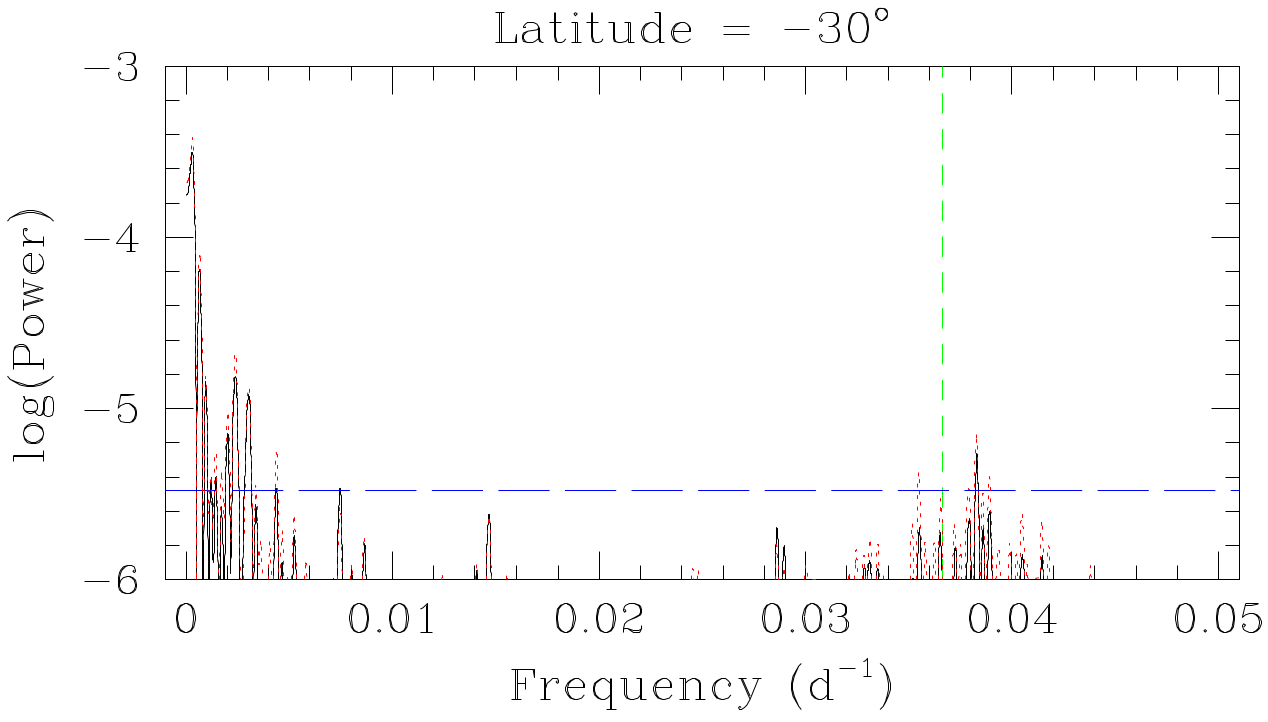}
    \hfill
    \includegraphics[width=0.4\hsize]{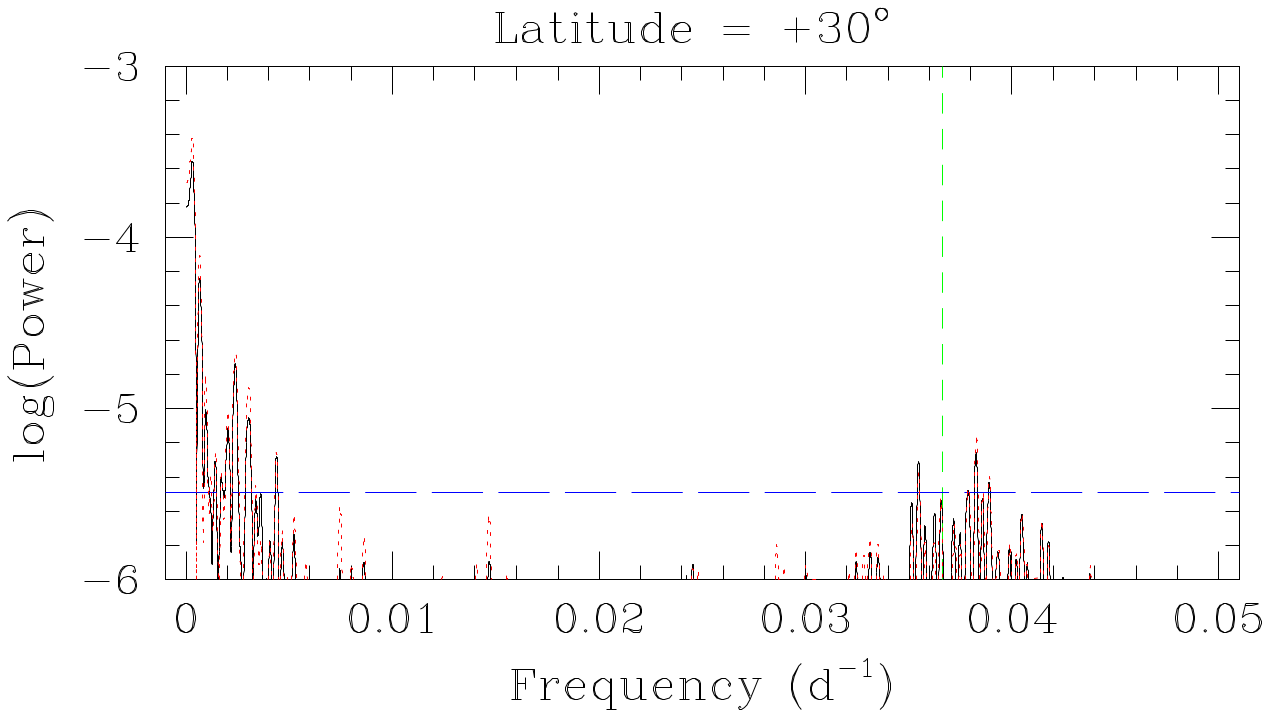}
    \hfill
    \includegraphics[width=0.4\hsize]{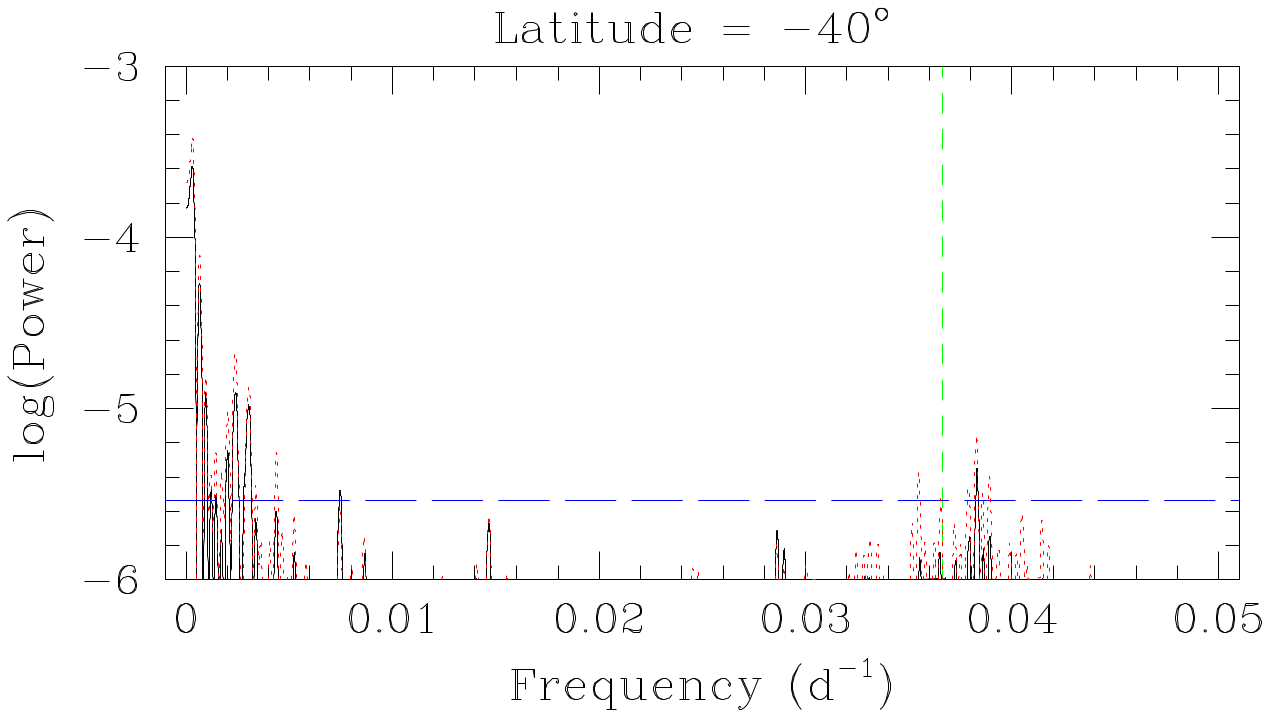}
    \hfill
    \includegraphics[width=0.4\hsize]{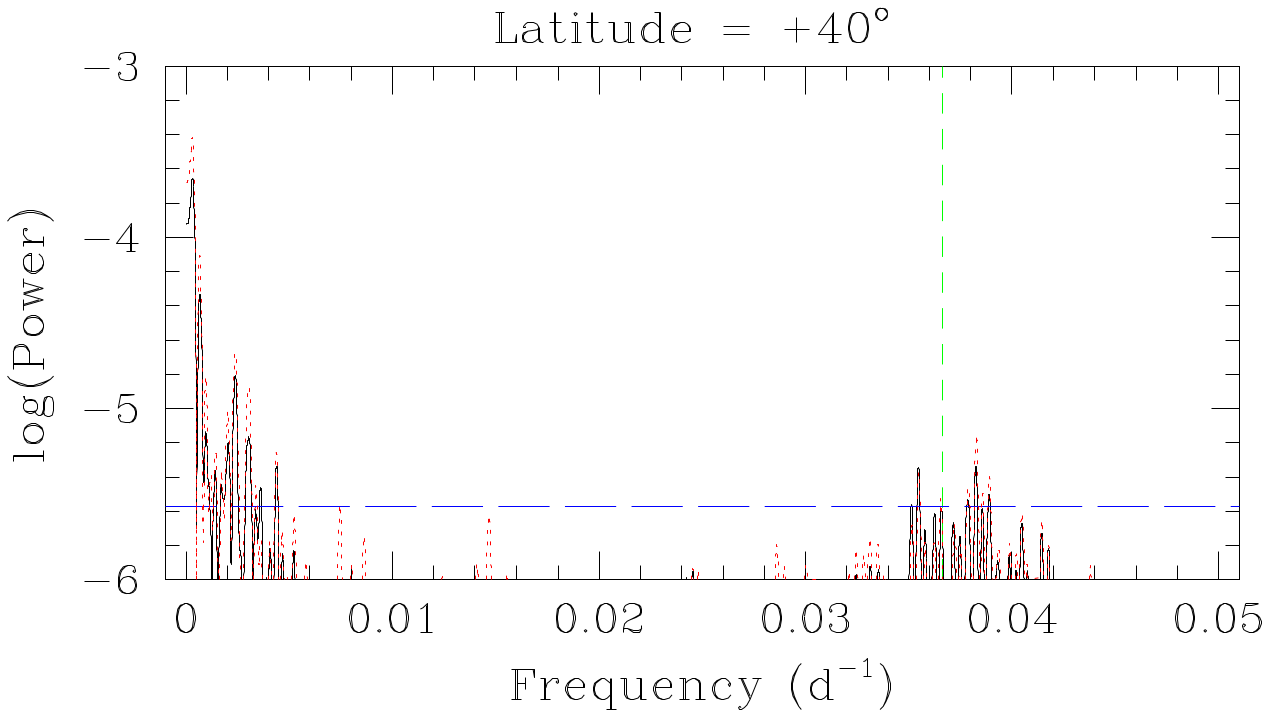}
    \caption{Fourier power spectra for different inclination angles and for frequencies below 0.05 d$^{-1}$. Inclination angles are specified at the top of the plots and represent the number of degrees relative to the Equator-on ($i = 0^{\circ}$). Left panels: inclinations to the South Pole view ; Right panels: inclinations to the North Pole view. In each panel, the actual power spectrum for that inclination is shown in black, whereas the red dotted curve replicates the power spectrum for $i = 0^{\circ}$. The green dashed line yields $\nu_{\rm Car}$ while the long-dashed blue horizontal line yields the 99\% significance level.}
\end{figure}

\begin{figure}[h]
    \ContinuedFloat
    \includegraphics[width=0.4\hsize]{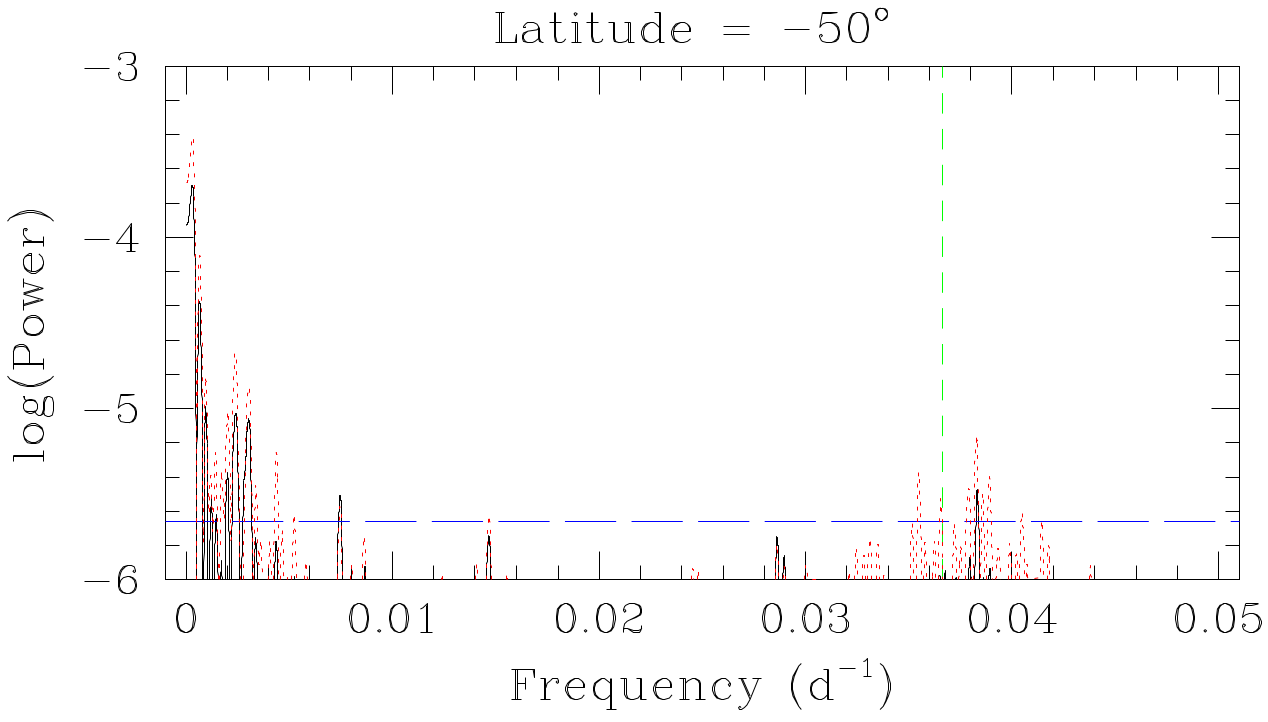}
    \hfill
    \includegraphics[width=0.4\hsize]{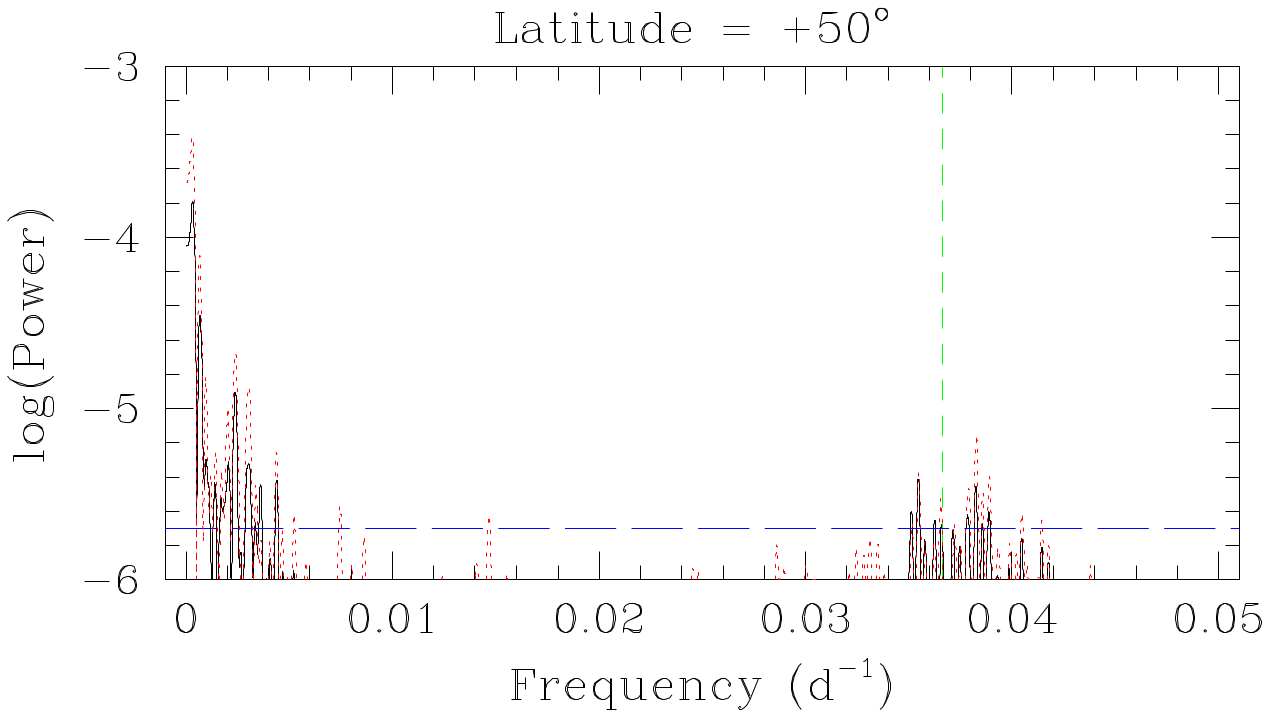}
    \hfill
    \includegraphics[width=0.4\hsize]{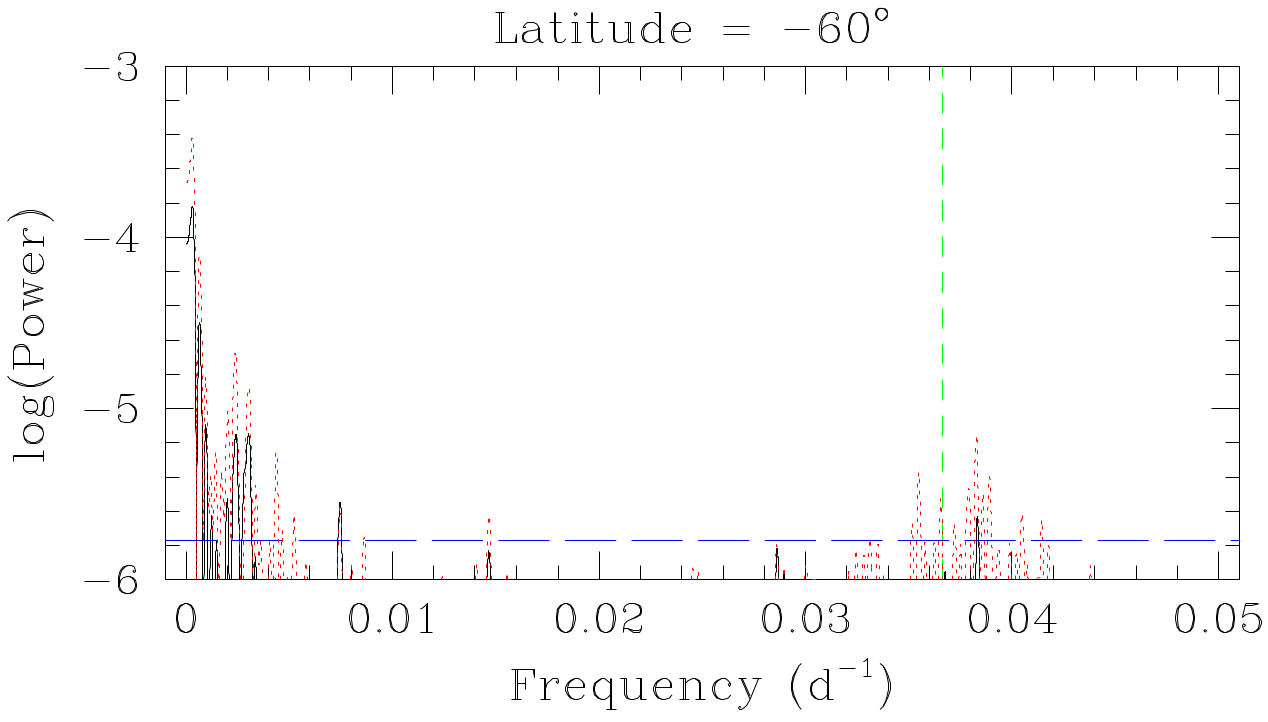}
    \hfill
    \includegraphics[width=0.4\hsize]{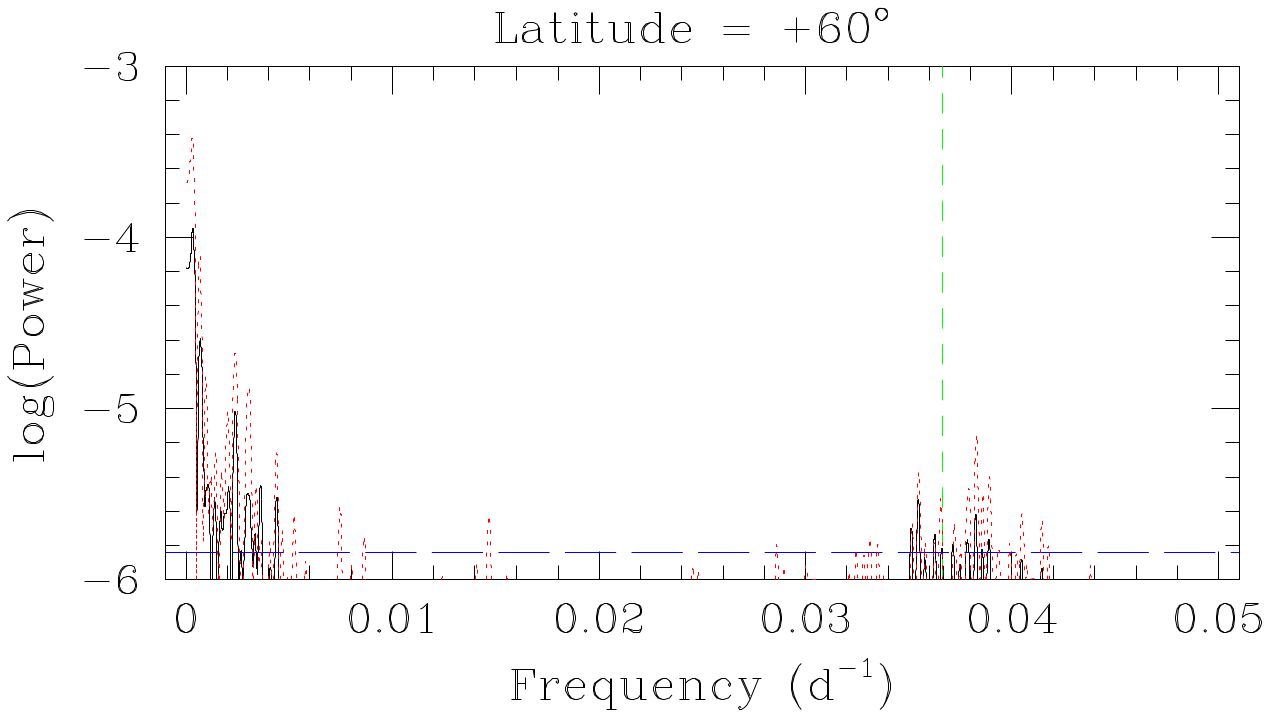}
    \hfill
    \includegraphics[width=0.4\hsize]{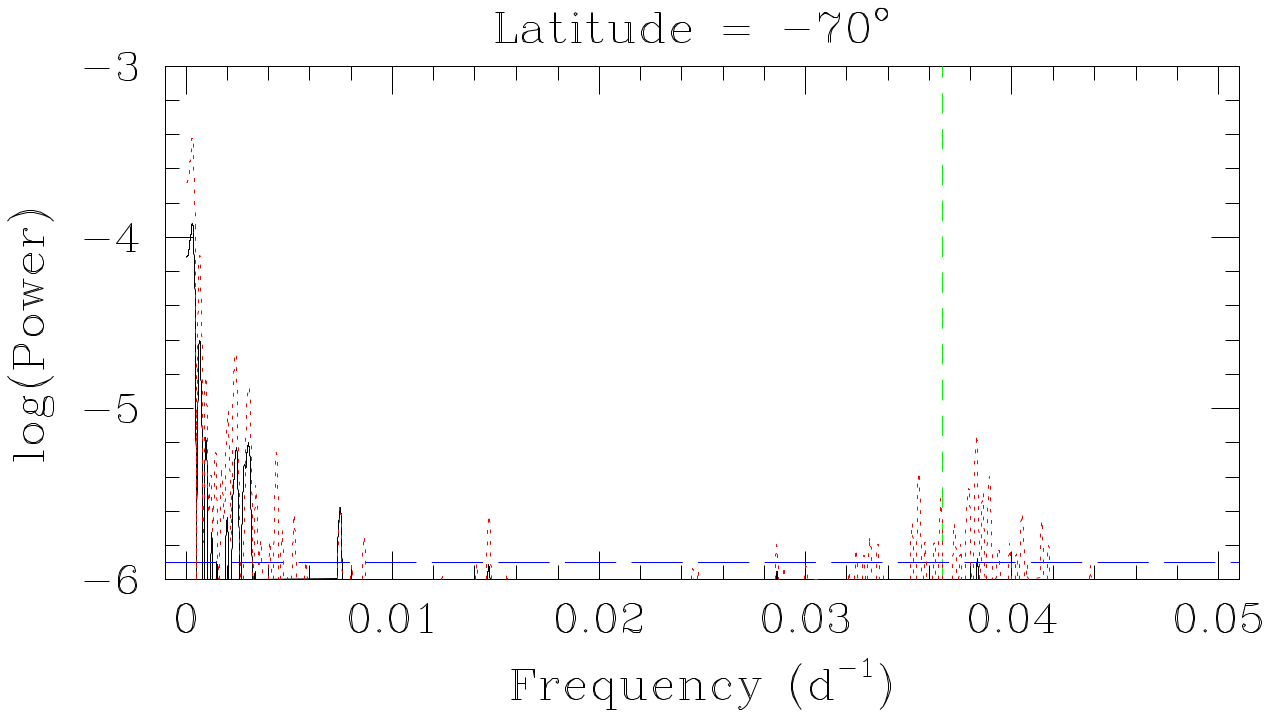}
    \hfill
    \includegraphics[width=0.4\hsize]{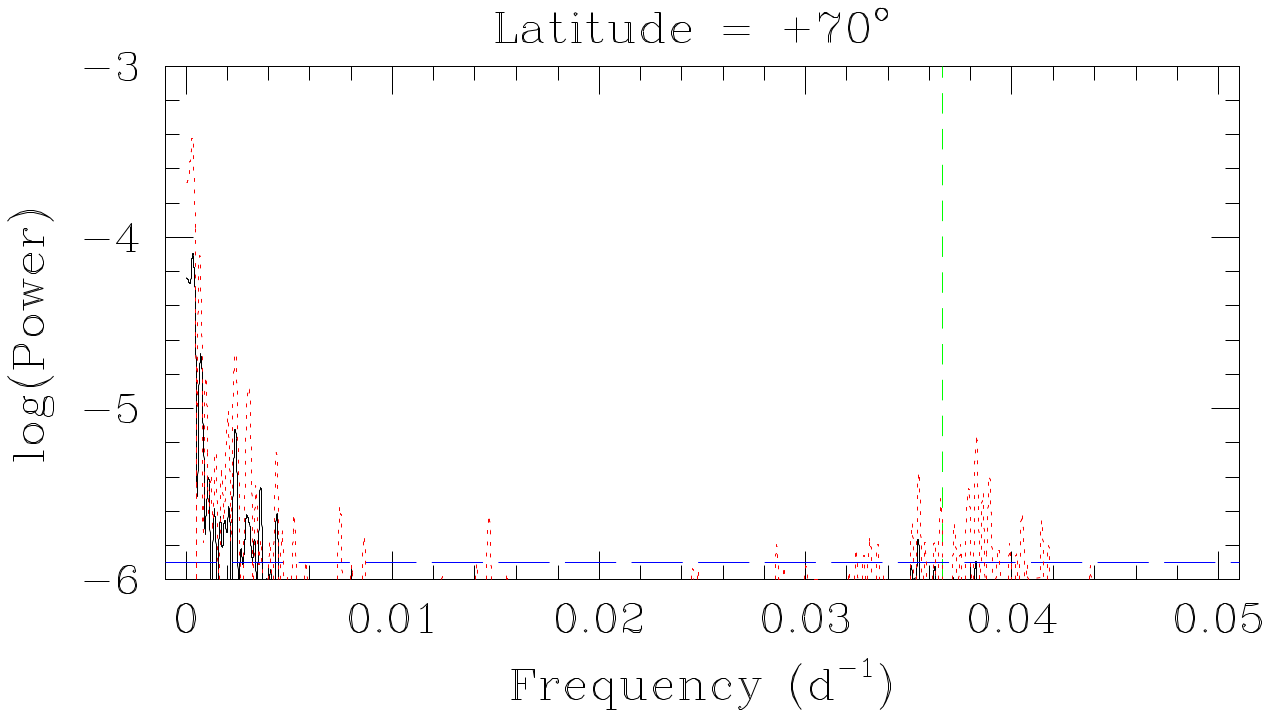}
    \hfill
    \includegraphics[width=0.4\hsize]{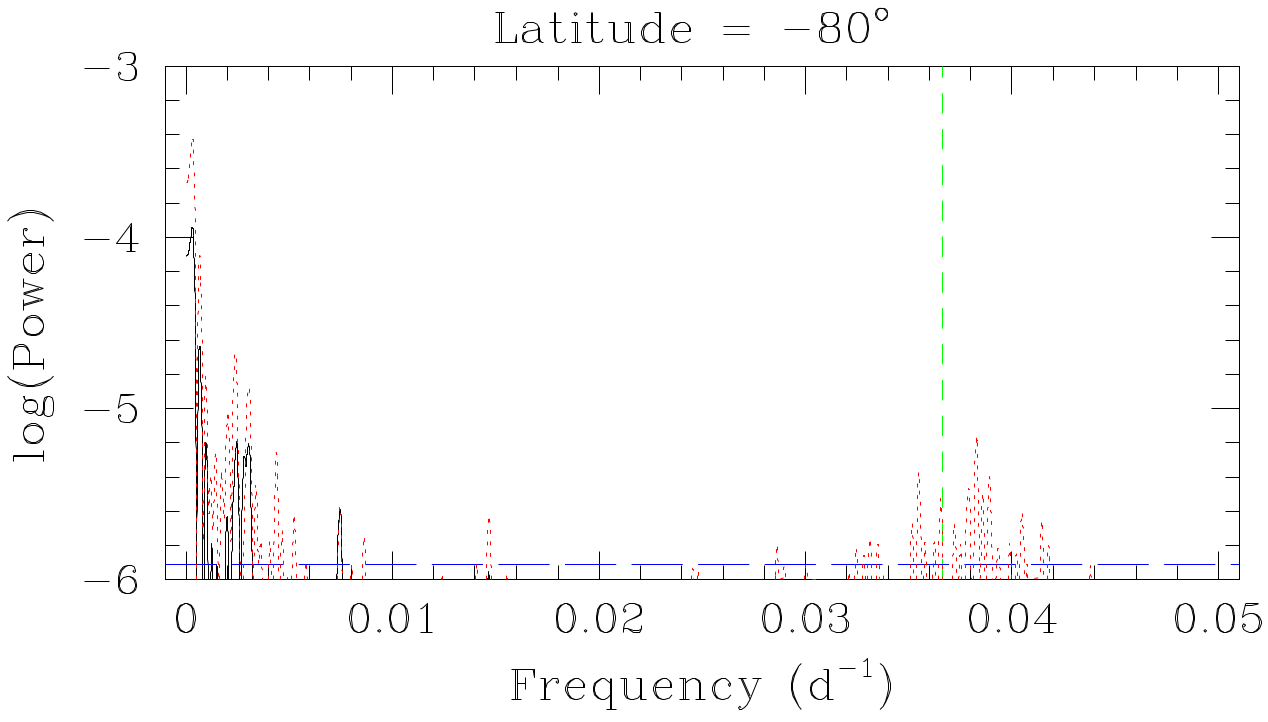}
    \hfill
    \includegraphics[width=0.4\hsize]{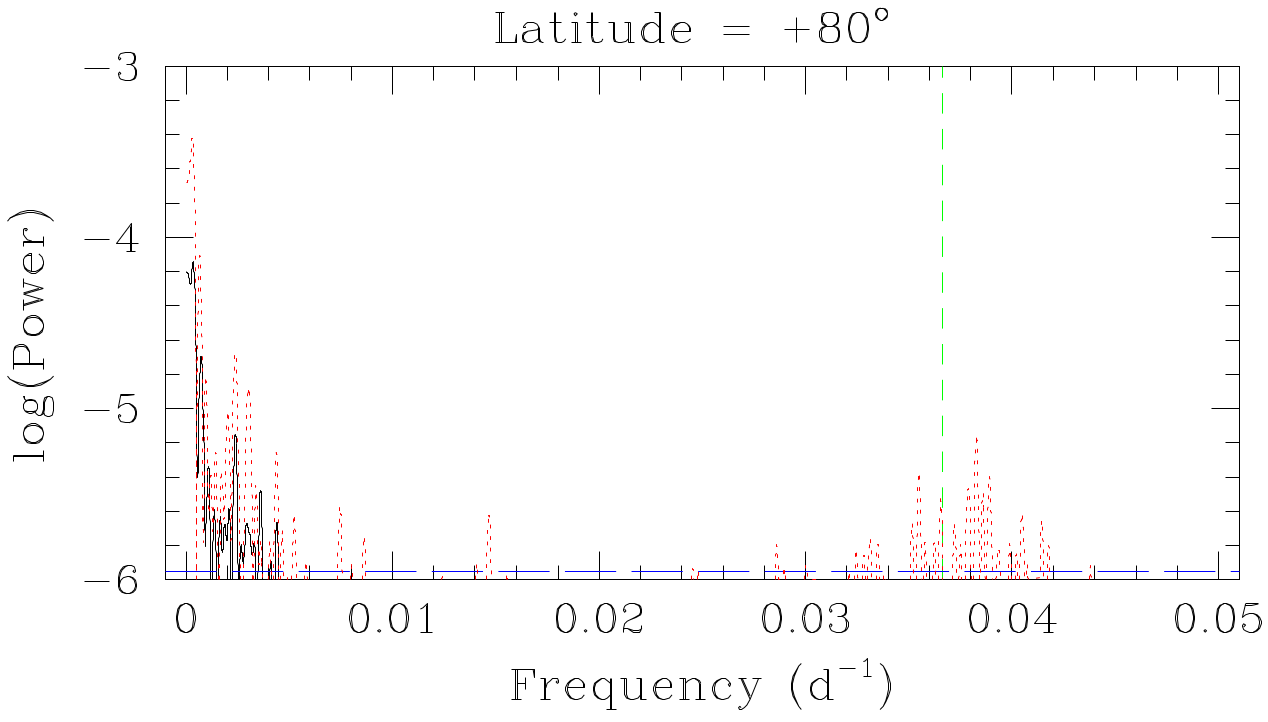}
    \hfill
    \includegraphics[width=0.4\hsize]{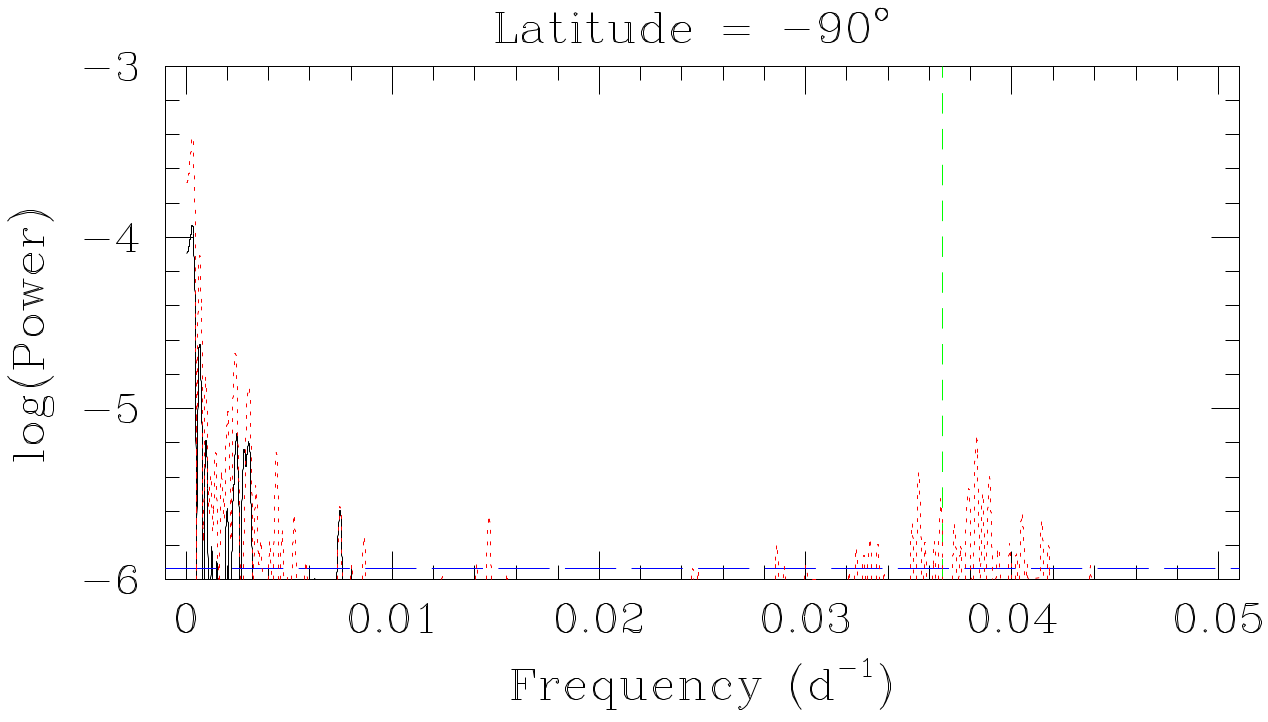}
    \hfill
    \includegraphics[width=0.4\hsize]{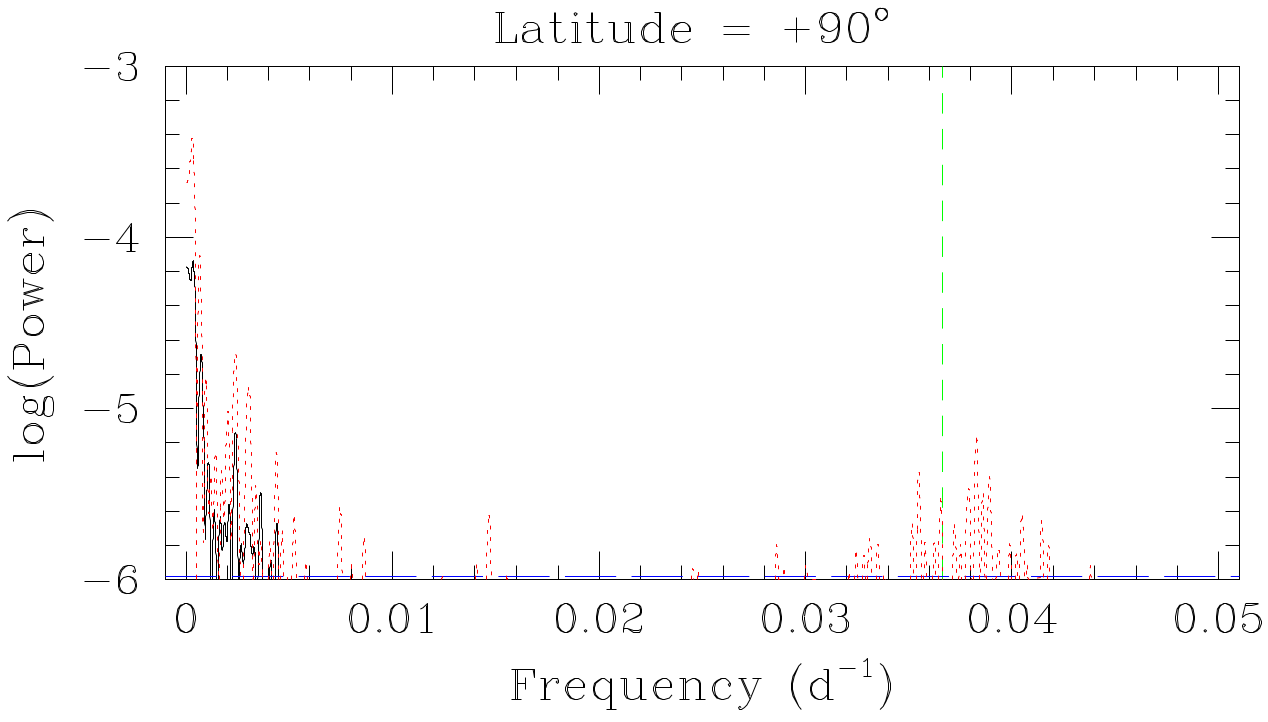}
    \caption{Continued}
    \label{fig:Fourier_power}
\end{figure}

    \onecolumn

\section{Fourier power spectra for different inclinations (near the solar rotation frequency)}\label{Annexe_C}

\begin{figure}[h]
    \vspace{-0.5cm}
    \centering 
    \includegraphics[width=0.435\hsize]{images/Fourier/rot0deg.pdf} \\
    \includegraphics[width=0.435\hsize]{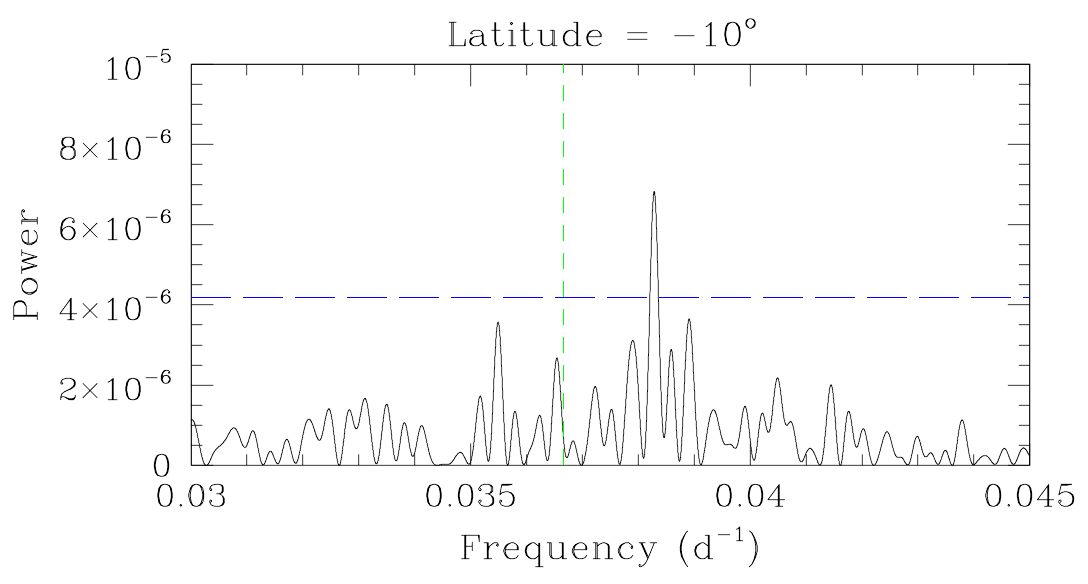}
    \hfill
    \includegraphics[width=0.435\hsize]{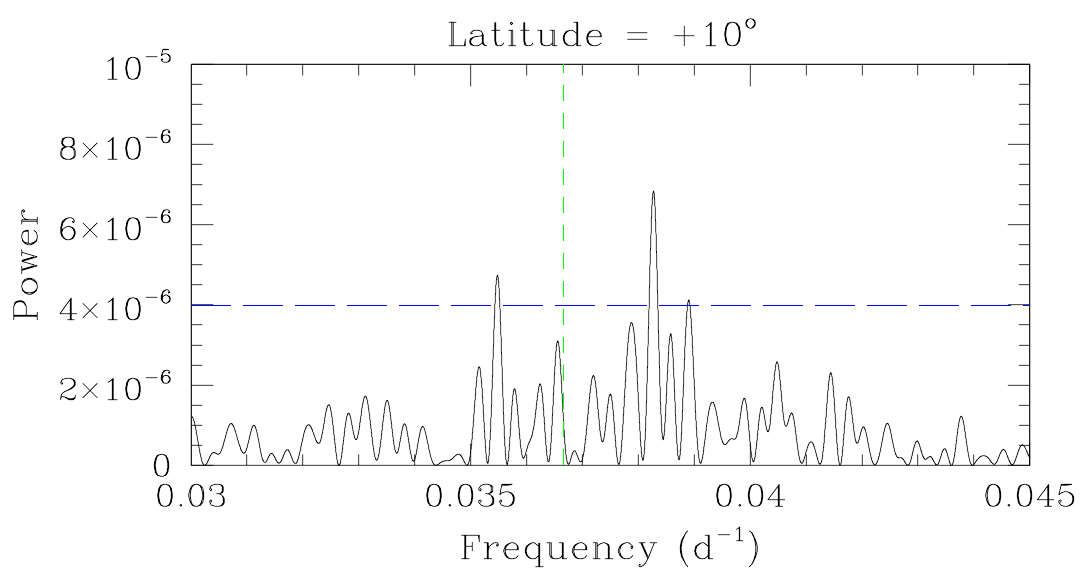}
    \hfill
    \includegraphics[width=0.435\hsize]{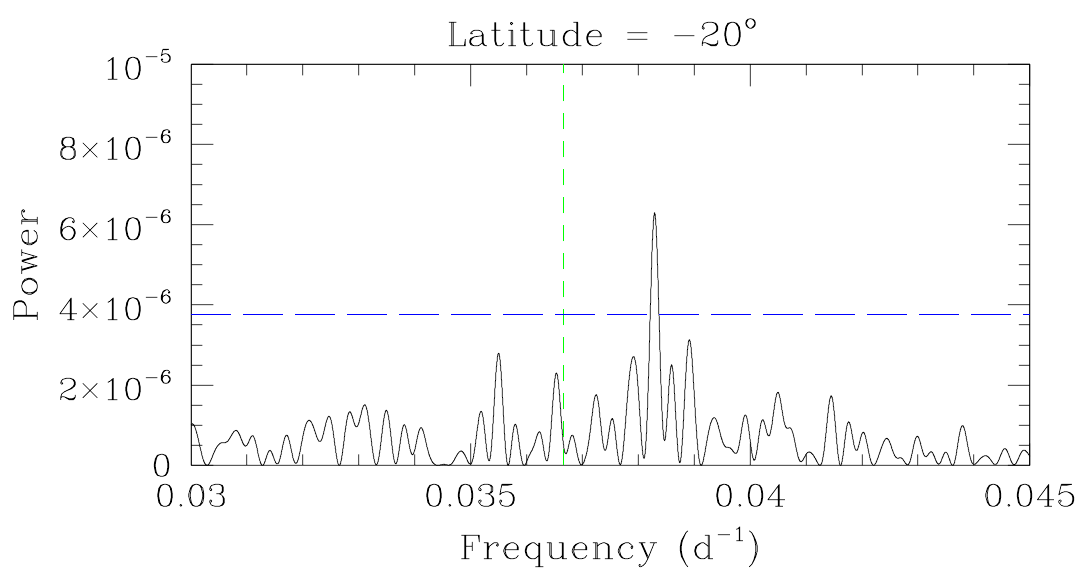}
    \hfill
    \includegraphics[width=0.435\hsize]{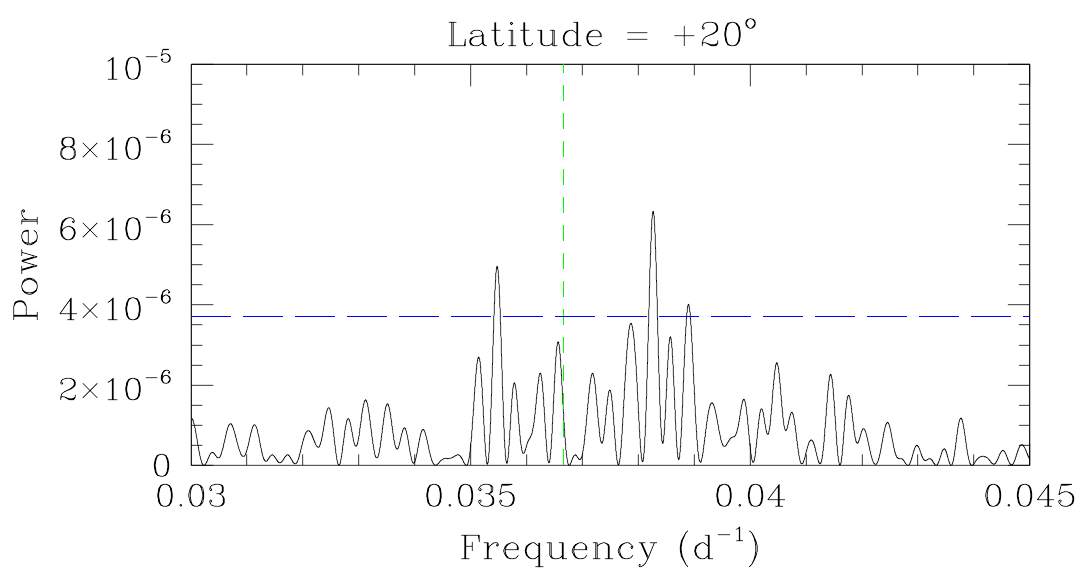}
    \hfill
    \includegraphics[width=0.435\hsize]{images/Fourier/rot30degs.pdf}
    \hfill
    \includegraphics[width=0.435\hsize]{images/Fourier/rot30degn.pdf}
    \hfill
    \includegraphics[width=0.435\hsize]{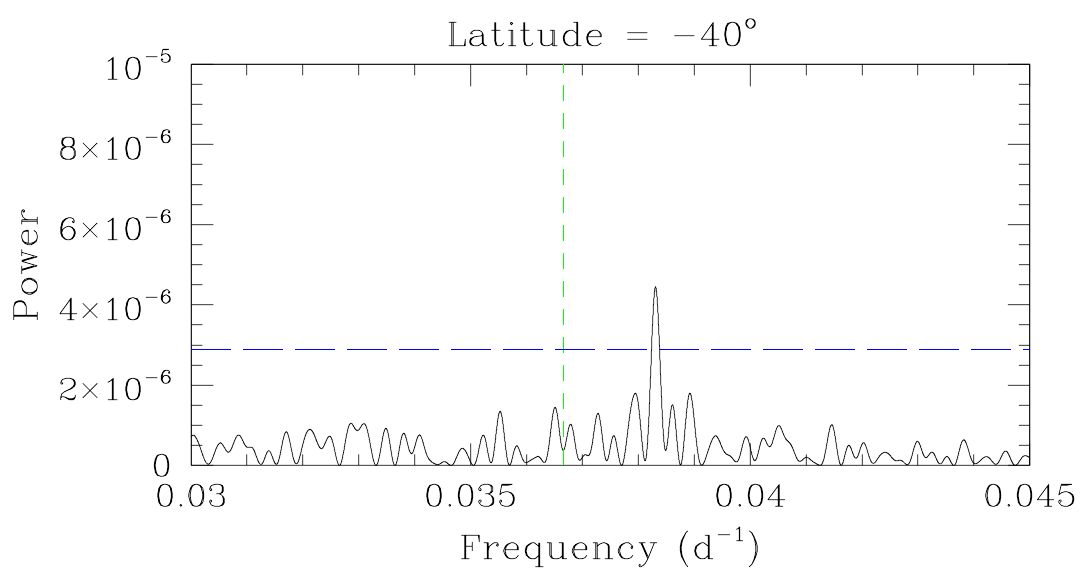}
    \hfill
    \includegraphics[width=0.435\hsize]{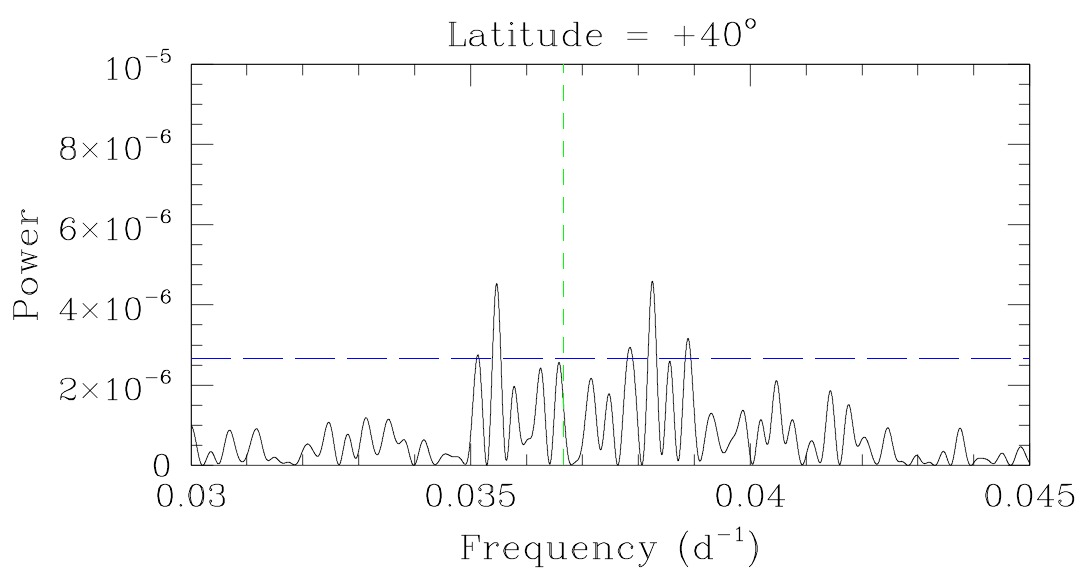}
    \caption{Evolution of the Fourier power spectrum near the rotation frequency of the Sun with inclinations specified at the top of the plot representing the number of degrees relative to the Equator-on ($i = 0^{\circ}$). Left panels: inclinations to the South Pole view ; Right panels: inclinations to the North Pole view. The green dashed line yields $\nu_{\rm Car}$, while the long-dashed blue horizontal line yields the 99\% significance level.}
\end{figure}

\begin{figure}[h]
    \ContinuedFloat
    \includegraphics[width=0.435\hsize]{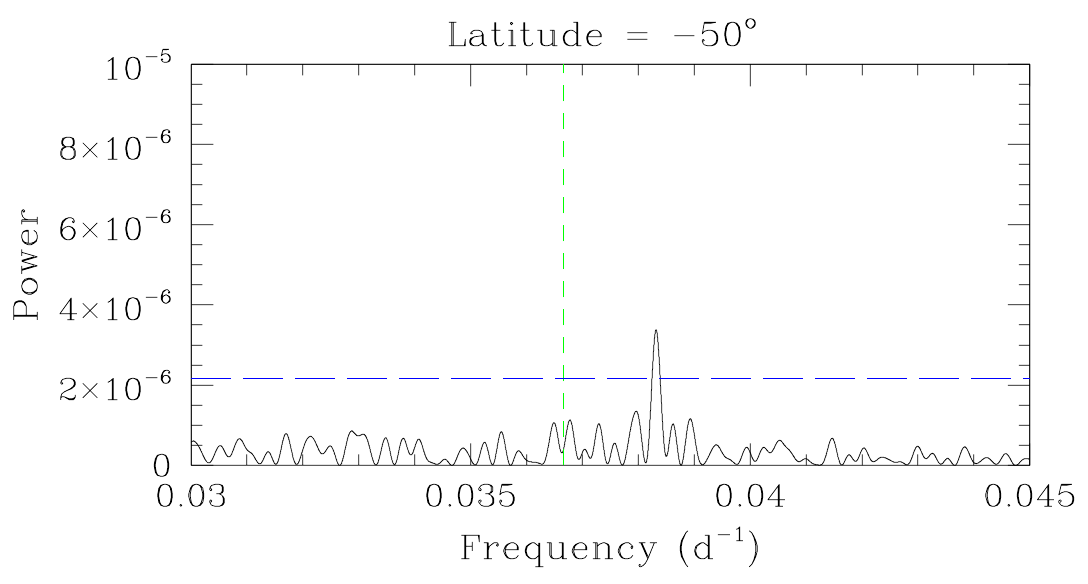}
    \hfill
    \includegraphics[width=0.435\hsize]{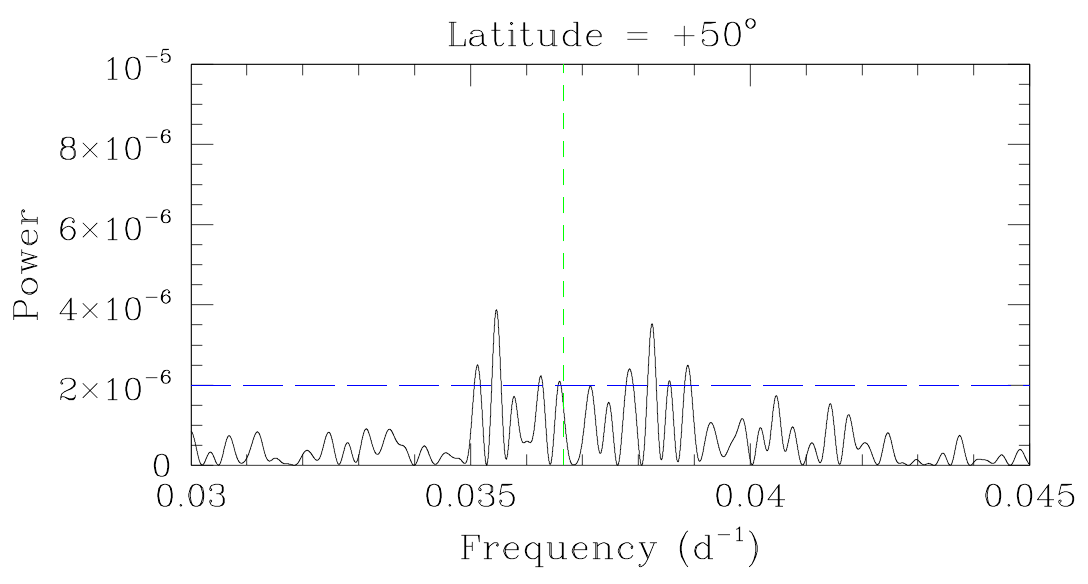}
    \hfill
    \includegraphics[width=0.435\hsize]{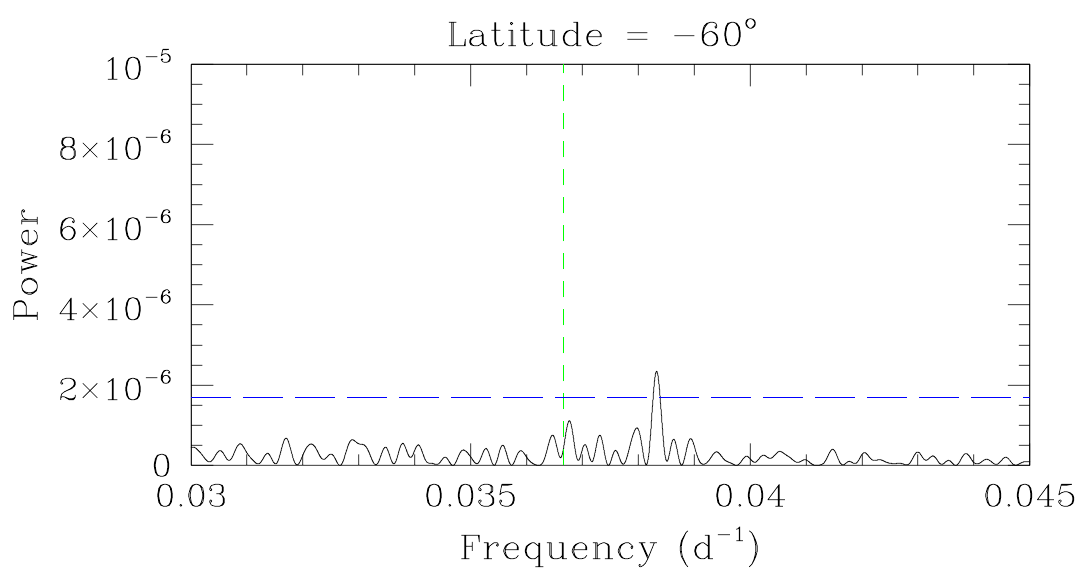}
    \hfill
    \includegraphics[width=0.435\hsize]{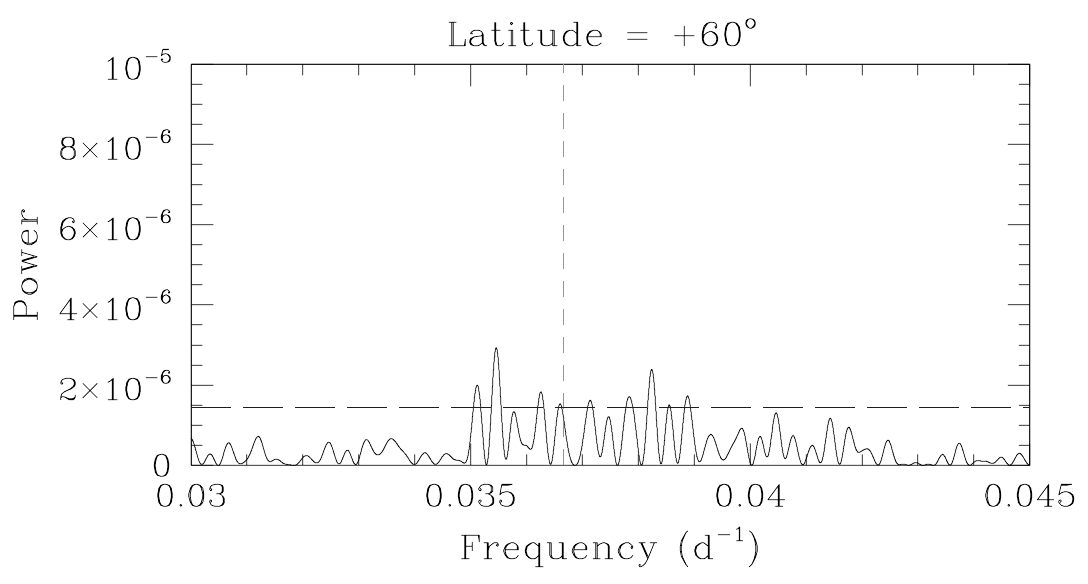}
    \hfill
    \includegraphics[width=0.435\hsize]{images/Fourier/rot70degs.pdf}
    \hfill
    \includegraphics[width=0.435\hsize]{images/Fourier/rot70degn.pdf}
    \hfill
    \includegraphics[width=0.435\hsize]{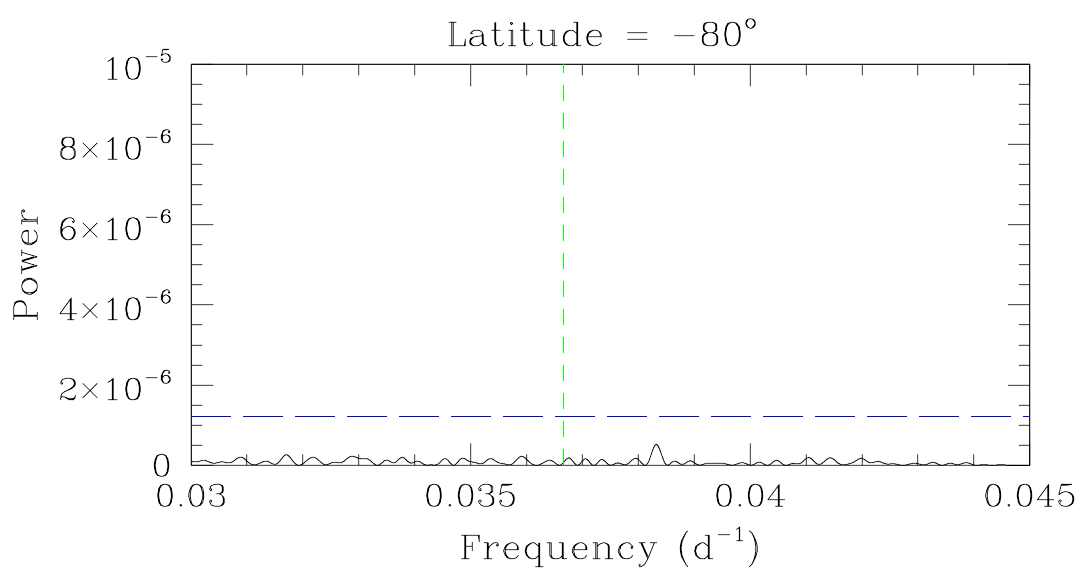}
    \hfill
    \includegraphics[width=0.435\hsize]{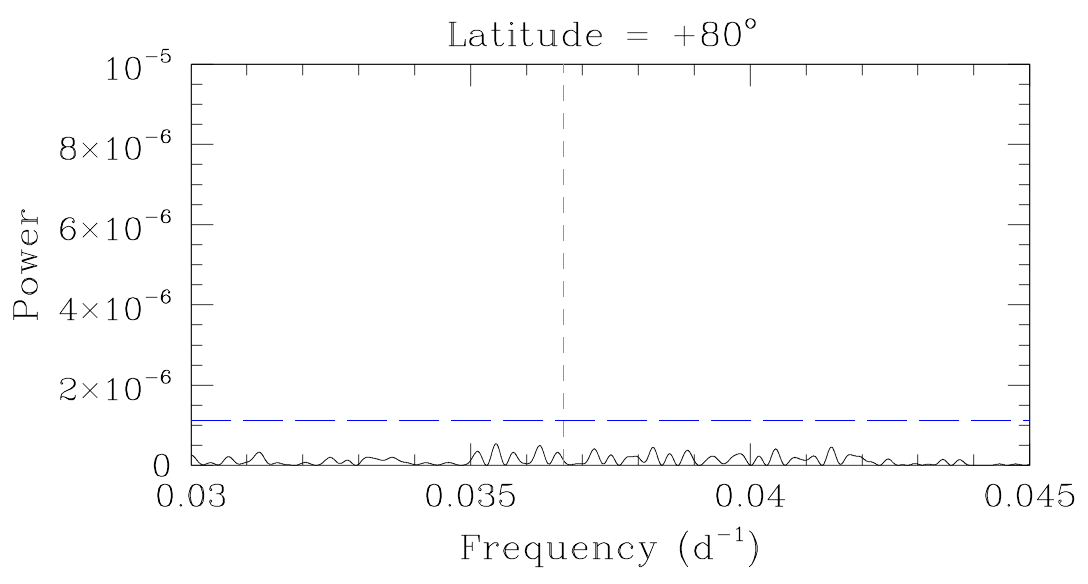}
    \hfill
    \includegraphics[width=0.435\hsize]{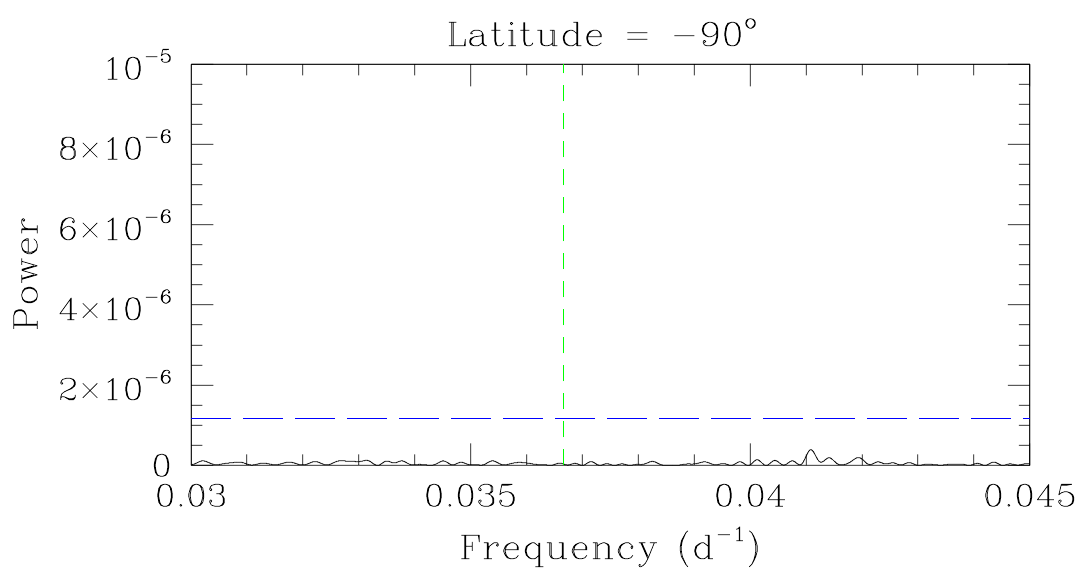}
    \hfill
    \includegraphics[width=0.435\hsize]{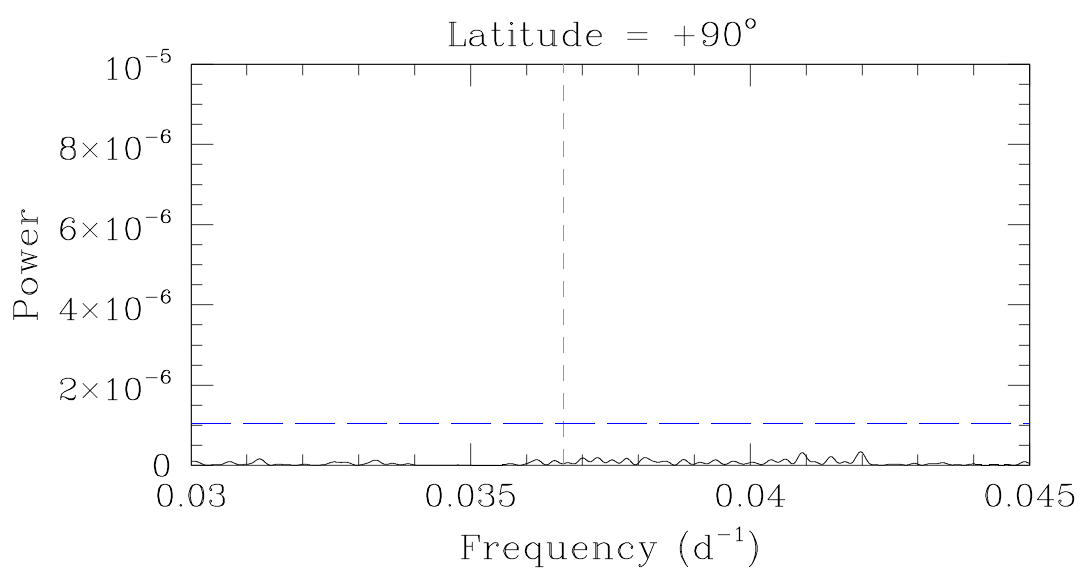}
    \caption{Continued}
    \label{fig:Fourier_rot}
\end{figure}

    \onecolumn

\section{Fourier power spectra for different inclinations (near the solar activity cycle frequency)}\label{Annexe_D}

\begin{figure}[h]
    \vspace{-0.5cm}
    \centering 
    \includegraphics[width=0.44\hsize]{images/Fourier/cyc0deg.pdf} \\
    \includegraphics[width=0.44\hsize]{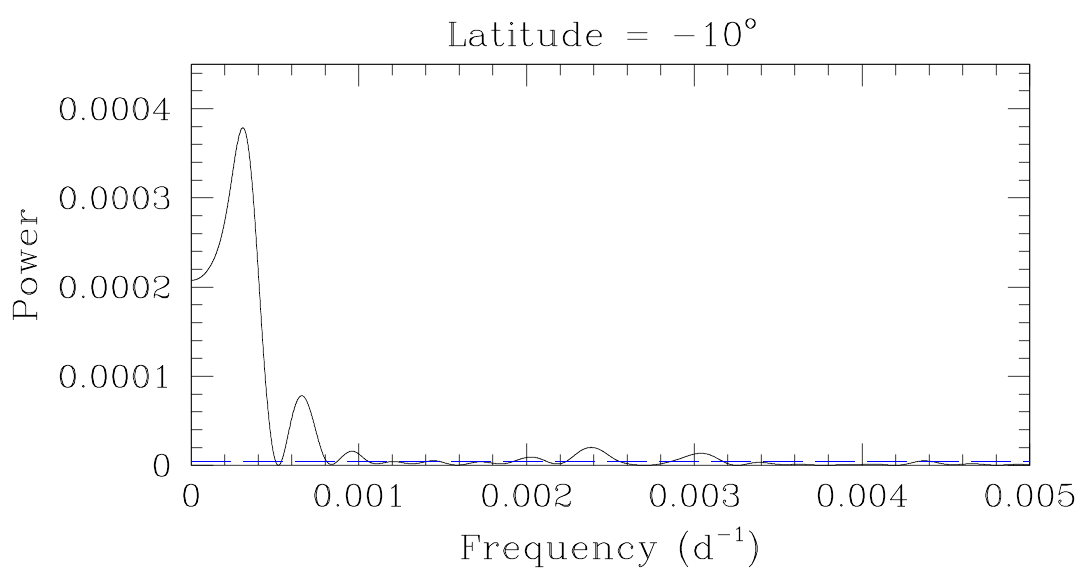}
    \hfill
    \includegraphics[width=0.44\hsize]{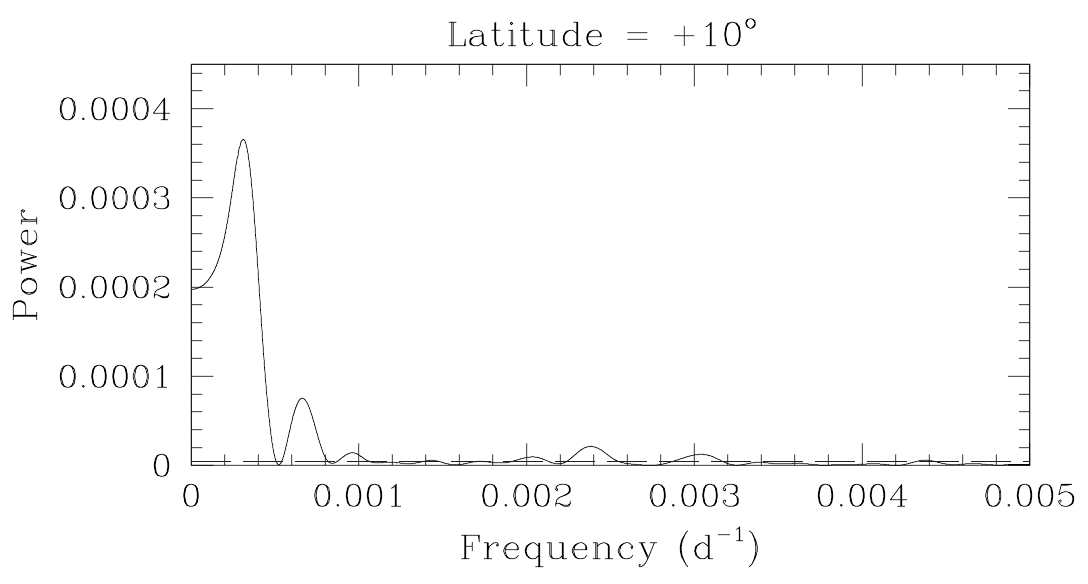}
    \hfill
    \includegraphics[width=0.44\hsize]{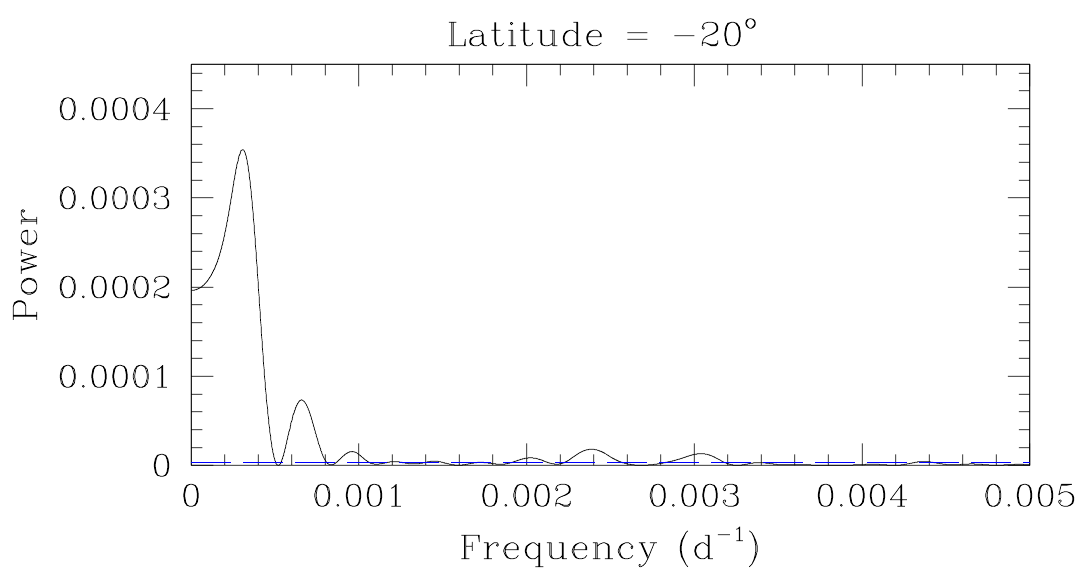}
    \hfill
    \includegraphics[width=0.44\hsize]{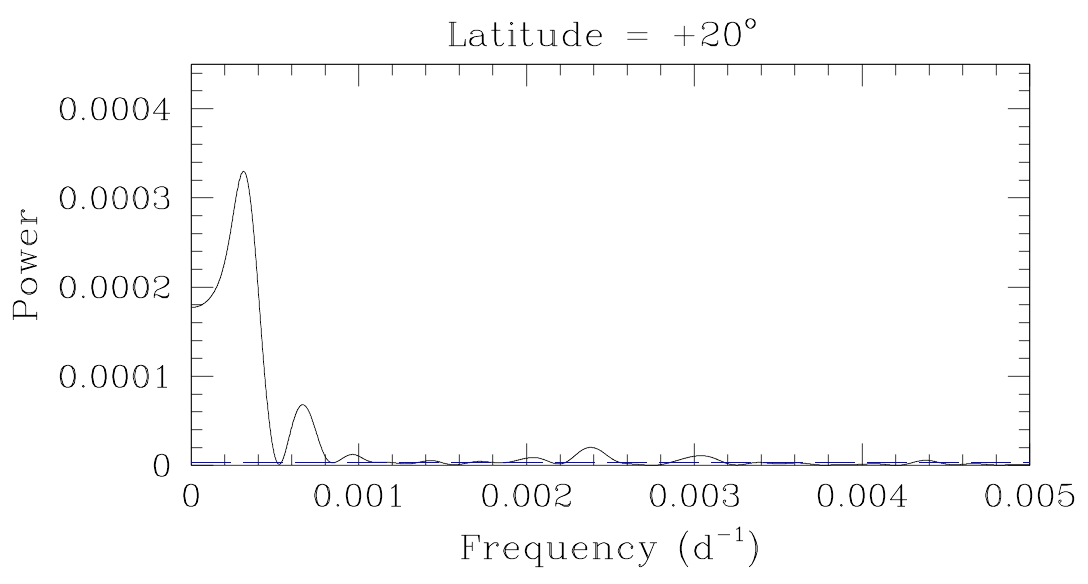}
    \hfill
    \includegraphics[width=0.44\hsize]{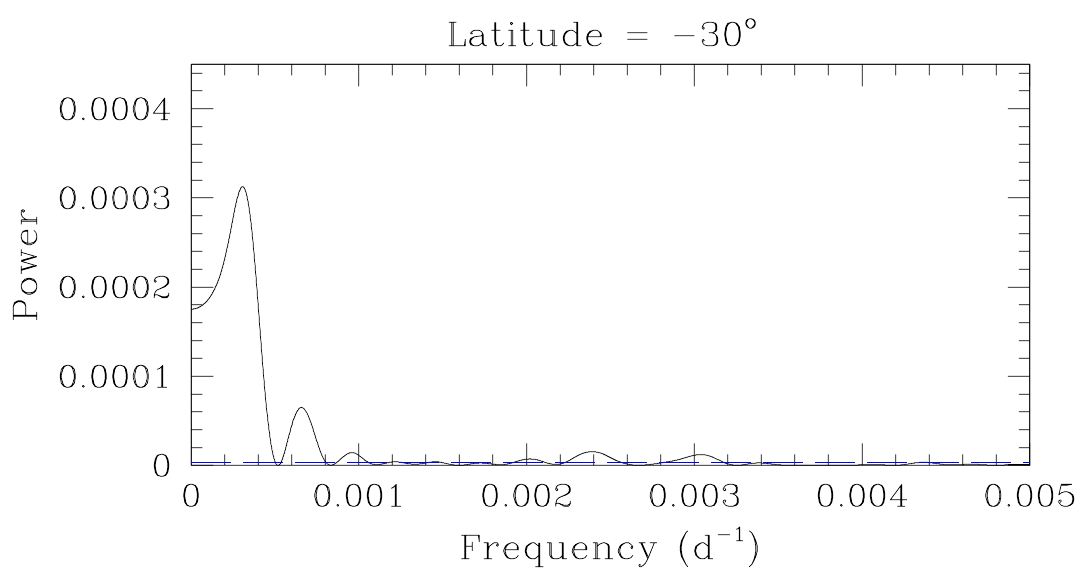}
    \hfill
    \includegraphics[width=0.44\hsize]{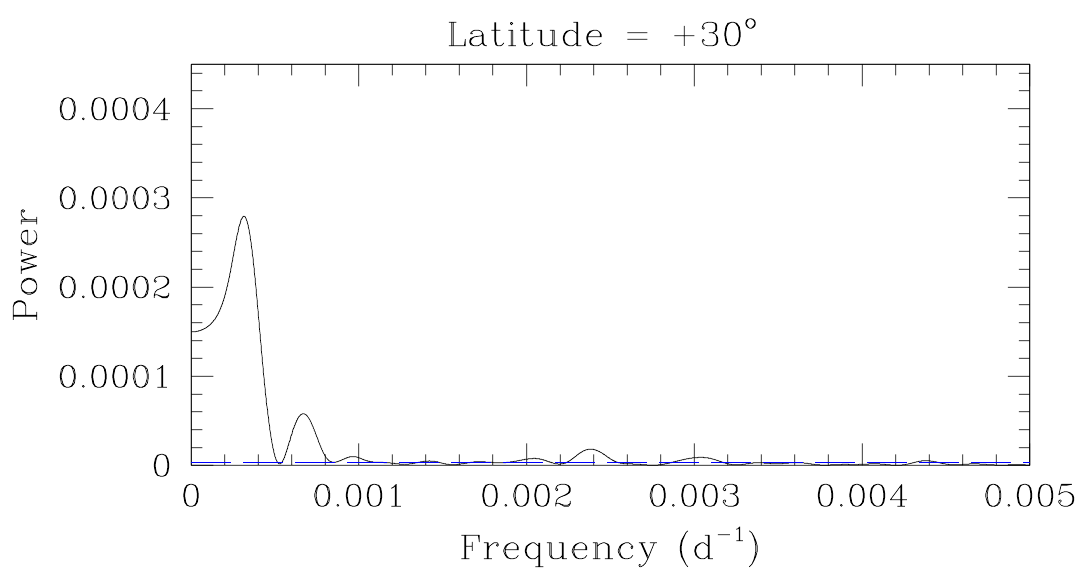}
    \hfill
    \includegraphics[width=0.44\hsize]{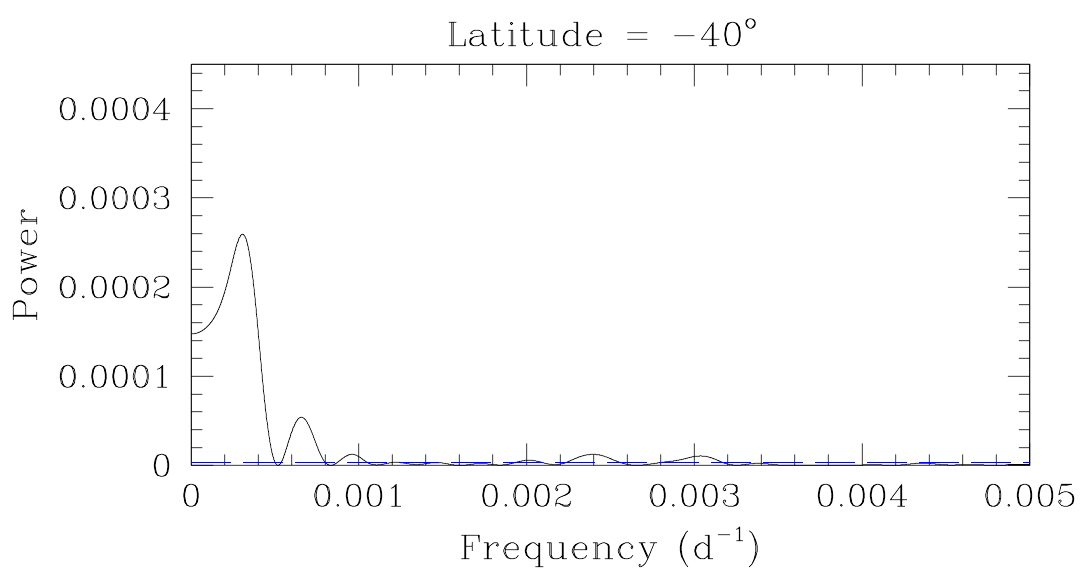}
    \hfill
    \includegraphics[width=0.44\hsize]{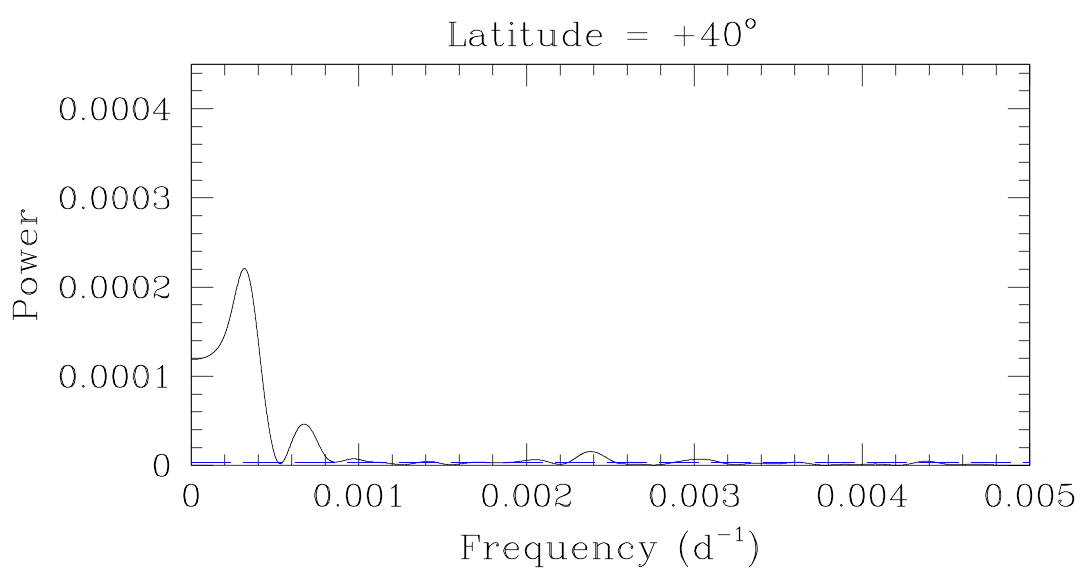}
    \caption{Evolution of the Fourier power spectrum near the frequency of solar activity cycle with inclinations specified at the top of the plot representing the number of degrees relative to the Equator-on ($i = 0^{\circ}$). Left panels: inclinations to the South Pole view ; Right panels: inclinations to the North Pole view.}
\end{figure}

\begin{figure}[h]
    \ContinuedFloat
    \includegraphics[width=0.44\hsize]{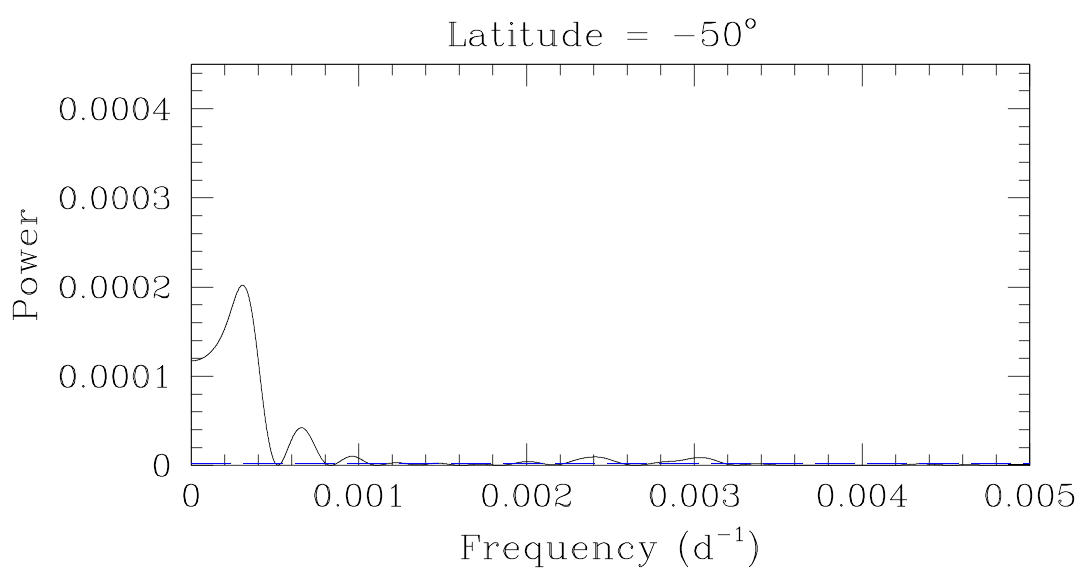}
    \hfill
    \includegraphics[width=0.44\hsize]{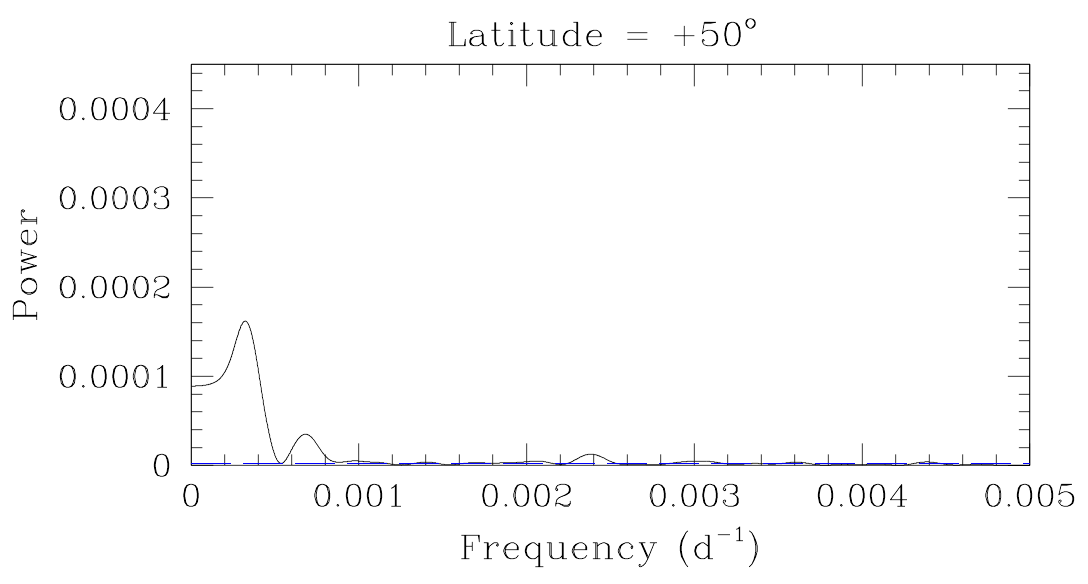}
    \hfill
    \includegraphics[width=0.44\hsize]{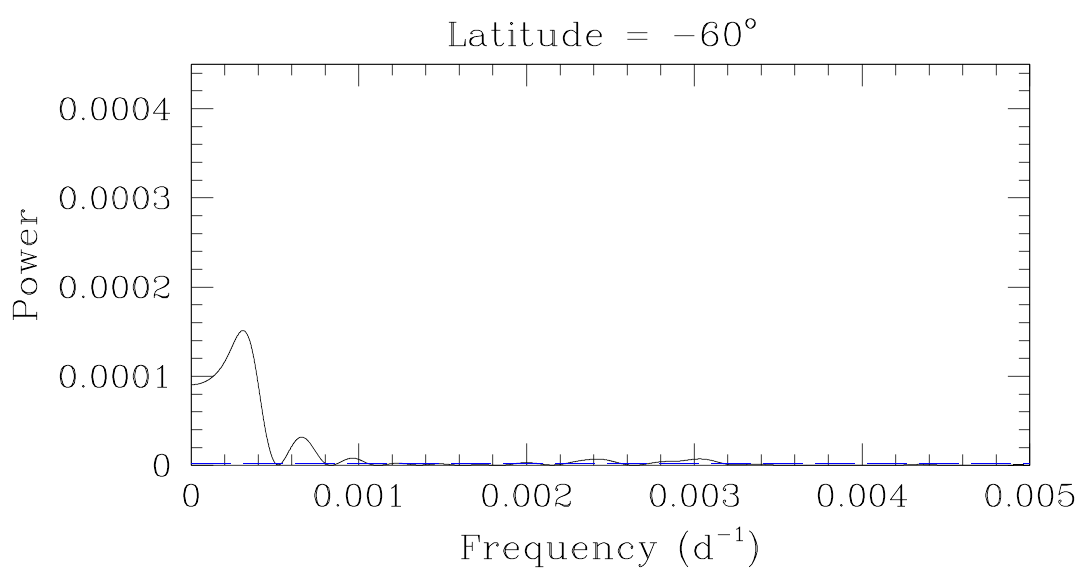}
    \hfill
    \includegraphics[width=0.44\hsize]{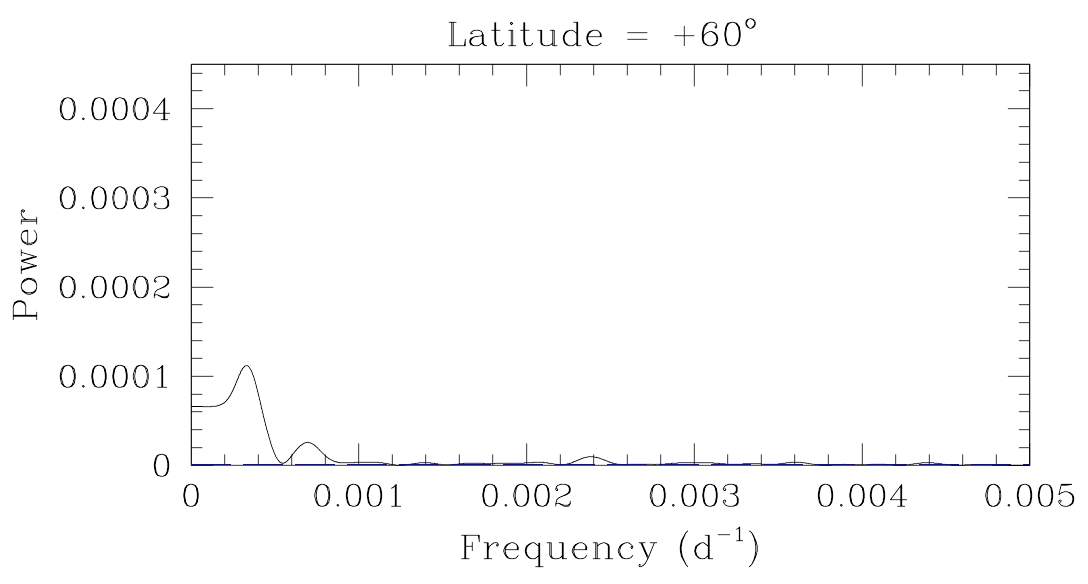}
    \hfill
    \includegraphics[width=0.44\hsize]{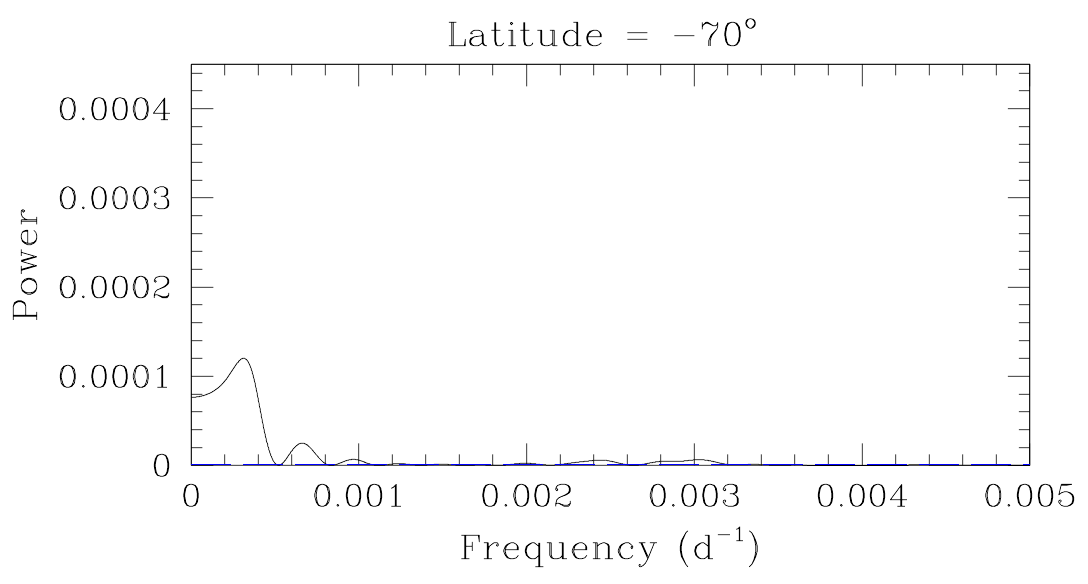}
    \hfill
    \includegraphics[width=0.44\hsize]{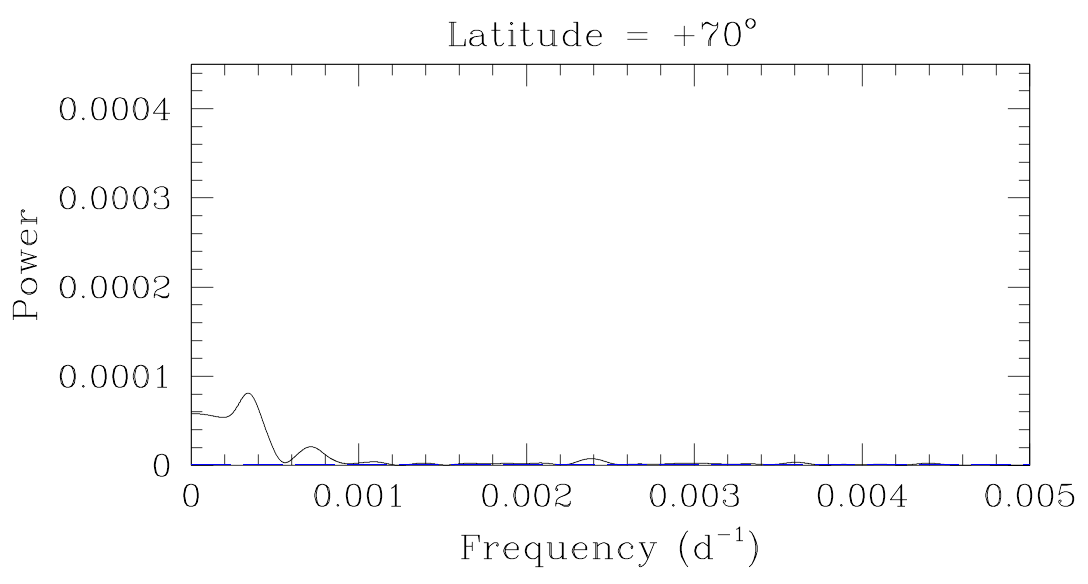}
    \hfill
    \includegraphics[width=0.44\hsize]{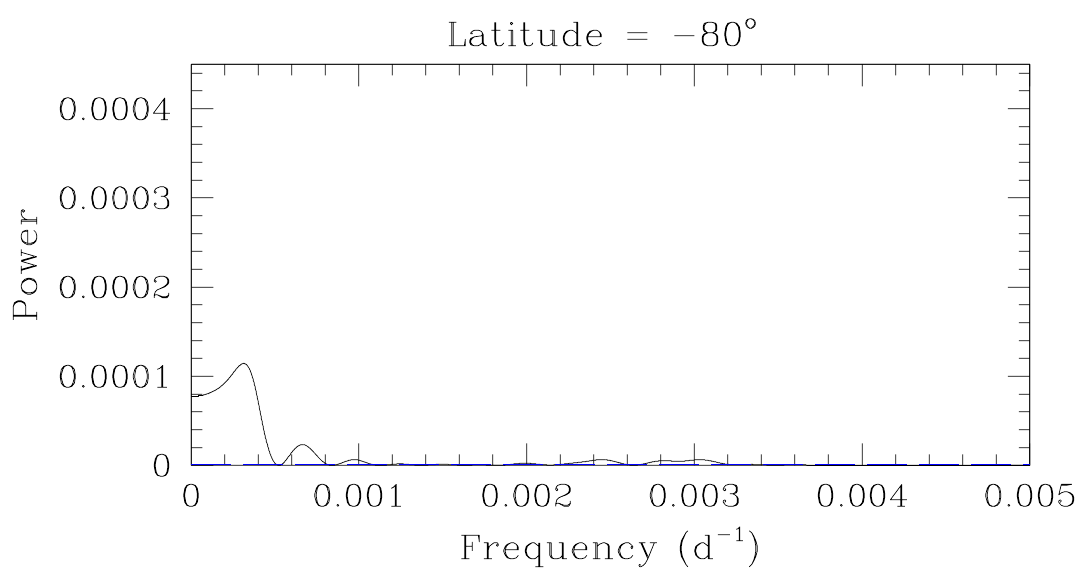}
    \hfill
    \includegraphics[width=0.44\hsize]{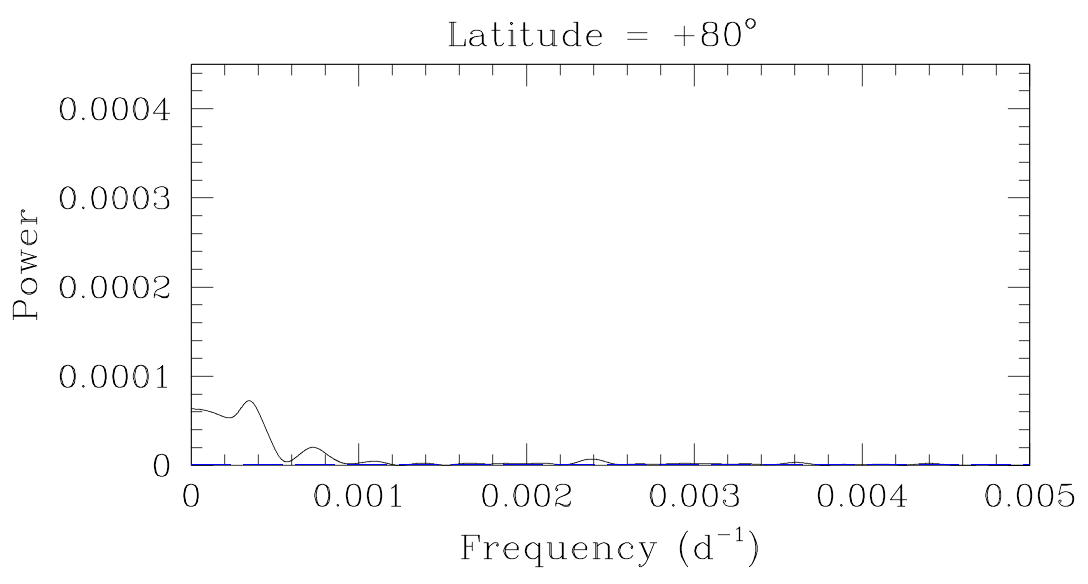}
    \hfill
    \includegraphics[width=0.44\hsize]{images/Fourier/cyc90degs.pdf}
    \hfill
    \includegraphics[width=0.44\hsize]{images/Fourier/cyc90degn.pdf}
    \caption{Continued}
    \label{fig:Fourier_cyc}
\end{figure}

    \onecolumn

\section{Fourier power spectrum of the resampled USET time series assuming 206 observations spread over ten years for different viewing angles}\label{Annexe_E}

\begin{figure}[h]
    \vspace{-0.6cm}
    \centering
    \includegraphics[width=0.4\hsize]{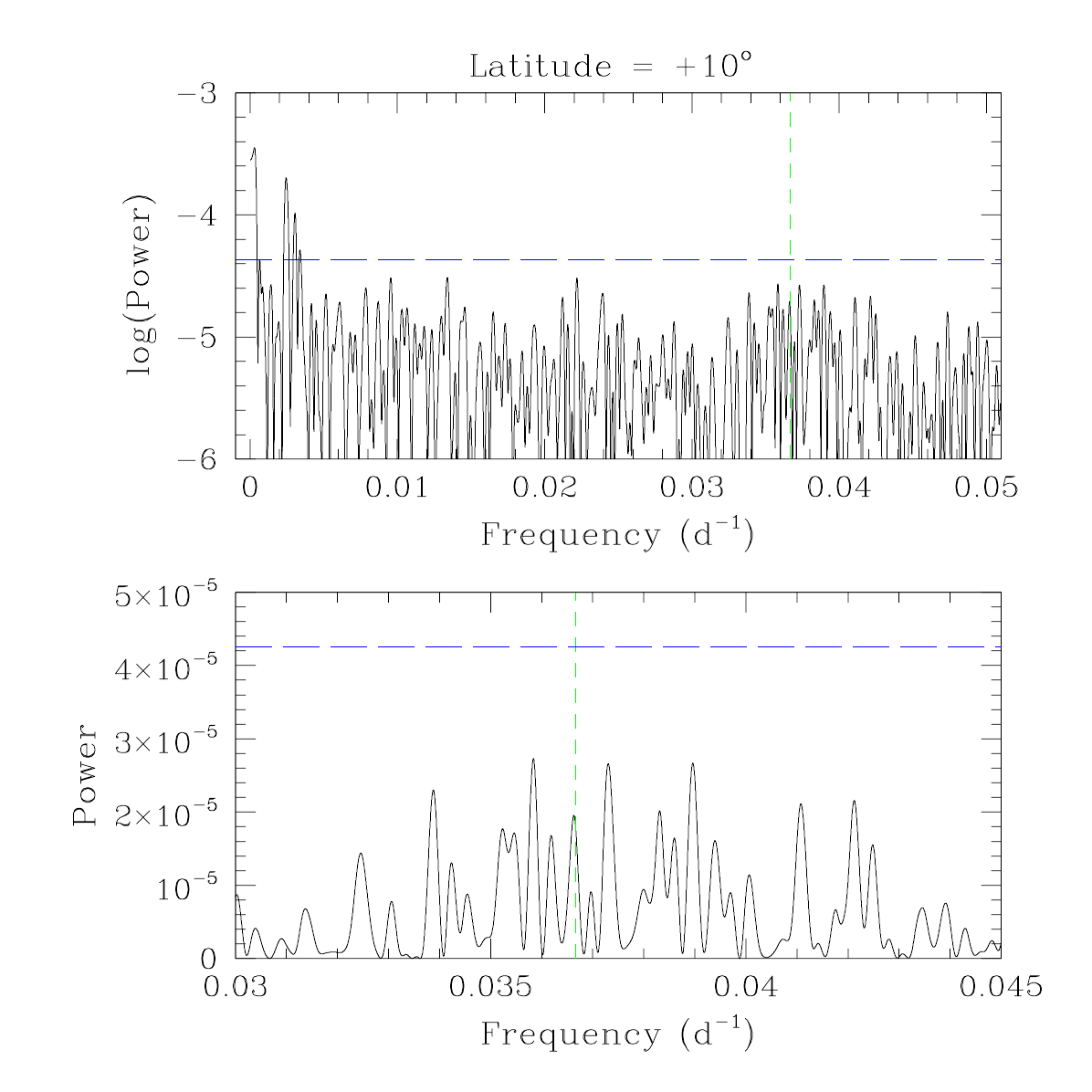}
    \includegraphics[width=0.4\hsize]{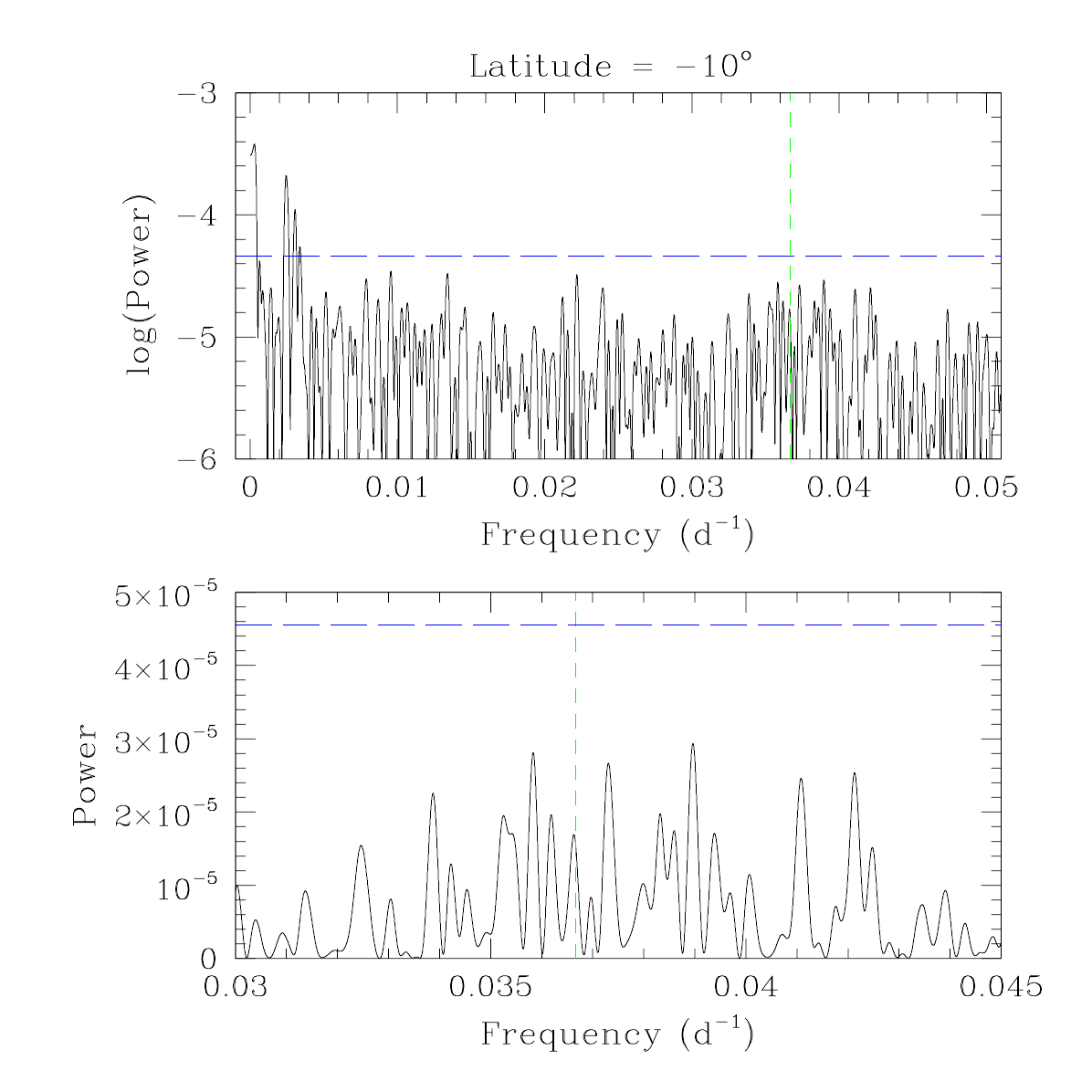}
    \includegraphics[width=0.4\hsize]{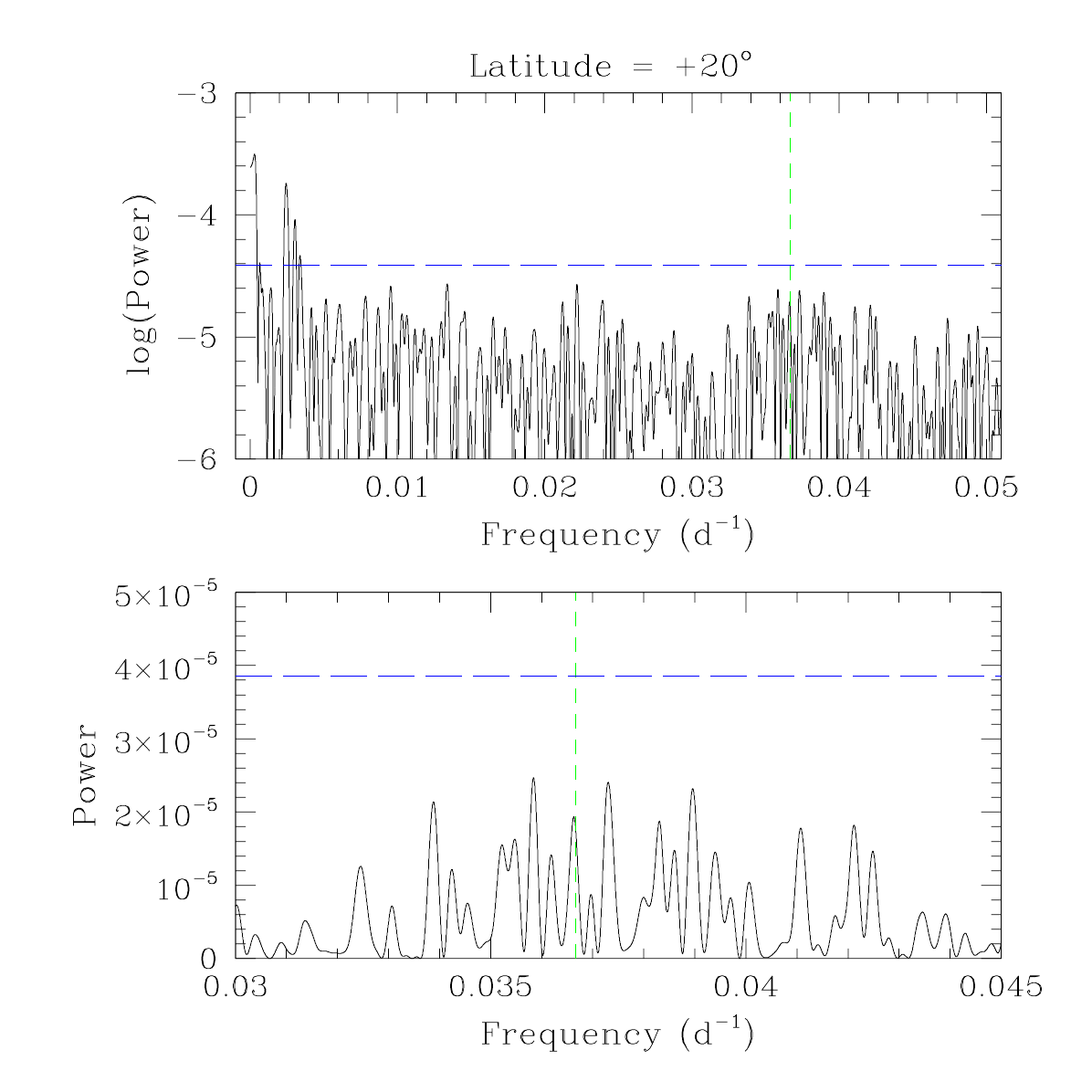}
    \includegraphics[width=0.4\hsize]{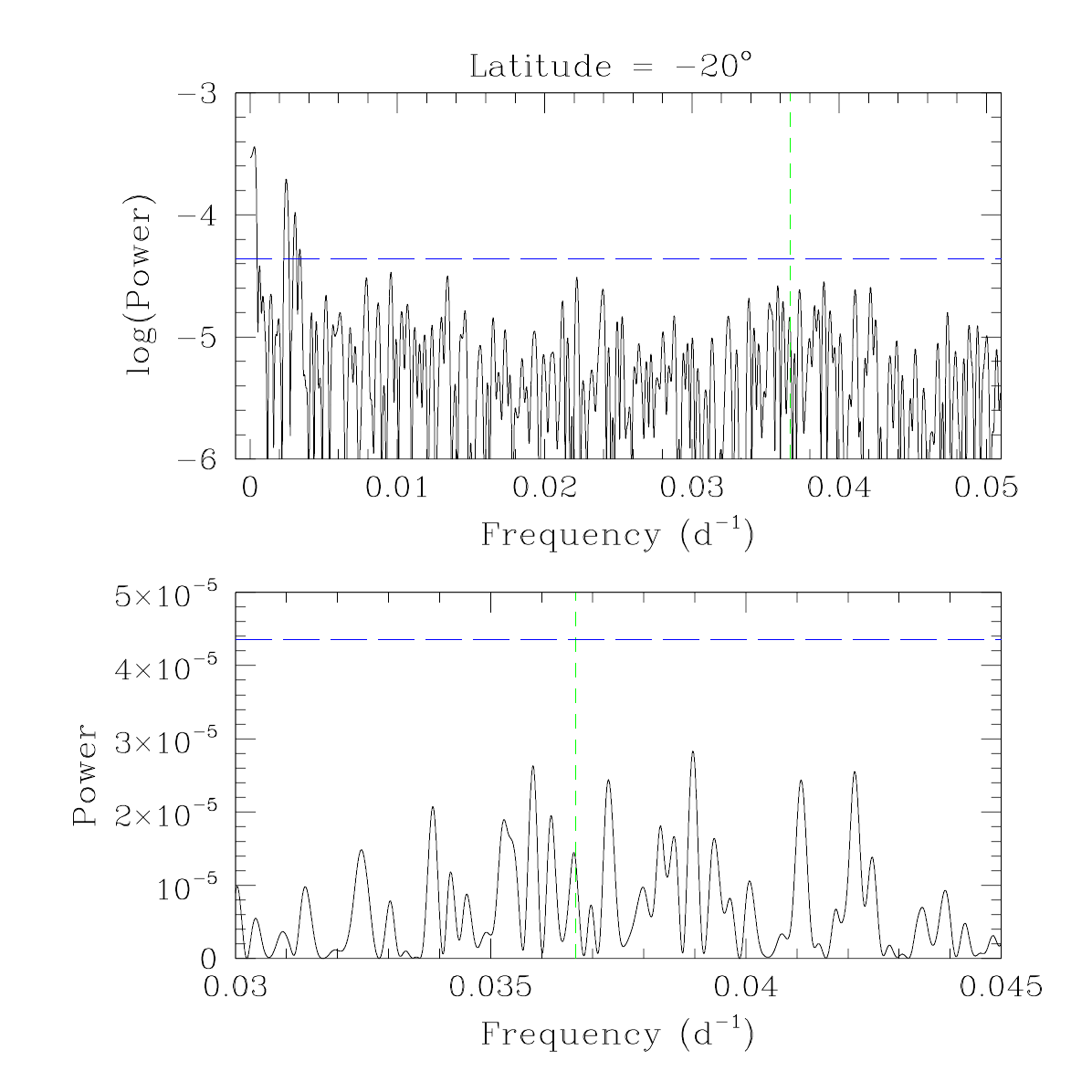}
    \includegraphics[width=0.4\hsize]{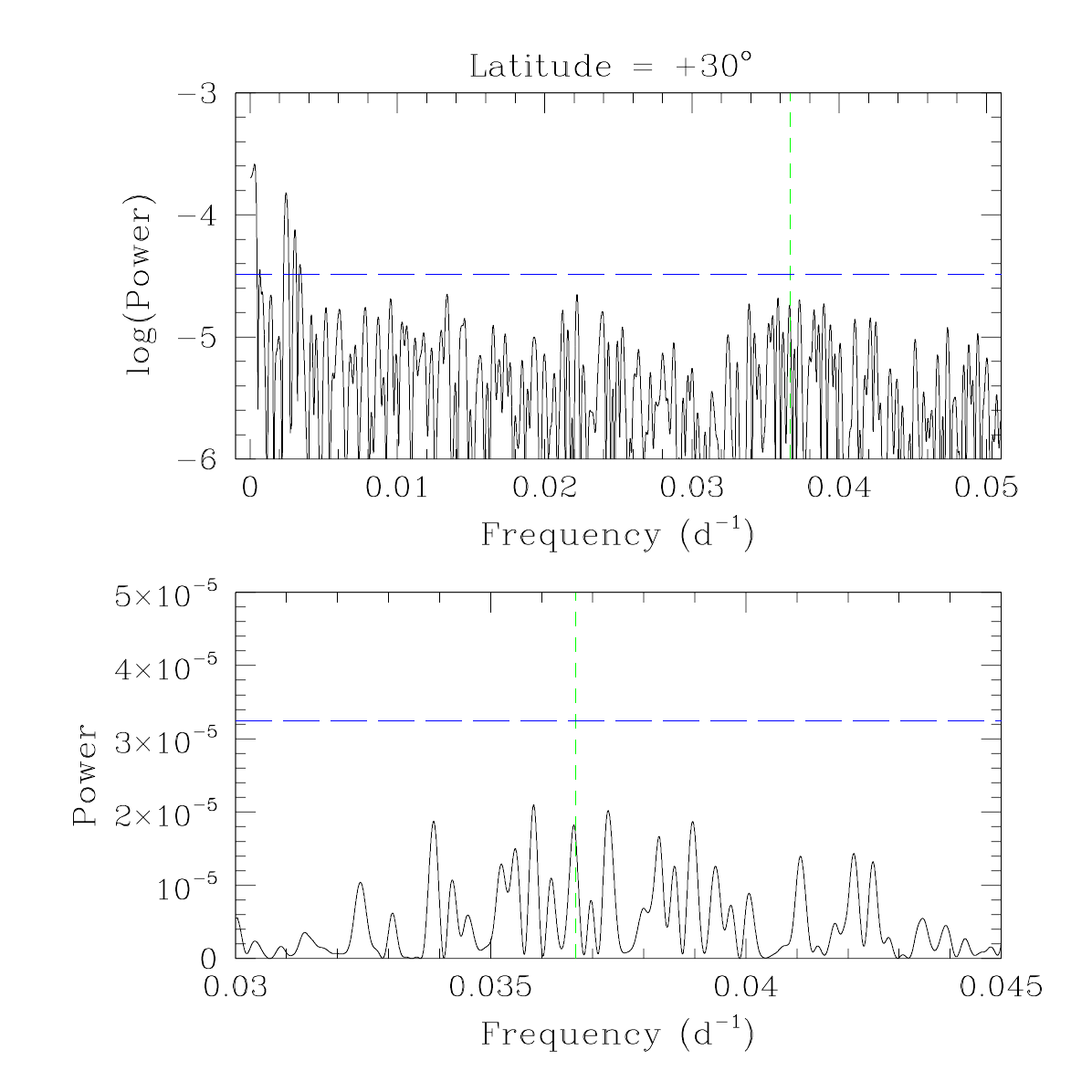}
    \includegraphics[width=0.4\hsize]{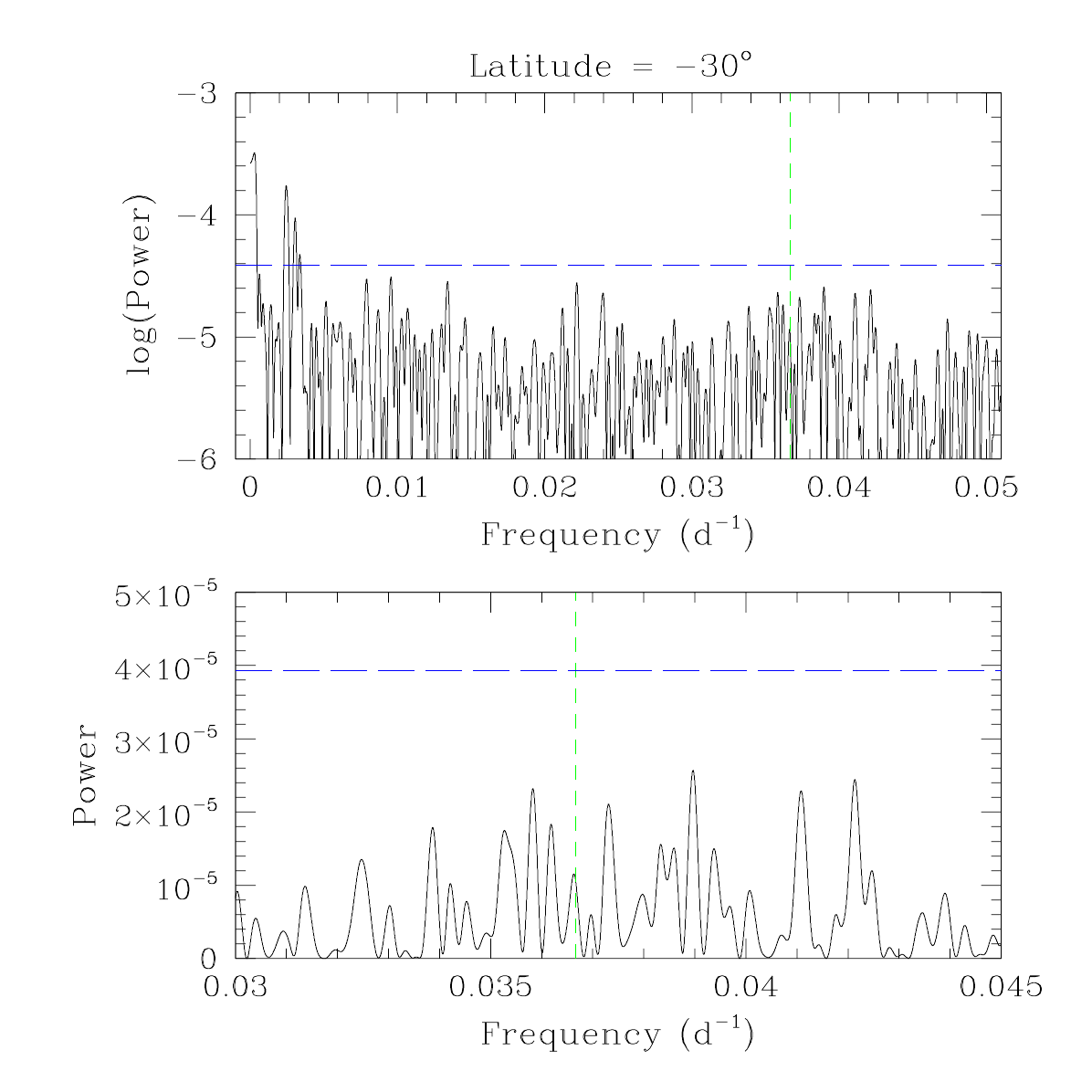}
    \caption{The top panels illustrate the logarithm of the power spectrum for frequencies below 0.05 d$^{-1}$. The bottom panels provide a zoom on the power around $\nu_{\rm Car}$ (given by the short-dashed green vertical line. The long-dashed blue horizontal line yields the 99\% significance level.}
    \label{fig:starpower1}
\end{figure}

\begin{figure}[h]
    \ContinuedFloat
    \centering
    \includegraphics[width=0.4\hsize]{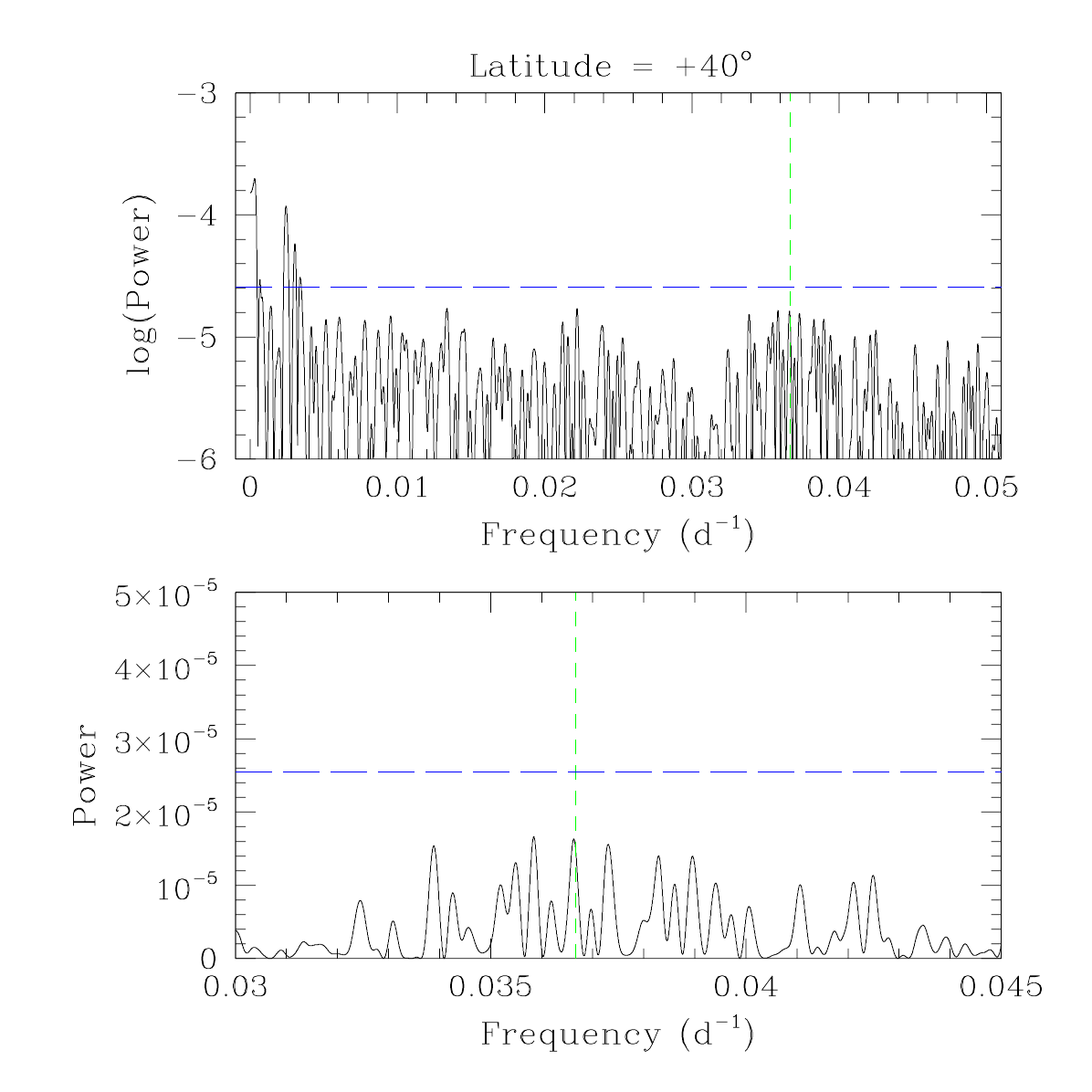}
    \includegraphics[width=0.4\hsize]{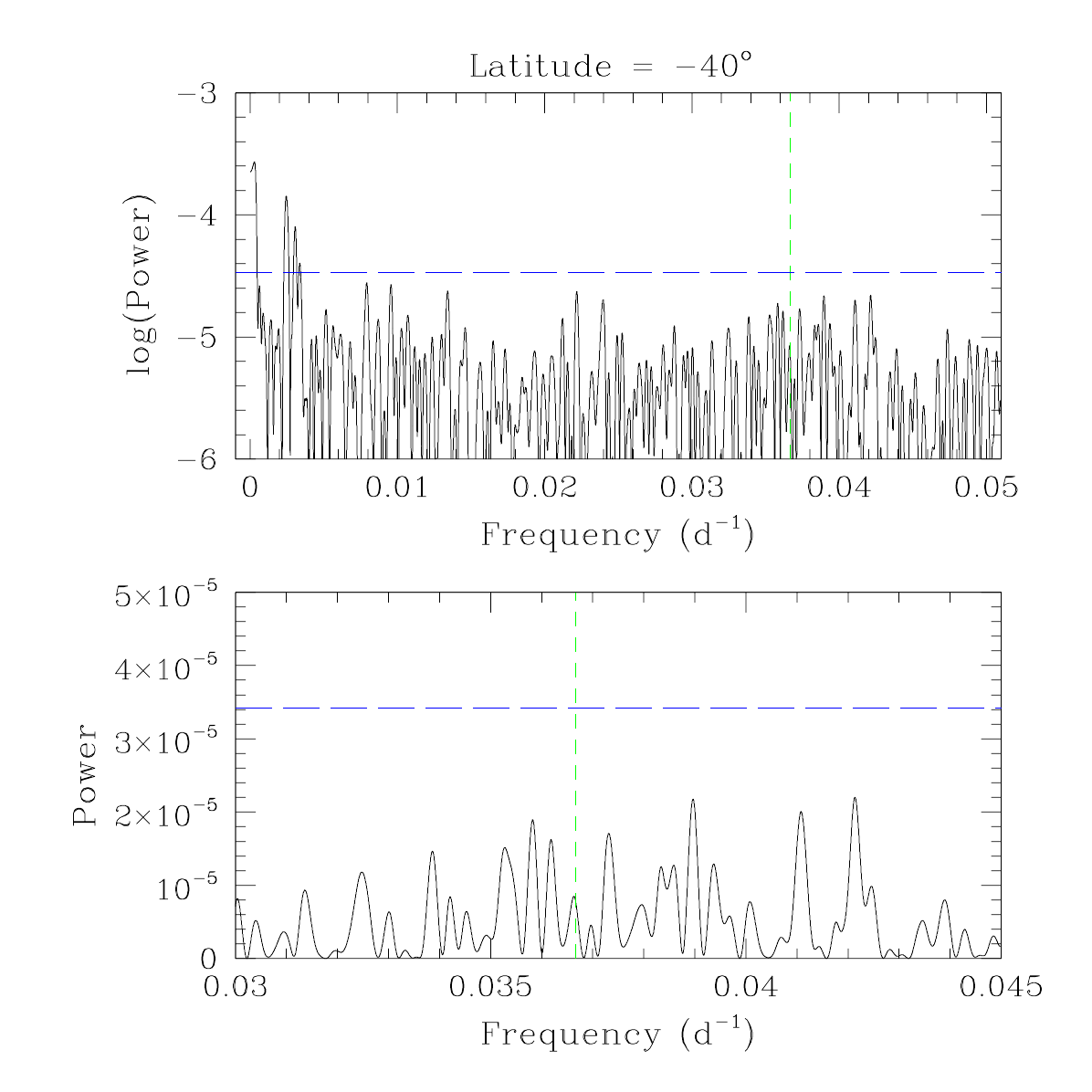}  
    \includegraphics[width=0.4\hsize]{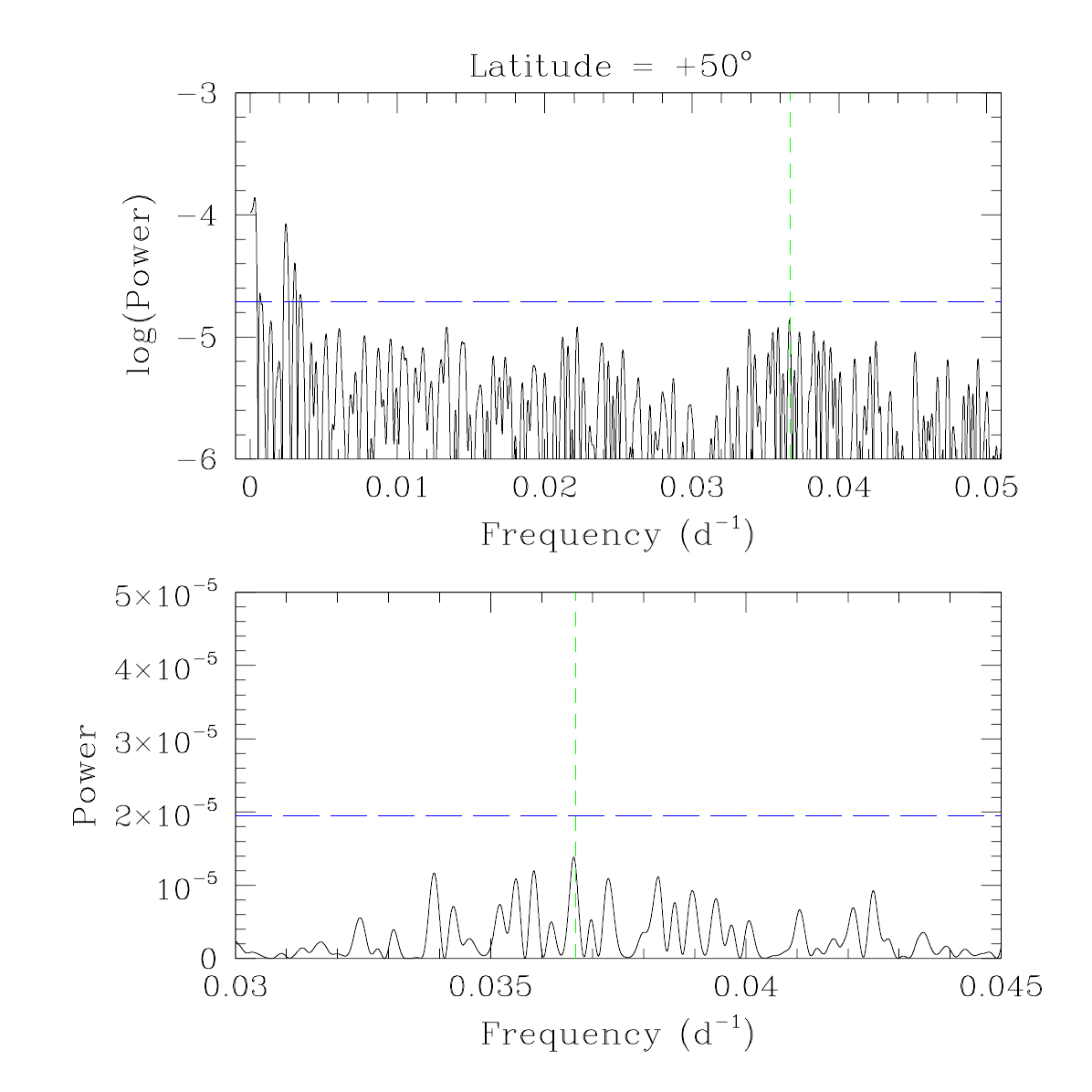}
    \includegraphics[width=0.4\hsize]{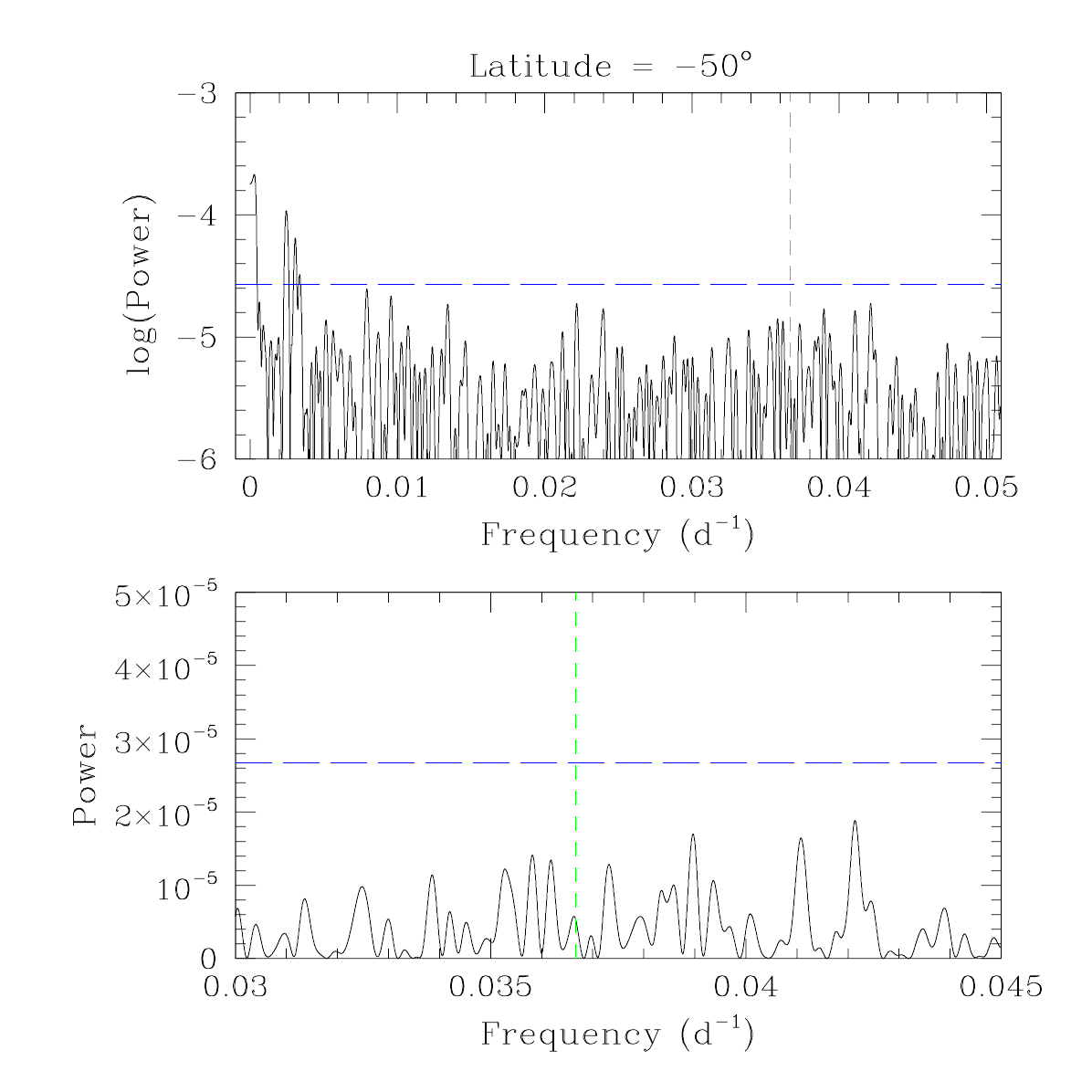}
    \includegraphics[width=0.4\hsize]{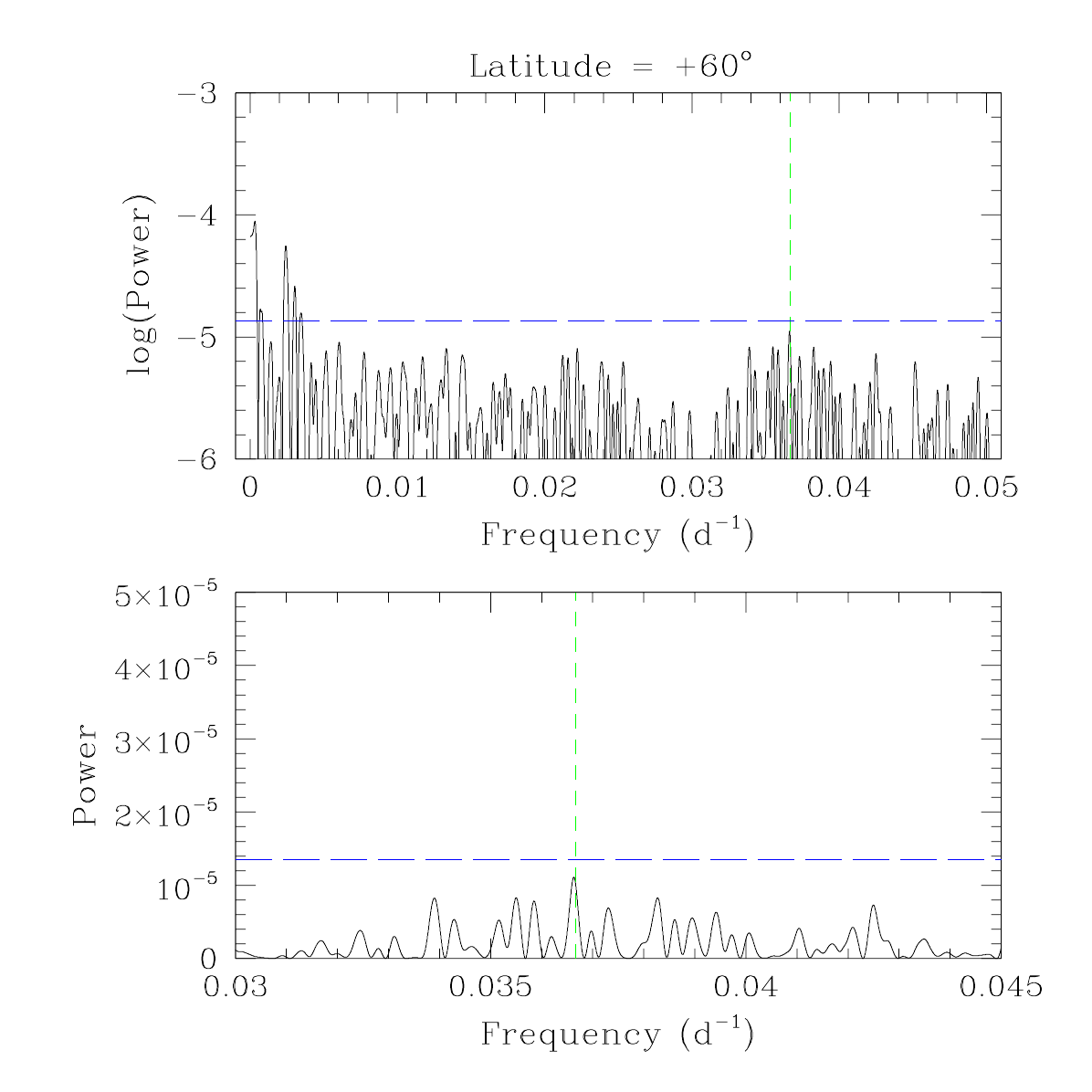}
    \includegraphics[width=0.4\hsize]{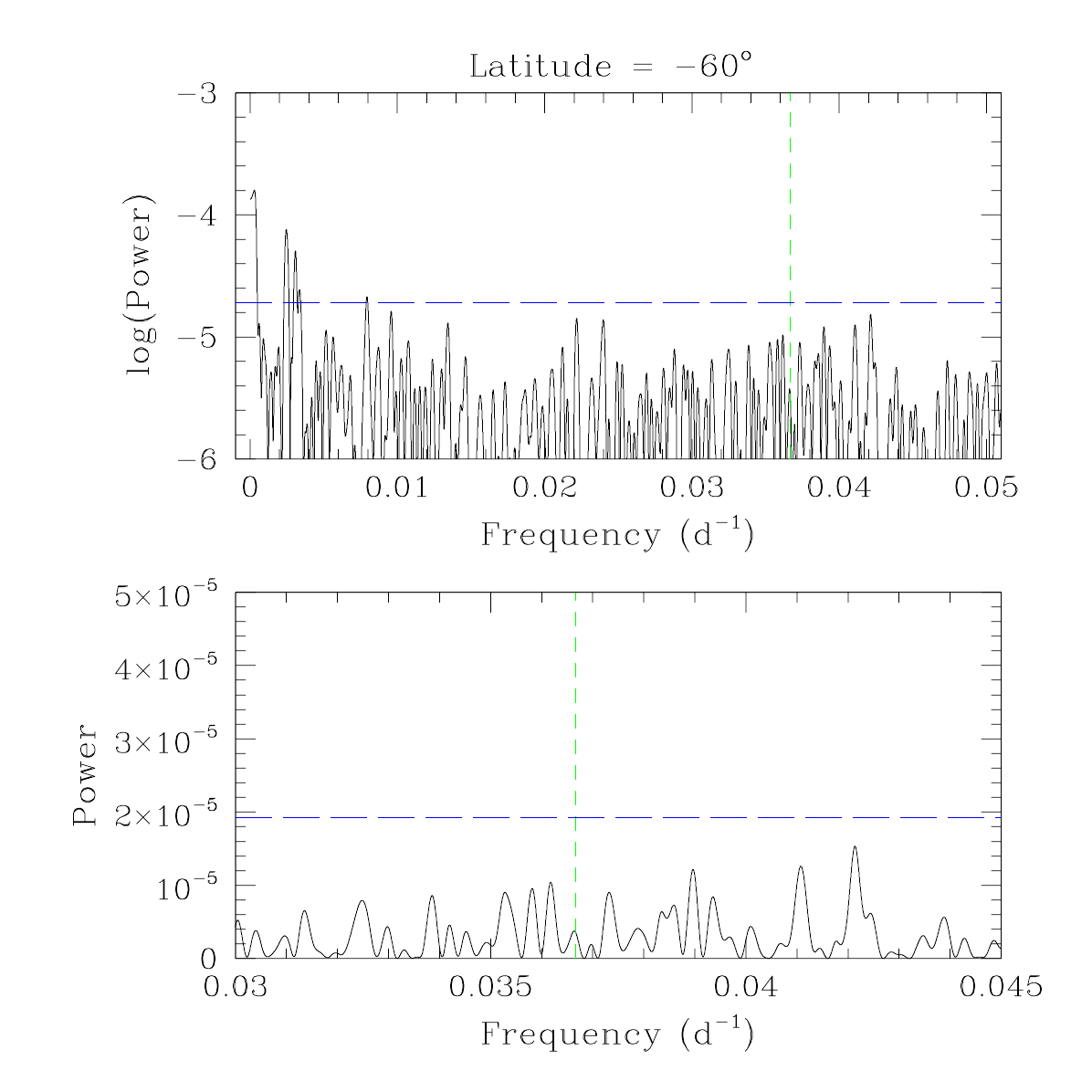}
    \caption{Continued.}
\end{figure}

\begin{figure}[h]
    \ContinuedFloat
    \centering
    \includegraphics[width=0.4\hsize]{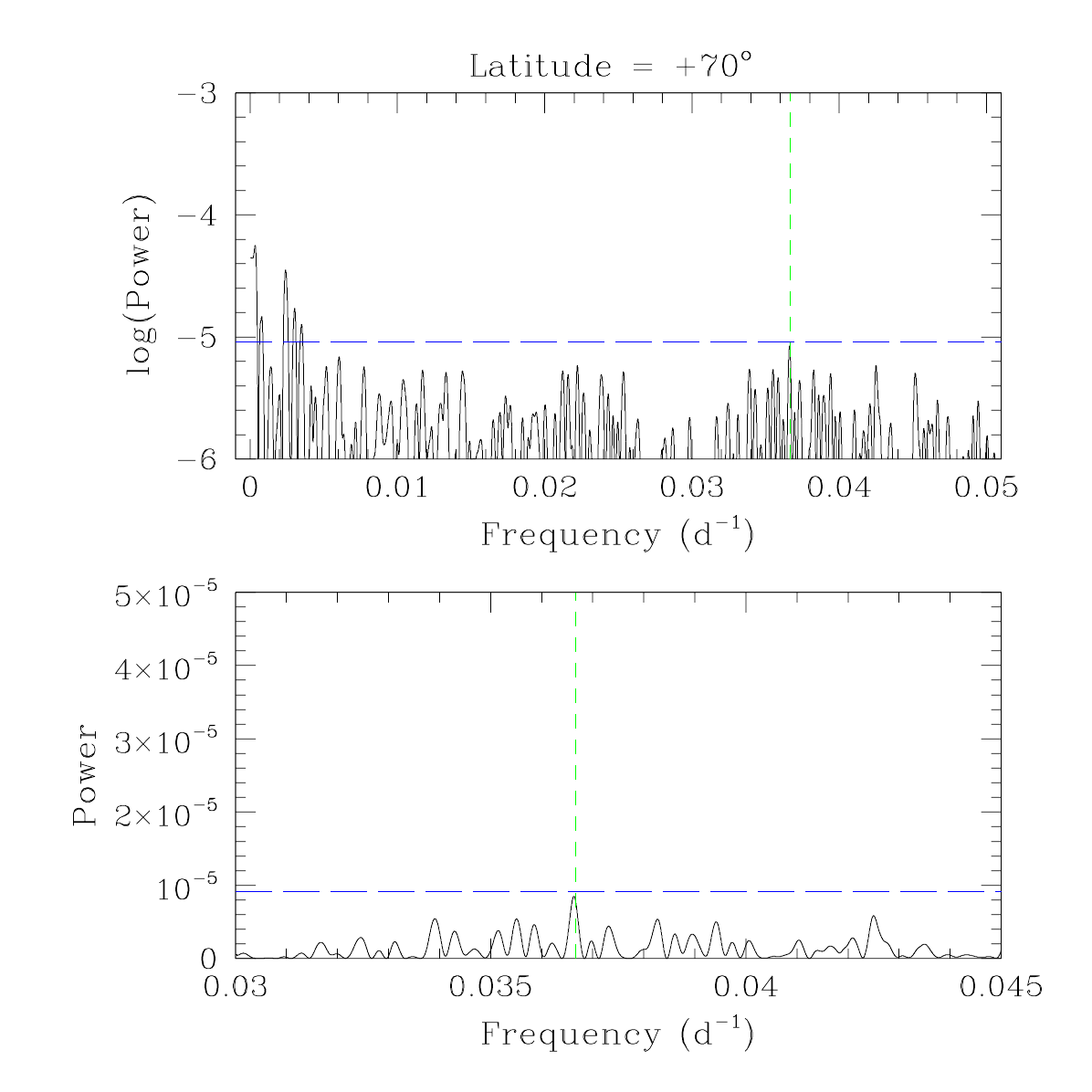}
    \includegraphics[width=0.4\hsize]{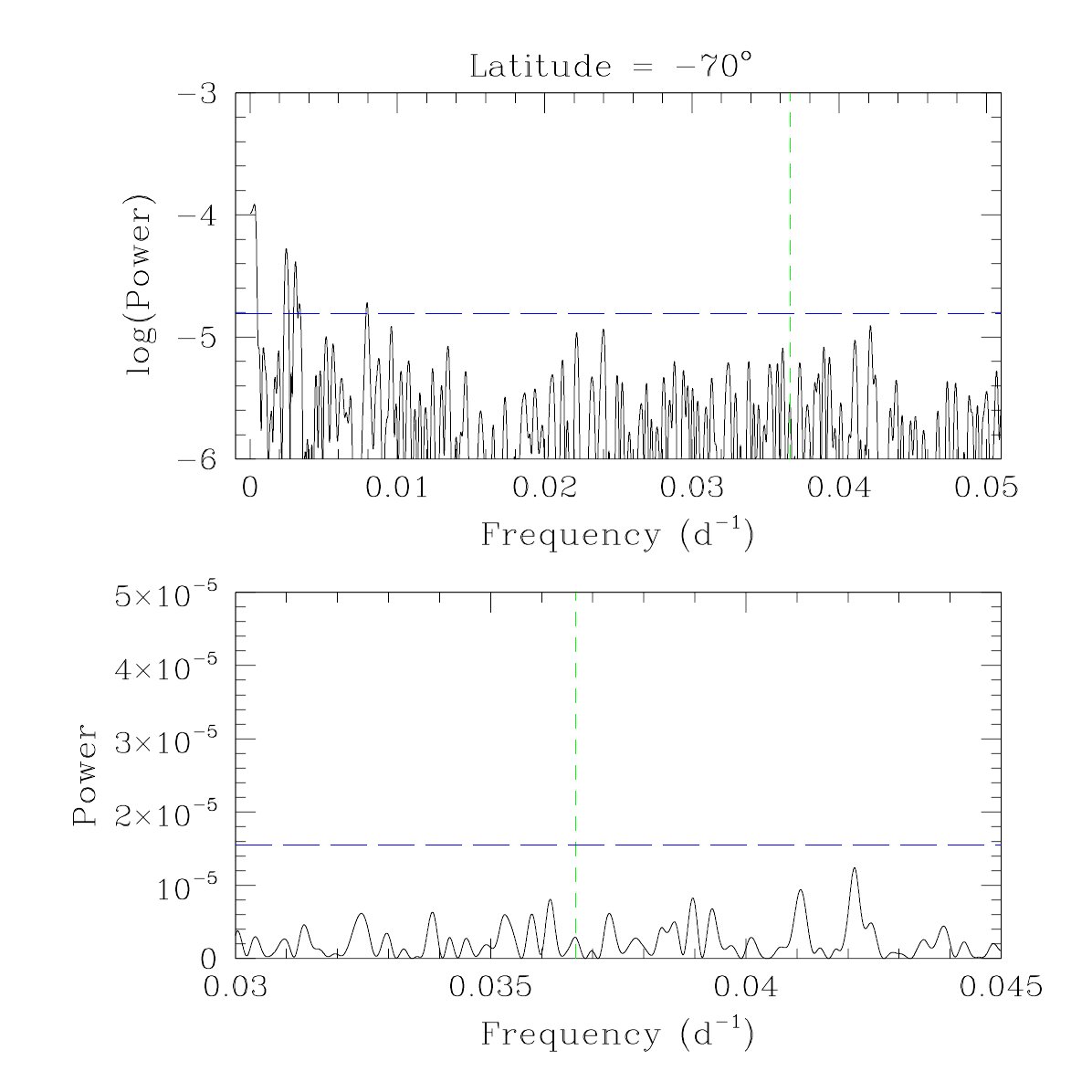}
    \includegraphics[width=0.4\hsize]{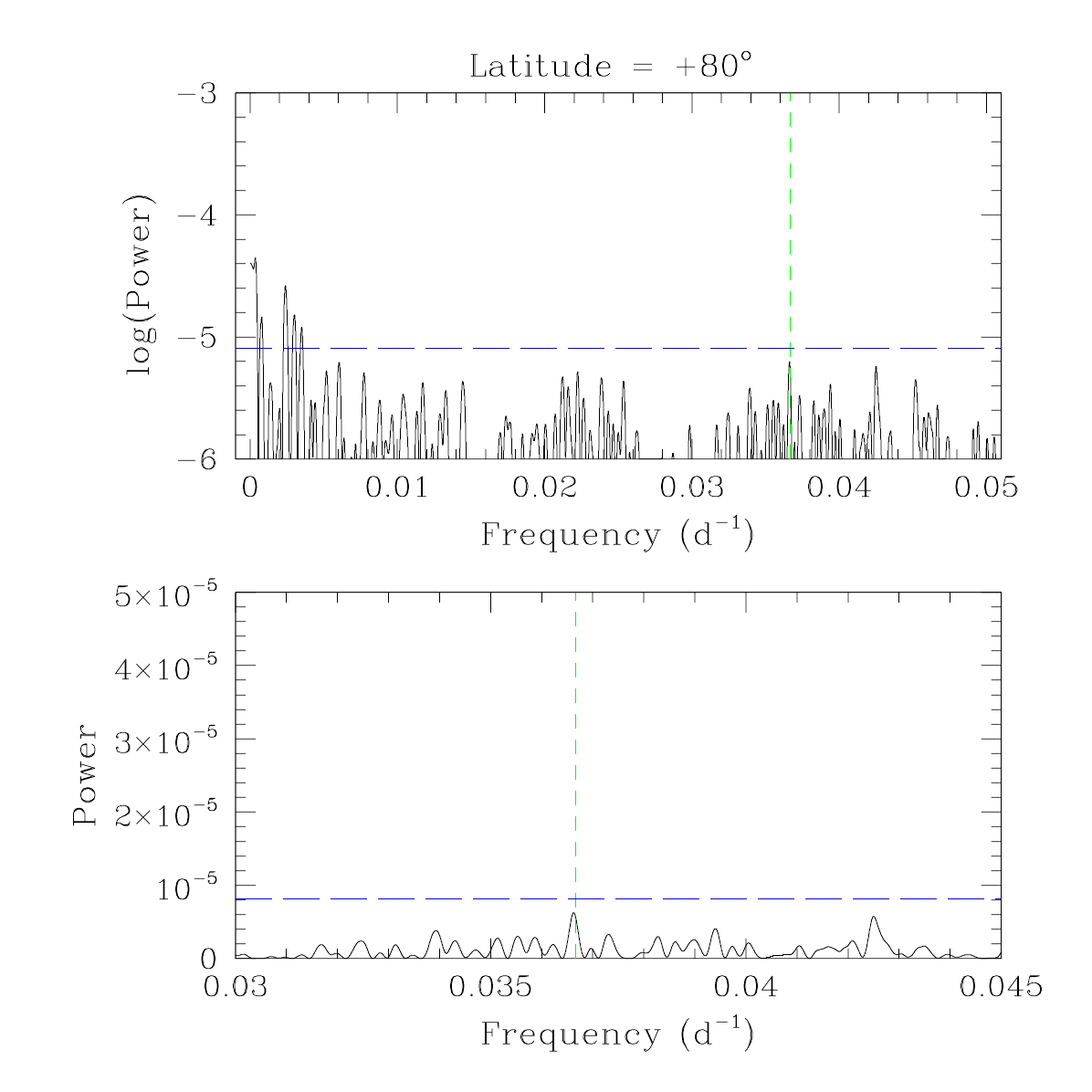}
    \includegraphics[width=0.4\hsize]{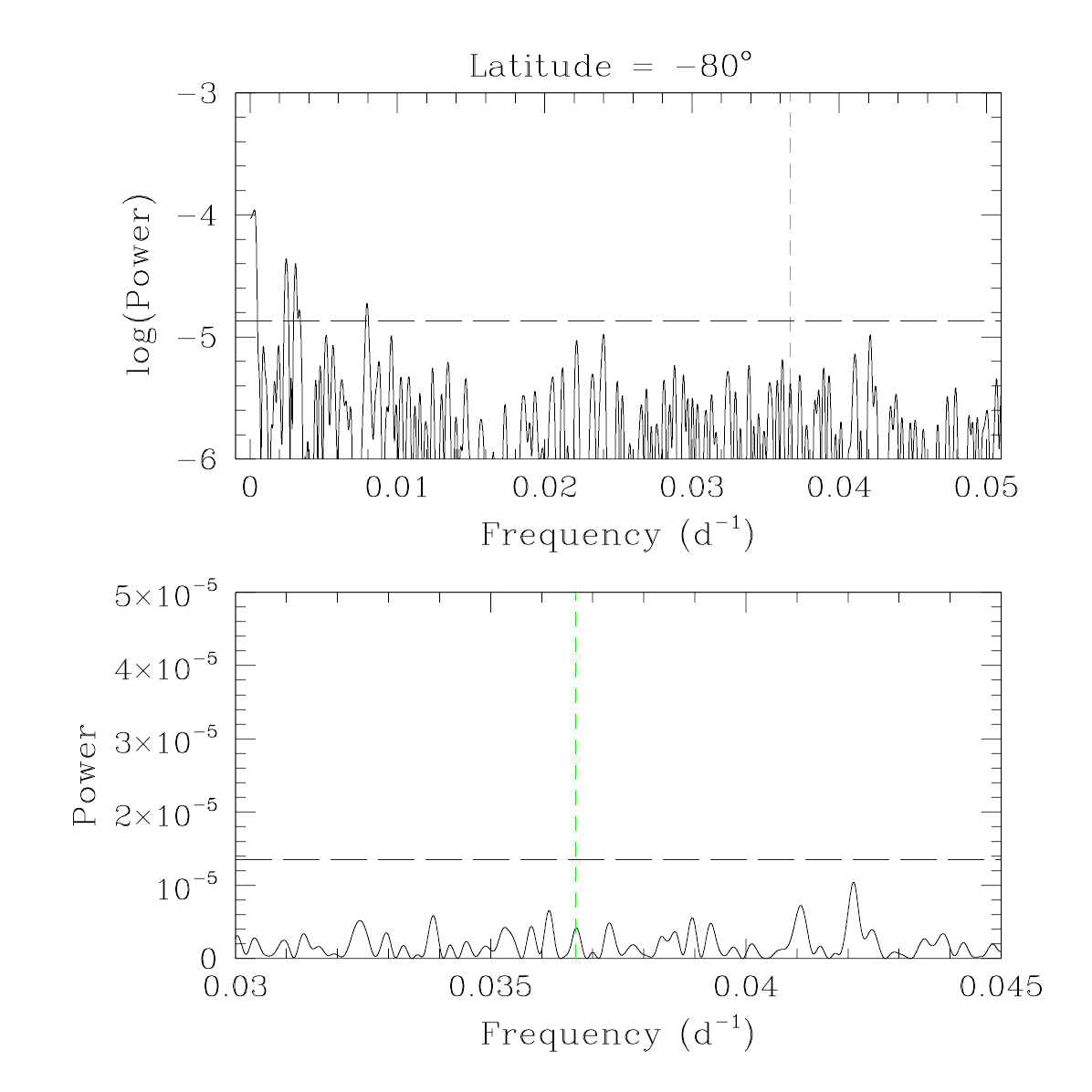}
    \includegraphics[width=0.4\hsize]{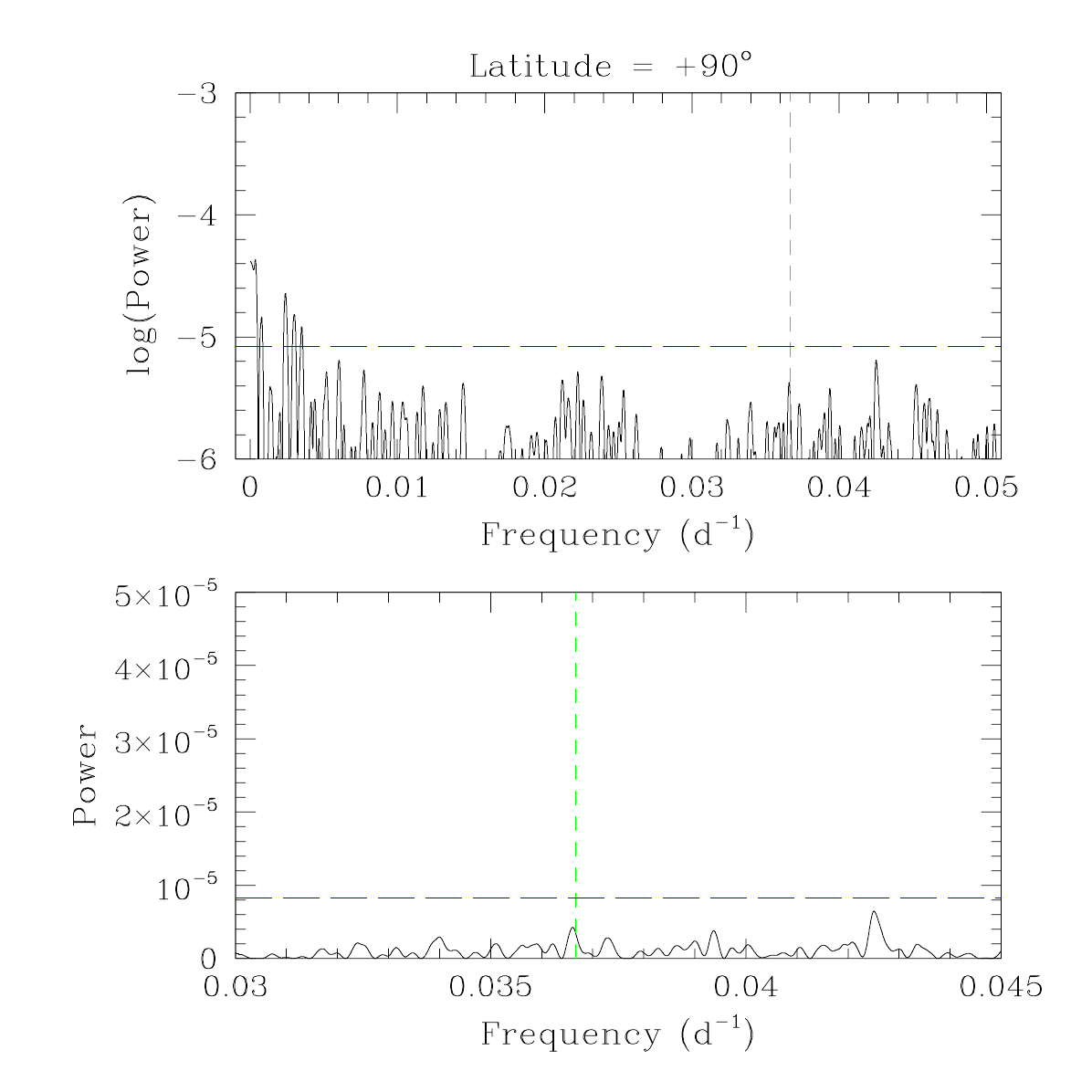}
    \includegraphics[width=0.4\hsize]{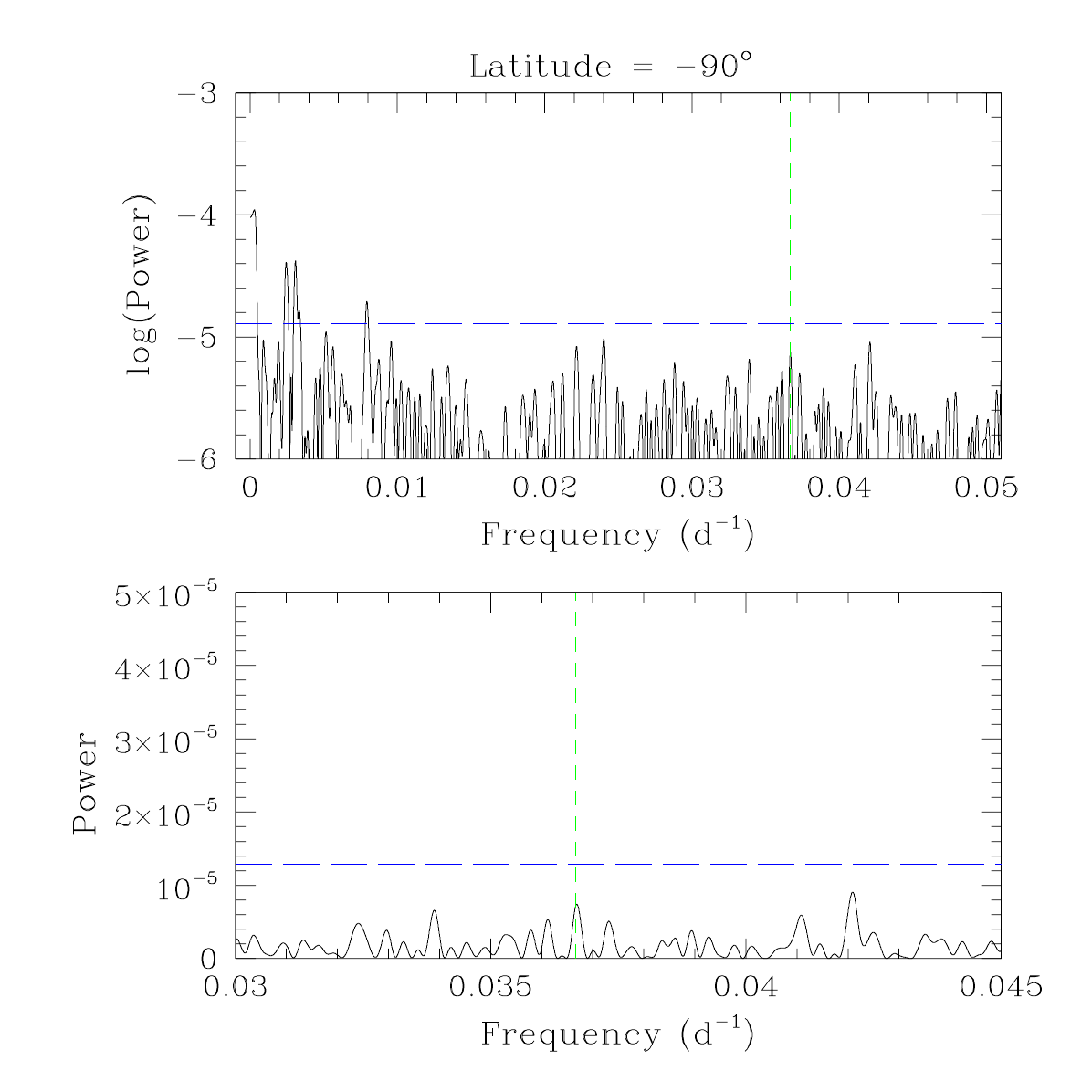}
    \caption{Continued.}
\end{figure}
\end{appendix}

\end{document}